\documentclass[fleqn,usenatbib]{mnras}
\usepackage{newtxtext,newtxmath}
\usepackage[T1]{fontenc}
\DeclareRobustCommand{\VAN}[3]{#2}
\let\VANthebibliography\thebibliography
\def\thebibliography{\DeclareRobustCommand{\VAN}[3]{##3}\VANthebibliography}

\usepackage{graphicx}	

\newcommand{\MSUN}{{\rm M}_{\sun}}

\title[Quenched fractions, gas content \& SFHs of TNG50 MW/M31-like satellites]{Satellites of Milky Way- and M31-like galaxies with TNG50: quenched fractions, gas content, and star formation histories}

\author[C. Engler et al.]{Christoph Engler$^{1,2}$\thanks{E-mail: engler@mpia.de}, 
Annalisa Pillepich$^{1}$, Gandhali D. Joshi$^{3,1}$, Anna Pasquali$^{2}$, Dylan Nelson$^{4}$, 
\newauthor 
and Eva K. Grebel$^{2}$ 
\\
$^{1}$Max-Planck-Institut f\"{u}r Astronomie, K\"{o}nigstuhl 17, 69117 Heidelberg, Germany\\
$^{2}$Astronomisches Rechen-Institut, Zentrum f\"{u}r Astronomie der Universit\"{a}t Heidelberg, M\"{o}nchhofstra\ss e 12-14, 69120 Heidelberg, Germany\\
$^{3}$Department of Physics and Astronomy, University College London, Gower Street, London WC1E 6BT, UK\\
$^{4}$Institut f\"{u}r theoretische Astrophysik, Zentrum f\"{u}r Astronomie der Universit\"{a}t Heidelberg, Albert-Ueberle-Stra\ss e 2, 69120 Heidelberg, Germany
}

\date{Accepted 2023 April 30. Received 2023 April 26; in original form 2022 June 17}

\pubyear{2023}

\begin{document}
\label{firstpage}
\pagerange{\pageref{firstpage}--\pageref{lastpage}}
\maketitle

\begin{abstract}
We analyse the quenched fractions, gas content, and star formation histories of $\sim$1200 satellite galaxies with $M_* \geq 5 \times 10^6~\MSUN$ around 198 Milky Way- (MW) and Andromeda-like (M31) hosts in TNG50, the highest-resolution simulation of IllustrisTNG. Satellite quenched fractions are larger for smaller masses, for smaller distances to their host galaxy, and in the more massive M31-like compared to MW-like hosts. As satellites cross their host’s virial radius, their gas content drops: most satellites within $300~{\rm kpc}$ lack detectable gas reservoirs at $z = 0$, unless they are massive like the Magellanic Clouds and M32. Nevertheless, their stellar assembly exhibits a large degree of diversity. On average, the cumulative star formation histories are more extended for brighter, more massive satellites with a later infall, and for those in less massive hosts. Based on these relationships, we can even infer infall periods for observed MW and M31 dwarfs: e.g.~$0-4~{\rm Gyr}$ ago for the Magellanic Clouds and Leo~I, $4-8$ and $0-2~{\rm Gyr}$ ago for M32 and IC~10, respectively. Ram pressure stripping (in combination with tidal stripping) deprives TNG50 satellites of their gas reservoirs and ultimately quenches their star formation, even though only a few per cent of the present-day satellites around the 198 TNG50 MW/M31-like hosts appear as jellyfish. The typical time since quenching for currently quenched TNG50 satellites is $6.9\substack{+2.5\\-3.3}$~Gyr ago. The TNG50 results are consistent with the quenched fractions and stellar assembly of observed MW and M31 satellites, however, satellites of the SAGA survey with $M_* \sim 10^{8-9}~\MSUN$ exhibit lower quenched fractions than TNG50 and other, observed analogues.
\end{abstract}




\section{Introduction}
\label{sec:intro}

While environment affects satellite galaxies in hosts across the mass spectrum, from massive galaxy clusters to group hosts, the Local Group (LG) is thought to have left a particularly strong imprint on its satellite population. Low-mass dwarf galaxies with stellar masses of $M_* < 10^8~{\rm M}_\odot$ found within the virial radius of both the Milky Way (MW) or Andromeda (M31) exhibit a distinct transition compared to their analogues at larger distances regarding many properties: their morphologies are more spheroidal, they contain little atomic/neutral gas, and their star formation activity is mostly shut down \citep{Einasto1974, Grcevich2009, McConnachie2012, Slater2014, Phillips2015, Wetzel2015}. There is a clear correlation between the gas content of dwarfs and the distance to their host galaxy: more isolated dwarfs in the outskirts are gas-richer \citep{Blitz2000, Grebel2003, Spekkens2014, Putman2021}. Isolated dwarfs outside of the LG further support this picture by exhibiting a larger gas content and more active star formation \citep{Bradford2015, Stierwalt2015}.

As galaxies become satellites of a more massive host, a range of environmental effects are thought to start acting on them and on their mass components. Galaxy harassment by high-velocity encounters \citep{Moore1996, Moore1998} or tidal shocking at pericentric passages \citep{Gnedin1999, Mayer2007, Joshi2020} may drive morphological transformation. Tidal stripping by the host's gravitational potential removes mass from the outside in: from the dark matter haloes surrounding the satellite galaxies to the gas they contain and to the stars in their stellar halo or ultimately in their main body, which may result in features like tidal tails or even their complete disruption \citep{Merritt1983, Barnes1992}. After infall, satellites typically cease to form new stars and become quenched due to the loss of their reservoirs of cold, star-forming gas. The accretion of new gas may simply be cut off resulting in a starvation scenario, in which the satellite galaxy exhausts its remaining gas reservoirs over an extended period of time \citep{Larson1980, Balogh2000, Kawata2008, Wetzel2013}. Ram pressure stripping, on the other hand, may quench satellites more rapidly as their cold, star-forming gas is driven out while they pass through the intracluster or intragroup medium of their host environment \citep{GunnGott1972, Tonnesen2007, Bekki2014, Fillingham2016, Simpson2018, Yun2019}.

In order to learn more about the star formation processes of dwarfs and their build-up of stellar mass, cumulative star formation histories (SFH) have been reconstructed in numerous studies. In observations with resolved stellar populations, SFHs can be derived directly from colour-magnitude diagrams of individual stars. By looking for the oldest main-sequence turn-off stars, several single age stellar populations can be separated from each other \citep{Dolphin2002, Brown2014, Skillman2017, Cignoni2018, Sacchi2018}. Comparisons of the SFHs of dwarfs within and outside the LG in the nearby Universe have emphasised the impact their environment has on their evolution. While individual SFHs exhibit a large diversity, the mean SFHs of dwarfs of different morphological types are -- for the most part -- remarkably similar \citep{Weisz2011a}. Differences are mostly found in the last few Gyr of their evolution as they transform from gas-rich to gas-poor. The ACS Nearby Galaxy Survey Treasury (ANGST) has found dwarfs in the LG to be representative of other dwarfs in the local Universe \citep{Weisz2011b}. The diversity of SFHs of satellites in the LG is related to both their own stellar mass and their environment, resulting in differences even between the populations of the MW and M31: less massive dwarfs in higher density environments exhibit a more rapid evolution at earlier times than more massive dwarfs or those in lower-density environments \citep{Weisz2014a, Gallart2015, Weisz2019}.

As it is unclear how representative the MW and M31 or their respective satellite systems are within a cosmological context, observational surveys in recent years have expanded beyond the LG and even the Local Volume. The Exploration of Local VolumE Satellites survey \citep[ELVES,][]{Carlsten2022} has investigated 30~MW-like hosts and their satellite systems within the Local Volume  (i.e.~within $12~{\rm Mpc}$) and found 338~confirmed satellites with 105~more candidates that require further distance measurements. These satellite systems are complete to $M_{\rm V} \lesssim -9~{\rm mag}$ and $\mu_{\rm 0,V} \lesssim 26.5~{\rm mag~arcsec}^{-2}$. Both the general abundance of ELVES satellites and their star formation activity and quenched fractions are consistent with dwarfs around the MW and M31. On the other hand, the Satellites Around Galactic Analogs \citep[SAGA;][]{Geha2017, Mao2021} survey is aiming to construct the first large statistical sample of observed MW-like hosts outside of the Local Volume. The current second stage of SAGA has detected 127~satellites with $M_{\rm r} < - 12.3~{\rm mag}$ around 36~MW analogues. Whereas the satellite abundance of these hosts is consistent with the MW, the satellite quenched fractions are much lower than -- and therefore at odds with -- observations of the LG: most SAGA-II satellites ($> 80$~per~cent) are still actively forming stars. Their star formation activity, estimated in SAGA-II via H$\alpha$, has since been confirmed independently using near- and far-UV data by \citet{Karunakaran2021}, who have hence found an inconsistency with the results of the APOSTLE \citep{Fattahi2016} and Auriga \citep{Grand2017} cosmological simulations of LG- and MW/M31-like systems, respectively.

Cosmological hydrodynamical simulations of galaxy formation and evolution of recent years have indeed been utilised to study the characteristics of the satellite systems of MW analogues: see a summary in table~1 of \citet{Engler2021b}. These include the aforementioned Auriga and APOSTLE simulations, as well as Latte \citep{Wetzel2016} and the other 13~FIRE-2 simulations of MW-mass hosts \citep{Hopkins2018, GarrisonKimmel2019, Wetzel2022}. As a high level of numerical resolution is required to model the low-mass dwarfs that orbit MW-like galaxies, both the host sample size and the satellite mass range have been limited. More recently, the ARTEMIS simulations have produced a sample of 45~MW-like hosts \citep{Font2020, Font2021} at a baryonic particle mass of $2 \times 10^4~\MSUN$, whereas the DC Justice League project has simulated the entire range of classical satellites, even reaching into the ultrafaint dwarf regime for 4~MW-like hosts at a baryonic mass resolution of $994~\MSUN$ \citep{Applebaum2021, Akins2021} and \cite{Grand2021} have determined the full satellite population of a MW-mass host with a baryonic particle mass of $800~\MSUN$.

Below the approximate stellar mass scale of $10^8~\MSUN$, these simulations broadly agree that satellites are rapidly stripped of their gas reservoirs as they travel through the circumgalactic medium of their host. Most of them are predicted to quench over the course of a few Gyr after infall \citep{Fillingham2015, Fillingham2016, Simpson2018, Akins2021}, with multiple factors influencing the length of this quenching period: the satellite's own stellar mass, the ram pressure it experiences at infall, and its orbit through its host's gaseous halo \citep{Simons2020}. In fact, galaxies may already be affected by ram pressure stripping outside of their host's virial radius or while they are technically still considered to be centrals \citep{Ayromlou2021}.

In terms of quenched fractions, the FIRE-2 simulations recently found broadly consistent results between the LG and the SAGA-II hosts \citep{Samuel2022}, in that the simulation average lies in between them, the host-to-host scatter is large, and the simulation suite comprises a limited number of 14~hosts. On the other hand, the ARTEMIS simulations \citep{Font2022} have reconciled some of the tensions regarding the observed satellite galaxies of the LG and SAGA-II only by identifying previously-unnoticed details in the selection function of the observed external galaxies. Whereas the satellites detected in SAGA-II are subject to an explicit absolute $r$-band magnitude limit, there is also an {\it implicit} limitation on their surface brightness of $\mu_{\rm r} \sim 25~{\rm mag~arcsec}^{-2}$. According to the ARTEMIS simulations, the SAGA-II satellites are not fully sampled at a given $M_{\rm r}$ but miss a number of low-surface brightness satellites \citep{Font2022}. Since ARTEMIS predicts these low-surface brightness satellites to be mostly quenched, the SAGA-II satellite sample could be biased towards actively star-forming galaxies with brighter $\mu_{\rm r}$. The ELVES survey supports this explanation and finds that the SAGA-II hosts lack a number of red, elliptical satellites compared to similar-mass hosts in the Local Volume \citep{Carlsten2022}. In fact, a recent re-analysis of both the SAGA-II and the ELVES surveys by \cite{Karunakaran2022} finds agreement between the quenched fractions of the two samples once the same UV-derived and specific star formation rate-based quenched definition is adopted and after applying either the SAGA–II absolute magnitude cut or a surface brightness cut in addition to the SAGA–II absolute magnitude cut to the ELVES sample.

In this study, we further tackle this debate with another model for galaxy formation and investigate the star formation activity of satellite galaxies around MW and M31 analogue hosts with the high-resolution TNG50 simulation of the IllustrisTNG project \citep{Nelson2019b, Pillepich2019}. TNG50 contains the first, full statistical sample of 198 high-resolution MW/M31-like hosts \citep{Pillepich2023}. The majority of these galaxies have stellar bars; some of them exhibit stellar disk lengths and heights similar to those inferred for the Galaxy and Andromeda \citep{Sotillo2022, Sotillo2023}. Moreover, a large fraction of these MW/M31-like galaxies display X-ray emitting bubbles and shells reminiscent of the eROSITA bubbles observed around the MW \citep{Pillepich2021}; the formation scenarios of their bulges have been connected to their environment, merger histories, and bars \citep{Gargiulo2022}; and they have been used to constrain the presence and dynamics of intermediate black holes accreted from merging dwarfs \citep{Weller2022}.

The satellite populations of these 198 TNG50 MW/M31-like hosts can be reliably identified and characterised at least down to stellar masses of $5 \times 10^6~\MSUN$ (i.e.~with at least $60-100$ stellar particles), corresponding to the stellar mass of the MW's Leo~I dwarf and thus covering the more massive half of the classical satellite regime. If we only consider a single stellar particle instead, luminous satellites are present down to stellar masses of as small as $10^{4.1-4.5}~\MSUN$ in TNG50. In a previous paper \citep{Engler2021b}, we have shown that, amid a remarkable degree of host-to-host diversity, the abundance of satellites around TNG50 MW/M31-like hosts is consistent with observations of the MW and M31, with the recent observational surveys ELVES and SAGA-II, as well as with previous hydrodynamical simulations of MW- and LG-like systems after carefully accounting for differing definitions and selection methodologies. The secular and environmental processes that shape the evolution of the general galaxy population in IllustrisTNG have previously been studied across the mass spectrum: this includes effects from AGN feedback \citep{Weinberger2017, Terrazas2020, Zinger2020}, morphological transformations after infall into dense environments \citep{Joshi2020}, ram pressure stripping resulting in jellyfish galaxies \citep{Yun2019}, tidal stripping and its effects on the galaxy-halo connection \citep{Engler2021a}, as well as how these processes and pre-processing in previous hosts influence the star formation activity and quenched fractions of galaxies \citep{Donnari2019, Donnari2021a, Donnari2021b}. Furthermore, the star formation histories of dwarfs have been studied across centrals and satellites in group and cluster environments \citep{Joshi2021}. Here, we expand upon these studies by focusing specifically on the satellite populations of MW/M31-like hosts to examine what the TNG50 simulation returns in terms of their star formation activity and gas content, their stellar assembly, and the co-evolution of their mass components after infall. 

This paper is structured as follows: in \S\ref{sec:methods}, we introduce the TNG50 simulation and the selection criteria for our fiducial TNG50 MW/M31-like hosts and their satellite populations, as well as for matched samples to selected observational surveys. We quantify the quenched fractions of satellites around MW/M31-like hosts and correlations with host and satellite properties in \S\ref{sec:quenchFracs}. In \S\ref{sec:gasContent}, we study their gas content and its relation to host distance, phase-space position and infall time. Furthermore, we examine their stellar assembly using cumulative star formation histories in \S\ref{sec:stellarAssembly}. In \S\ref{sec:discussion}, we discuss the co-evolution of satellite mass components (i.e.~dark matter, gas, and stars) after infall and the presence of jellyfish galaxies around MW/M31-like hosts, in addition to offering some remarks about the implications of our results on the underlying galaxy formation model in comparison to previous ones in the literature.
Finally, \S\ref{sec:summary} summarises the main findings of this study.

\section{Simulation data and galaxy selections}
\label{sec:methods}

\subsection{The TNG50 simulation}

The results presented in this study are based on the TNG50 simulation \citep{Nelson2019b, Pillepich2019}. This is the highest-resolution flagship run of the IllustrisTNG\footnote{\url{http://www.tng-project.org}} suite of cosmological magnetohydrodynamical simulations of galaxy formation in a $\Lambda$CDM framework \citep{Marinacci2018, Naiman2018, Nelson2018, Nelson2019a, Pillepich2018b, Springel2018} based on the moving mesh code {\sc Arepo} \citep{Springel2010}. 

The IllustrisTNG model of galaxy formation includes physical processes such as gas heating by a spatially uniform and time-dependent UV background, primordial and metal-line gas cooling, a subgrid model for star formation and for the unresolved structure of the interstellar medium. The evolution and chemical enrichment of stellar populations are tracked following ten individual elements from supernovae Ia, II, and AGB stars and the effects of feedback from star formation and super massive black hole feedback are accounted for. All details of the model are given in \cite{Weinberger2017} and \cite{Pillepich2018a} and have been discussed and explored in numerous previous studies\footnote{\url{https://www.tng-project.org/results/}}. Of relevance for this paper, the solution to the coupled equations of gravity, magnetohydrodynamics and the astrophysics of galaxies is such that phenomena like galaxy mergers, galaxy interactions, gravitational tides, ram pressure, gravitational heating, shocks, etc. are all emergent phenomena in the simulations.

In this paper, we focus exclusively on the TNG50 run as it combines a cosmological volume of $(50~{\rm Mpc})^3$ and a statistically significant sample of galaxies (about 900 galaxies with $M_* \gtrsim 10^{10}~{\rm M}_\odot$) with a zoom-in-like level of resolution: namely, a baryonic mass resolution of $m_{\rm b} = 8.5 \times 10^4~\MSUN$ and an average spatial resolution in the hydrodynamics of $50-100~{\rm pc}$ in the star-forming regions of galaxies. Other details regarding the numerical resolution of TNG50 are given in \citet{Pillepich2019, Pillepich2021} and \citet{Nelson2019b, Nelson2020}.

Galaxies, satellites, and subhaloes are identified as local overdensities within larger Friends-of-Friends (FoF) haloes according to the {\sc subfind} algorithm \citep{Springel2001, Dolag2009}. We remove subhaloes that correspond to clumps and fragmentations of non-cosmological origin according to \cite{Nelson2019b}. Whereas a few clumps remain nevertheless, they do not influence our results due to the statistical size of our host and satellite samples.

\subsection{MW/M31-like galaxies in TNG50}
\label{sec:hostSelection}

As discussed in \cite{Engler2021b} and \cite{Pillepich2023}, the choice of selection criteria that define Milky Way-~(MW) and Andromeda-like~(M31) galaxies is essential to ensure both a relevant sample of hosts, and realistic environments for their satellite populations. In the following, we summarise the selection criteria of our fiducial sample of MW/M31-like hosts, and several alternative, more specific subselections of hosts utilised throughout this paper to compare to the results of selected observational surveys.

\subsubsection{Fiducial sample of TNG50 MW/M31-like hosts}
\label{sec:MWM31hosts}

We define MW/M31-like galaxies according to \cite{Pillepich2023} and as done in \cite{Engler2021b} based on their mass, their morphology, and their environment at redshift $z=0$. 

\begin{enumerate}
    \item {\it Stellar mass:} MW/M31-like candidates are required to have a stellar mass within $30~{\rm kpc}$ in the range of $M_* = 10^{10.5} - 10^{11.2}~{\rm M}_\odot$.
    \item {\it Morphology:} TNG50 MW/M31-like candidates need to exhibit a disky stellar morphology. Their shape is determined based on either the minor-to-major axis ratio of their 3D stellar mass distribution ($s<0.45$; chosen as an observationally motivated indicator of their morphology) or by visual inspection of synthetic 3-band stellar light images in face-on and edge-on projection.
    \item {\it Environment:} a minimum isolation criterion is imposed at $z=0$. No other massive galaxies with $M_* > 10^{10.5}~{\rm M}_\odot$ are allowed to be located within $500~{\rm kpc}$ of the MW/M31-like candidate. Furthermore, the total mass of its host halo is limited to $M_{\rm 200c} < 10^{13}~{\rm M}_\odot$.
\end{enumerate}

The virial mass $M_{\rm 200c}$ of a host denotes the total mass of a sphere around the FoF halo centre with a mean density of 200~times the critical density of the Universe. 

These criteria result in a sample of 198~MW/M31-like hosts in TNG50, eight of which are, according to the FoF halo finder, actually satellite galaxies. The absolute $K$-band magnitude, stellar, and total halo mass ranges covered by these galaxies are consistent with the selection criteria for MW and Local Group (LG) analogues from both observations and simulations \citep[][see also their table~1 for a summary of the selection criteria adopted by various simulations of MW/M31- or LG-like hosts]{Engler2021b}.

\begin{figure*}
	\centering
	\includegraphics[width=.9\textwidth]{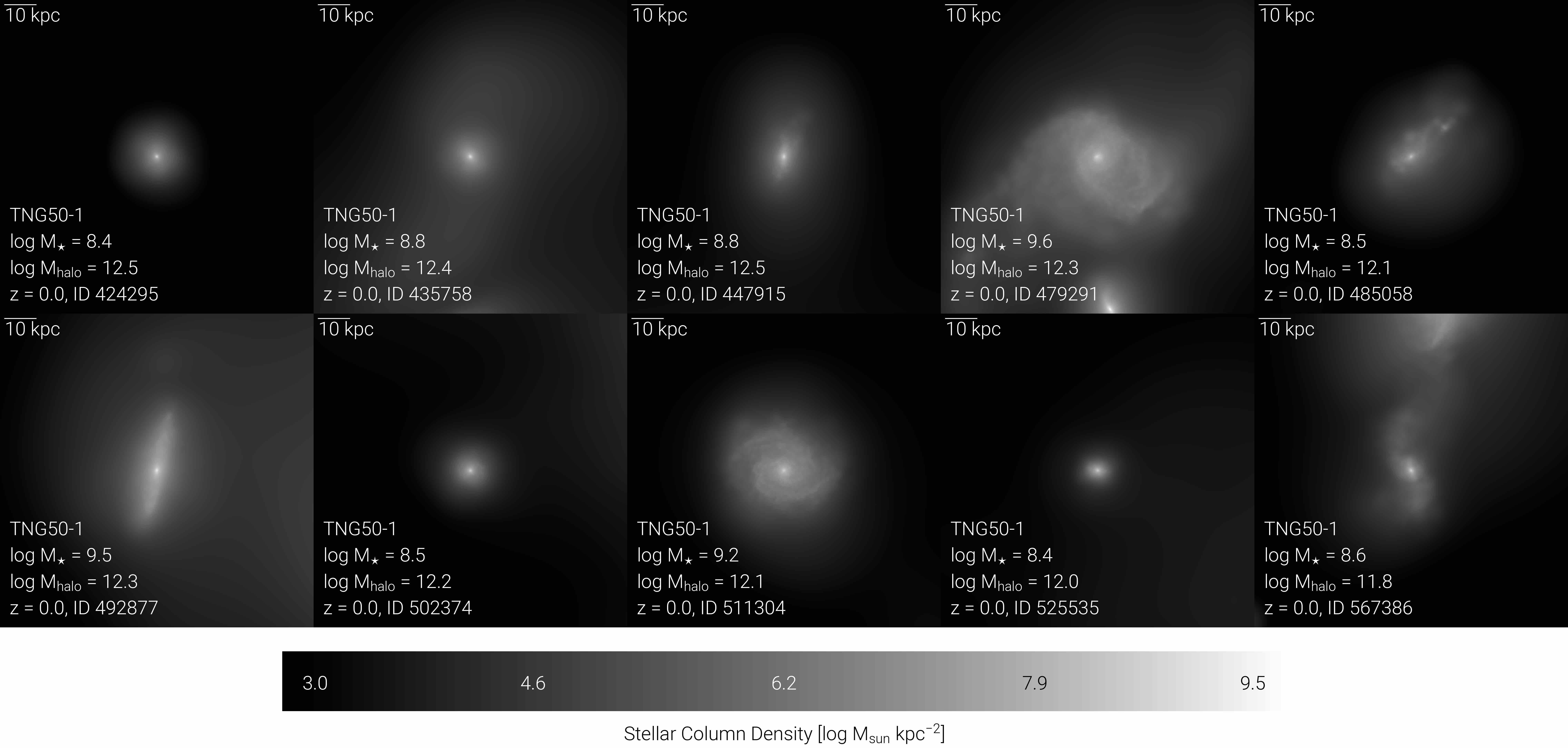}
	\includegraphics[width=.9\textwidth]{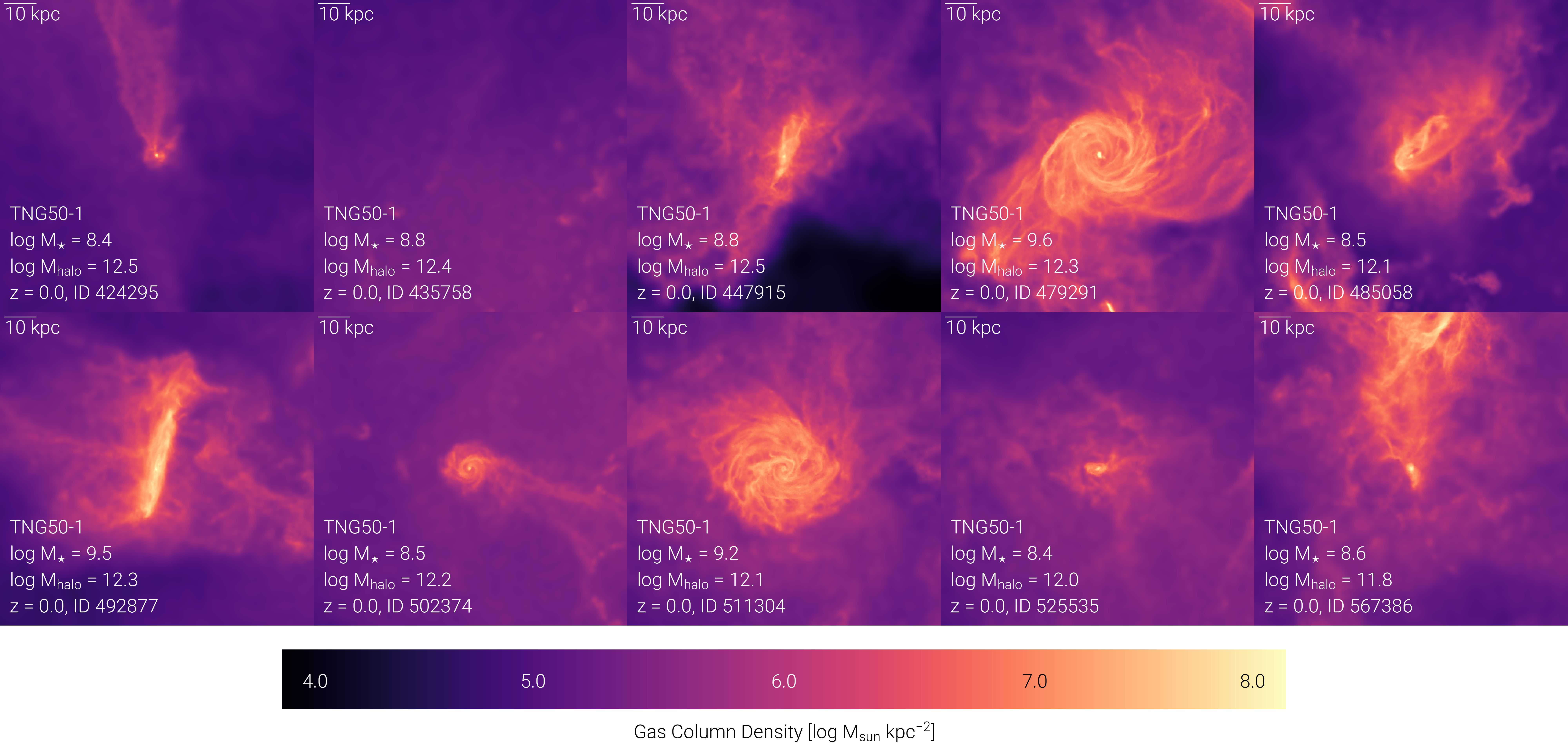}
	\caption{{\bf Stellar and gas column density maps of a random selection of massive satellites of MW/M31-like galaxies in TNG50 at $\mathbf{z=0}$.} The annotation $M_{\rm halo}$ refers to the total halo mass of the host MW/M31-like galaxy ($M_{\rm 200c}$) whereas $M_{\rm *}$ is the stellar mass of the depicted satellite. TNG50 MW/M31-like satellites exhibit a variety of stellar morphologies, from featureless ellipticals to spirals and irregulars. The gas content is also varied, including highly ram-pressured stripped tails (ID 424295): however, as shown in the subsequent sections, the majority of MW/M31-like satellites have little to no gas, similar to the case of the galaxy with ID 435758 for which only the background intra-halo gas of the host galaxy can be seen.}
    \label{fig:maps}
\end{figure*}

When suitable, we further divide our full sample of 198~TNG50 MW/M31-like galaxies into those more or less massive than a stellar mass of $10^{10.9}~\MSUN$ within $30~{\rm kpc}$. This returns 138~TNG50 MW-like hosts with $M_* = 10^{10.5} - 10^{10.9}~\MSUN$ and 60~TNG50 M31-like hosts with $M_* = 10^{10.9} - 10^{11.2}~\MSUN$ at $z=0$.

\subsubsection{TNG50 Local Group-like systems}
\label{sec:hostSelection_LGlike}

We divide our fiducial host sample into isolated MW/M31-like and LG-like hosts. In order to determine which hosts are part of an LG-like configuration, we examine their environment for the presence of another MW/M31-like galaxy. LG-like candidates are required to have one other disky galaxy of similar mass, i.e.~with $M_* = 10^{10.5} - 10^{11.2}~{\rm M}_\odot$, within a distance of $500 - 1000~{\rm kpc}$, and within the sample of 198 MW/M31 analogues -- the distance between the MW and M31 has been measured to $752 \pm 27~{\rm kpc}$ based on Cepheid period-luminosity relations \citep{Riess2012}. Furthermore, the nearby galaxy needs to exhibit a negative radial velocity with respect to the original MW/M31-like host. Hosts that meet these requirements are considered to be part of an LG-like system. This yields a sample of 3 MW+M31 pairs in LG-like configurations in TNG50 at $z=0$. Another TNG50 system has not only one but two other MW/M31 analogues with approaching radial velocities within $500 - 1000~{\rm kpc}$: we do not include it among the LG-like hosts. \S\ref{sec:quenchFracs_hostProps} examines the differences between star formation activities of satellite populations around isolated MW/M31-like and LG-like hosts.

\subsubsection{TNG50 SAGA-like host selection}
\label{sec:hostSelection_SAGAlike}

To compare the results of TNG50 to the SAGA survey, we adopt the host selection criteria according to \citet{Geha2017} and \citet{Mao2021} and construct a sample of TNG50 SAGA-like hosts. This selection is based on their $K$-band luminosity and their local environment: candidate galaxies need to exhibit a $K$-band luminosity in the range of $-23 > M_{\rm K} > -24.6$ and are not allowed to have another bright galaxy with a magnitude of $K < K_{\rm host} - 1.6$ within their virial radius (i.e.~within $300~{\rm kpc}$). Galaxies within massive hosts of $M_{\rm 200c} \geq 10^{13}~\MSUN$ are excluded. These host candidates are {\it not} required to be the centrals of their respective host halo. For each of the 36~observed SAGA hosts, we select the three TNG50 candidate galaxies with the most similar $K$-band luminosity, resulting in a sample of 108~TNG50 SAGA-like hosts (see figure~1 of \citealt{Engler2021b} for a comparison between the $K$-band luminosity ranges of our fiducial TNG50 MW/M31-like hosts, the TNG50 SAGA-like sample, and the observed host galaxies of the SAGA survey). 

\subsubsection{TNG50 ELVES-like host selection}
\label{sec:hostSelection_ELVESlike}

To compare the results of TNG50 to the ELVES survey, we construct a host sample similar to the nearby galaxies in the Local Volume studied by \citet{Carlsten2022}. The observed ELVES hosts are selected based on luminosity and spatial proximity, with no specific criteria for their environment or morphology. Thus, we adopt a similar $K$-band luminosity range of $-22 < M_{\rm K} < -25$ and further require candidates to be the central galaxy of their FoF halo and to exhibit a disky stellar shape (as in \S\ref{sec:MWM31hosts}). For each of the 30~ELVES hosts (we exclude NGC~3621 as it has no radial coverage for satellites), we select the three TNG50 galaxies with the closest $K$-band luminosity that meet all of the previous criteria. This results in a sample of 90~TNG50 ELVES-like hosts.

\subsection{Satellite samples}
\label{sec:satSelection}

We vary our satellite selection criteria depending on the comparison at hand: we define satellites as galaxies that are located within various physical apertures from their host at the time of inspection, e.g.~within $300~{\rm kpc}$ (our fiducial sample containing 1237~galaxies) or $600~{\rm kpc}$, or those that belong to their FoF host (FoF membership) when searching for satellites out to distances of $\gtrsim 1~{\rm Mpc}$, and to define their times of infall. Whereas these satellite samples overlap for the most part, there is no qualitative difference in our results between employing one or the other definition. Throughout our figures, we annotate the selection based on physical distance as e.g.~``Galaxies within $300~{\rm kpc}$'' and the selection based on FoF membership as e.g.~``Satellites within $300~{\rm kpc}$''. Furthermore, we require satellites to have a stellar mass within two stellar half-mass radii $R_{1/2}^*$ of at least $5 \times 10^6~{\rm M}_\odot$. This corresponds to the mass of the MW's own satellite Leo~I and ensures a reasonable level of numerical resolution with at least $60-100$~stellar particles each. As shown in \cite{Engler2021b}, the satellite galaxies within $300~{\rm kpc}$~(3D) of our 198~TNG50 MW/M31-like hosts form realistic scaling relations in agreement with various previous cosmological simulations, semi-empirical models, and observations of MW and M31 satellites. This includes their stellar-to-halo mass relation, as well as their maximum circular velocity, absolute $r$-band magnitude, half-light radius, and stellar velocity dispersion as a function of satellite stellar mass (see their figure~2).\\

{\it SAGA-like selection.} Furthermore, we employ satellite samples based on the selection criteria of SAGA-II \citep{Mao2021}. Their criteria require satellite galaxies to lie within a two-dimensional, randomly projected aperture of $300~{\rm kpc}$ from their host and to have a line-of-sight velocity of $\pm 275~{\rm km~s}^{-1}$ relative to their host galaxy. We apply these criteria to both our fiducial sample of TNG50 MW/M31-like hosts and the TNG50 SAGA-like hosts, and further augment the selection by matching specific TNG50 satellites to the observed SAGA-II satellites based on their absolute $r$-band magnitude throughout \S\ref{sec:quenchFracs} and specifically \S\ref{sec:fquench_compSAGA}. For example, in \S\ref{sec:quenchFracs} and Fig.~\ref{fig:fquench}, top panel, we match each observed SAGA-II satellite with {\it one} random TNG50 satellite around the SAGA-like hosts with an absolute $r$-band magnitude within $\pm 0.5~{\rm mag}$: whereas SAGA-II observes 127 satellites in total, we cannot find a suitable, simulated match for all of them. Thus, our fiducial sample of TNG50 SAGA analogue satellites consists of 113 galaxies. \\

{\it ELVES-like selection.} The radial coverage around the 31~ELVES hosts varies between $0 - 370~{\rm kpc}$, with 19~of them surveyed out to $300~{\rm kpc}$ \citep{Carlsten2022}. Nevertheless, we require analogue TNG50 satellites to be located within a projected aperture of $300~{\rm kpc}$ and within $\pm 500~{\rm kpc}$ along the line of sight of our ELVES-like hosts. While in the observations, the satellites' distances along the line of sight are measured using their redshift, the tip of the red giant branch, or surface brightness fluctuations, \cite{Carlsten2021} -- a predecessor study to the ELVES survey -- included a comparison with TNG100 where they adopted this spatial line-of-sight criterion.\\

Examples of satellites of TNG50 MW/M31-like galaxies are shown in Fig.~\ref{fig:maps}, in stellar (top) and gas (bottom) mass column density projections. The figure shows that, at least for massive satellites, TNG50 resolves also their internal structures and a diversity of stellar and gaseous morphologies is apparent.

\section{Satellite quenched fractions}
\label{sec:quenchFracs}

\begin{figure*}
	\centering
	\includegraphics[width=.8\textwidth]{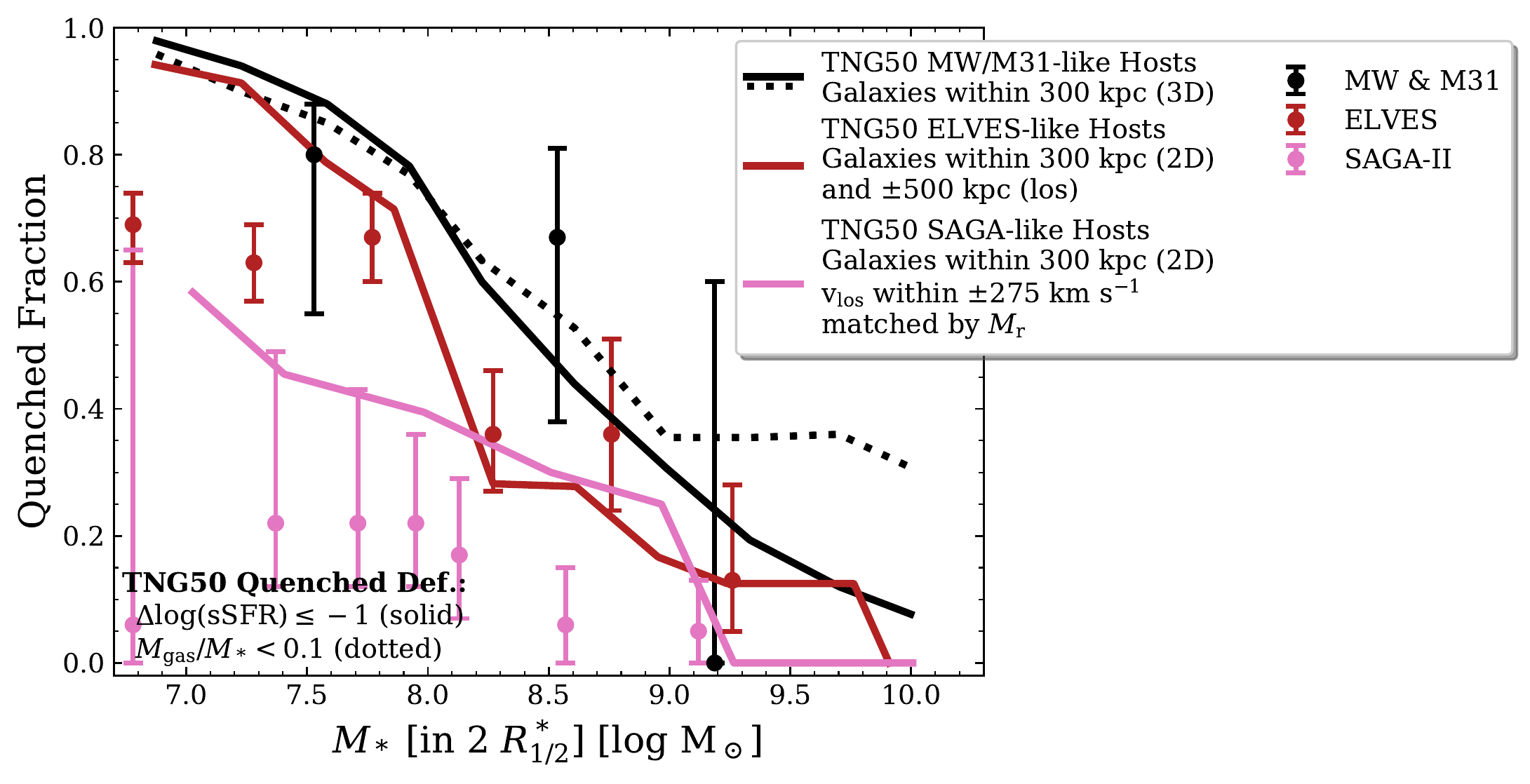}
	\includegraphics[width=\columnwidth]{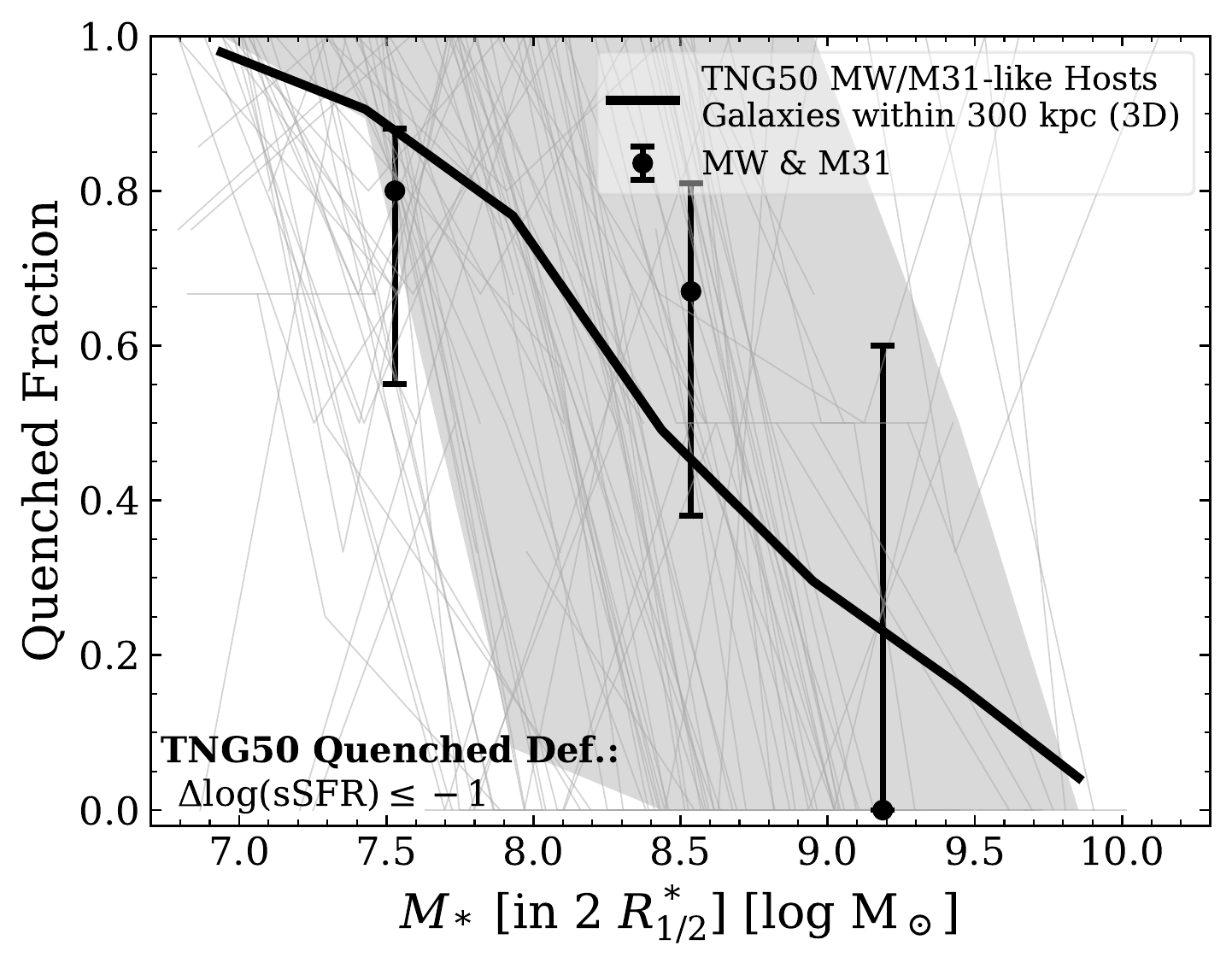}
	\includegraphics[width=\columnwidth]{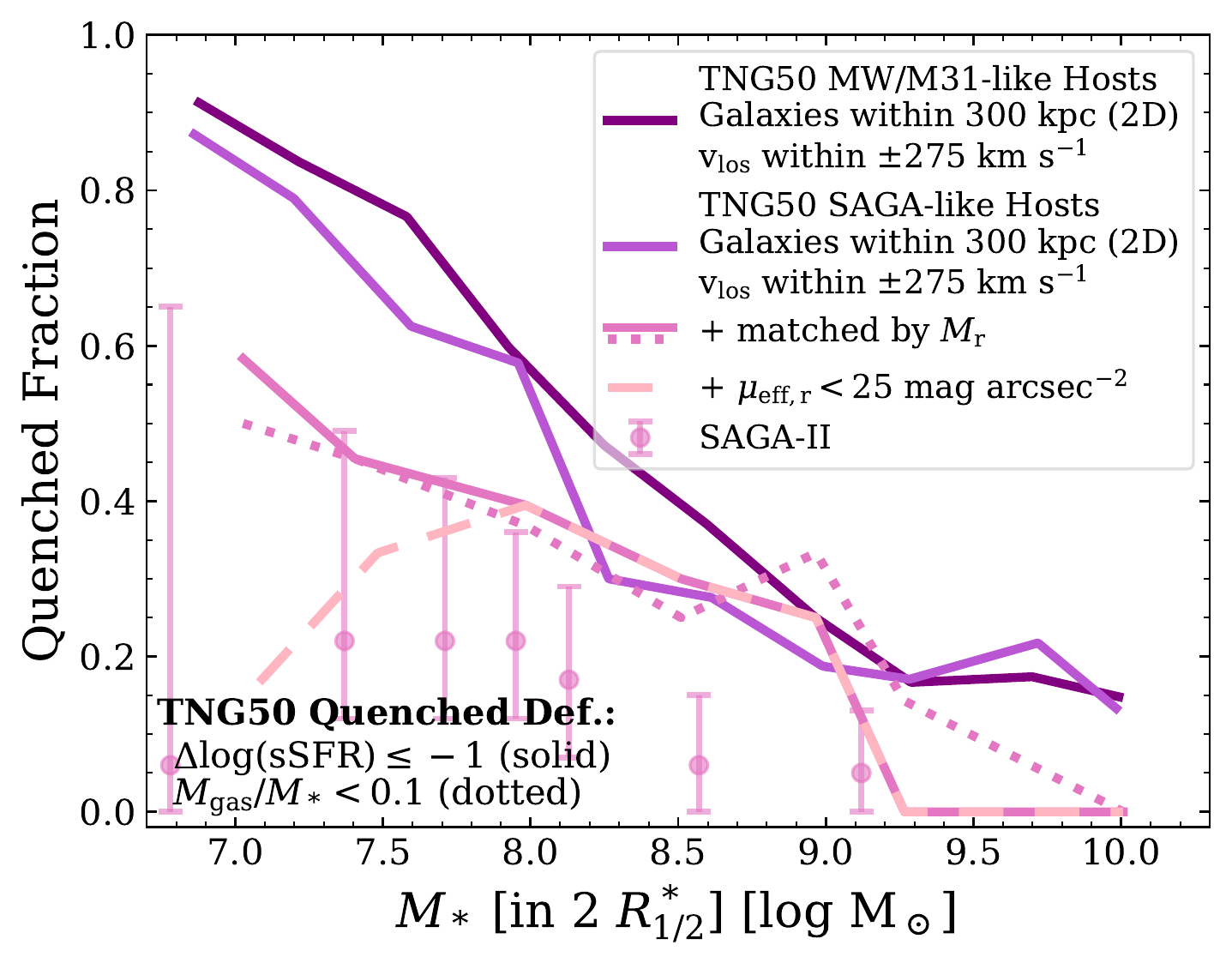}
    \caption{{\bf Quenched fractions of satellite galaxies with $\mathbf{M_{\boldsymbol{*}} \boldsymbol{\geq} 5 \boldsymbol{\times} 10^{6}~{\bf M}_{\boldsymbol{\odot}}}$ around MW/M31-like hosts in TNG50, the observed MW and M31, ELVES, and SAGA-II.} {\it Top panel:} median quenched fractions of TNG50 satellites stacked across all hosts (curves) and of observed satellites (points). In each comparison, we impose the respective observations' selection criteria on TNG50: galaxies within $300~{\rm kpc}$~(3D) of their host for the MW and M31 (black curve), galaxies within $300~{\rm kpc}$ (2D) and $\pm 500~{\rm kpc}$ along the line of sight for ELVES (red curve), and galaxies within $300~{\rm kpc}$ (2D) with line-of-sight velocities of $\pm 275~{\rm km~s}^{-1}$ that have been matched to the absolute $r$-band magnitude $M_{\rm r}$ of each observed SAGA-II satellite for the SAGA survey (pink curve). The black points denote the quenched fractions of the observed MW and M31 satellites \protect\citep{Wetzel2015} while the red and pink points correspond to the stacked satellites of ELVES \citep{Carlsten2022} and SAGA-II \protect\citep{Mao2021}. While our fiducial quenched definition is based on the satellites' distance to the SFMS (solid curves), \protect\cite{Wetzel2015} employ a gas fraction criterion for LG satellites, \protect\cite{Carlsten2022} determine star formation activity based on their satellites' morphology, and the quenched fractions of \protect\cite{Mao2021} are based on H$\alpha$ equivalent widths. Thus, the solid curves represent a comparison at face value. We adopt the gas fraction-based criterion of \protect\cite{Wetzel2015} (dotted, black curve) to more directly compare to their observations. {\it Bottom left panel:} satellite quenched fractions including host-to-host variations. The thin, grey curves in the background denote the quenched fractions of satellites around individual TNG50 MW/M31-like hosts while the thick, black curve and the grey shaded area correspond to their mean and scatter as $16^{\rm th}$ and $84^{\rm th}$ percentiles. {\it Bottom right panel:} effects of host and satellite selections on the quenched fractions of the TNG50 SAGA-II analogue sample using our fiducial quenched definition. We show the quenched fractions for satellites around both TNG50 MW/M31-like and SAGA-like hosts according to the spatial SAGA-II selection criteria (dark and medium purple solid curves, respectively), and for the smaller sample of TNG50 satellites that are matched to the SAGA-II satellites based on their absolute $r$-band magnitude (pink solid curve, as in the top panel). The dashed light pink curve gives the quenched fractions predicted by TNG50 for SAGA-like satellites including an implicit selection cut on their surface brightness as suggested by \protect\citet[][see text for more details]{Font2022}. We illustrate the effects of different quenched definitions by adopting the gas fraction-based definition for our matched SAGA-like satellite sample (dotted vs. solid pink curves).}
    \label{fig:fquench}
\end{figure*}

We quantify the star formation activity of satellite galaxies around MW/M31-like hosts in TNG50, i.e.~their quenched fractions as a function of satellite stellar mass $M_*$, in Fig.~\ref{fig:fquench}. 

In the top panel, we compare the median of the {\it stacked} satellites of all MW/M31-like hosts in TNG50 (solid curves) to the observations of the MW and M31 \citep[black dots,][]{Wetzel2015}, as well as to the ELVES \citep[red dots,][]{Carlsten2022} and SAGA-II surveys \citep[pink dots,][]{Mao2021} by adopting their respective selection criteria for TNG50 hosts and satellites. For the MW and M31, this simply corresponds to our fiducial satellite selection of all galaxies within $300~{\rm kpc}$~(3D, irrespective of FoF halo assignment) around our 198~MW/M31-like hosts, as chosen in \S\ref{sec:MWM31hosts} and \S\ref{sec:satSelection} (black curves). For comparisons to ELVES and SAGA-II, we adjust the host sample to the 90~ELVES-like and the 108~SAGA-like TNG50 hosts identified in \S\ref{sec:hostSelection_ELVESlike} and \S\ref{sec:hostSelection_SAGAlike} and select satellites as described in \S\ref{sec:satSelection}. The red and pink solid curves represent the TNG50-predicted quenched fractions for 441 ELVES-like satellites around ELVES-like hosts and 113 SAGA-II-like satellites around SAGA-II like hosts.

Unless otherwise stated, we use a fiducial quenched definition based on the specific star formation rate~(sSFR) of satellite galaxies and their distance to the star forming main sequence~(SFMS), whereby galaxies with $\Delta \log({\rm sSFR}) \leq -1$ are considered quenched \citep{Pillepich2021, Donnari2021b}. Therefore, this is a comparison {\it at face value}, i.e.~without adjusting for the respective quenched definitions of the observational analyses (solid curves). In fact, \cite{Wetzel2015} consider satellites with a gas fraction of $M_{\rm gas}/M_* < 0.1$ to be quenched, whereas the definition of \cite{Mao2021} is based on H$\alpha$ equivalent widths with ${\rm EW(H}\alpha) < 2~\text{\r{A}}$ and \cite{Carlsten2022} define star formation activity based on morphology. We adopt the gas fraction-based criterion of \cite{Wetzel2015} for a matched comparison of the TNG50 MW/M31-like satellite quenched fractions (dotted, black curve), whereby satellites with gas fractions of $M_{\rm gas}/M_* < 0.1$ are considered quenched. Therefore, solid vs. dotted curves assess at least part of the systematic uncertainty due to different definitions of ``quenched''.

Qualitatively, the medians of the stacked satellite galaxies in Fig.~\ref{fig:fquench}, top panel, all display the same trend, regardless of their sample selection or quenched definition: less massive satellites exhibit larger quenched fractions. As less massive galaxies have shallower gravitational potentials, they are more prone to be affected by environmental effects, and e.g.~to be stripped of their gas -- and subsequently quenched -- in the halo of their host galaxy. This is consistent with previous results on the quenched fractions of IllustrisTNG satellites across the host mass spectrum (see \citealt{Donnari2021a} for TNG300 and TNG100 satellites in $M_{\rm 200c} = 10^{12} - 10^{15.2}~\MSUN$ hosts, and \citealt{Joshi2021} for TNG50 satellites in $M_{\rm 200c} = 10^{12} - 10^{14.3}~\MSUN$ hosts).

Quantitatively, however, the quenched fractions of the three host and satellite selections (solid curves in the top panel of Fig.~\ref{fig:fquench}) are very distinct from each other, even if the underlying galaxy population and galaxy formation model are the same. While our fiducial choice of satellites within $300~{\rm kpc}$~(3D, black solid line) range from $90-100$~per~cent quenched at the low-mass end to $30-40$~per~cent quenched for $>10^{9}~{\rm M}_\odot$ satellites, the average quenched fractions predicted by TNG50 for the SAGA-like selection (pink solid line) is overall lower, with $50-60$~per~cent at the low-mass end and no massive quenched satellites. The criteria of the observational SAGA-like selection decrease the satellite quenched fractions predicted by TNG50 significantly, since they allow for the inclusion of star-forming field galaxies in the fore- and background of host galaxies, which ``contaminate'' the satellite sample (see also \S\ref{sec:quenchFracs_hostProps}). The TNG50 ELVES-like satellites, on the other hand, fall between the two previous selections in terms of their star formation activity (red curve). Whereas their quenched fractions at $M_* > 10^{8.2}~\MSUN$ remain relatively low and reach only up to $25-30$~per~cent -- similar to the observed ELVES satellites --, they exhibit a sharp increase towards smaller stellar masses of up to $90-95$~per~cent. 

Compared to the observations (curves vs. points in Fig.~\ref{fig:fquench}, top panel), we find a reasonable level of agreement between the TNG50 simulation and the observed satellites within $300~{\rm kpc}$ of the MW and M31 across almost the whole range of stellar mass. The stacked TNG50 quenched fractions -- from a total of about 1200 satellites around 198 hosts -- are compatible with the observational results of the Galaxy's and Andromeda's systems, given their errors and barring somewhat larger quenched fractions in TNG50 than observed at $M_* \sim 10^{7.5}~\MSUN$. This is the case whether we use an sSFR-based or gas fraction-based definition of quenching (solid vs. dotted black curves). In fact, we have verified that, barring massive galaxies ($M_*\gtrsim 10^{9-10}~\MSUN$, see also \citealt{Donnari2021b}), the quenched definition is not crucial: only 3~per~cent of our satellites are quenched with the SFMS but are not with the gas fraction definition; vice versa, it is only 2~per~cent. We examine the host-to-host variations in quenched fractions across individual TNG50 MW/M31-like systems in \S\ref{sec:fquench_indHosts}. 

Satellites from SAGA-II, on the other hand, exhibit significantly lower quenched fractions than the satellites of the MW and M31 (pink vs. black points) and than our fiducial TNG50 MW/M31-like satellites (pink points vs. black curve), reaching quenched fractions of maximum $60-65$~per~cent when taking into account their scatter and their uncertainties at the low-mass end. Here, the error bars around the SAGA-II quenched fractions include both shot noise as well as the combined correction for incompleteness and interlopers \citep[the latter being subdominant;][]{Mao2021}. Whereas the sample of TNG50 SAGA-like satellites exhibit relatively lower quenched fractions compared to our fiducial selection (black vs. pink curves), they are still higher than the average (stacked) quenched fractions of SAGA-II (pink curve vs. pink points) and beyond the error bar limits in the $10^{8-9}~\MSUN$ range. These inconsistencies between the SAGA-II results and simulations predictions, on the one hand, and MW+M31 results, on the other, are consistent with previous claims (see Introduction): we further expand on this in \S\ref{sec:fquench_compSAGA}.

The quenched fractions of the observed ELVES satellites and their TNG50 analogues are consistent at $M_* \gtrsim 10^{7.7}~\MSUN$; for less massive satellites, on the other hand, the TNG50 quenched fractions are larger by up to 25~percentage points. The ELVES survey's smaller quenched fractions might result from near-field dwarfs in the fore- and background since the ELVES survey has an expected contamination fraction of $15-20$~per~cent. Furthermore, the differences between the observed and simulated satellites at the low-mass end can be attributed to the varying radial coverage around the 30~ELVES hosts considered here (excluding NGC~3621 with a radial coverage of $0~{\rm kpc}$). While 19~ELVES hosts include satellites within $300~{\rm kpc}$, two hosts have an extended radial coverage out to $330$ and $370~{\rm kpc}$. For the remaining eight ELVES hosts, satellites are confined to smaller apertures of $200$ or $150~{\rm kpc}$. Finally, as is the case for SAGA-II, the top panel of Fig.~\ref{fig:fquench} only provides a comparison made at face value since the ELVES survey's quenched definition is based on their satellites' morphology. Recently, \cite{Karunakaran2022} have shown that the quenched fractions of the ELVES satellites get even lower towards the low-mass end once a UV+sSFR quenched definition is applied, possibly increasing the difference with the TNG50 predictions. However, replicating such operational definitions is beyond the scope of this paper and shall be postponed to future work.

It should be noted that the findings described here are independent from possible limitations due to the finite numerical resolution of the TNG50 simulation. As shown by \cite{Joshi2021}, the quenched fractions of TNG50 satellite galaxies in the considered mass range are not subject to resolution effects and are in fact well converged between TNG50-1 (i.e.~TNG50, the highest resolution run employed here) and its lower-resolution analogues. Within the IllustrisTNG model implemented in AREPO, the satellites' evolution is dominated by the environmental effects imparted by their more massive and well-resolved hosts so that the quenched fractions of satellites are robust across many levels of resolution: see appendix~A of both \cite{Joshi2021} and \cite{Donnari2021a}, and appendix~C of \cite{Donnari2021b} for resolution effects on the quenched fractions of the general galaxy populations of TNG100, TNG300, and TNG50 both at $z=0$ and at earlier times.

\subsection{Quenched fractions around individual MW/M31-like hosts}
\label{sec:fquench_indHosts}

Whereas the top of Fig.~\ref{fig:fquench} shows the quenched fractions of TNG50 satellites stacked across all 198~MW/M31-like hosts, we examine the satellites of individual systems in the bottom left panel in order to quantify the predicted underlying host-to-host variations. Considering all galaxies within $300~{\rm kpc}$~(3D) around any given host, the thin, grey curves in the background give the mean satellite quenched fractions in bins of satellite mass of individual hosts from the simulation. The thick, black curve and grey shaded area correspond to the mean across hosts and the scatter as $16^{\rm th}$ and $84^{\rm th}$ percentiles. Furthermore, we include, as in the top panel, the quenched fractions of the MW and M31 \citep[black dots,][]{Wetzel2015}.

Most satellite systems of individual TNG50 MW/M31-like hosts exhibit a steep increase in quenched fractions below stellar masses of $10^9~\MSUN$. However, there are some hosts with massive satellites that are exclusively quenched or low-mass satellites with $M_* \sim 10^7~\MSUN$ that are still star forming in large fractions. This results in a significant degree of scatter above satellite stellar masses of $10^{7.5}~\MSUN$. Nevertheless, the mean curve across hosts is actually more similar to the median of the stacked satellites in the top panel of Fig.~\ref{fig:fquench}, continuously increasing towards smaller satellite stellar masses. The median curve across hosts, which we do not show here, would instead simply show a sharp increase in satellite quenched fractions from zero to 100~per~cent between stellar masses of $10^9$ to $10^{8}~\MSUN$.

Overall, the TNG50 mean across hosts is well in agreement with the observations of MW and M31 satellites across all stellar masses; however, the bottom left panel of Fig.~\ref{fig:fquench} shows that the host-to-host variation can be very large and could in fact make comparisons across small numbers of systems meaningless or impossible to interpret.

\subsection{Comparison to the SAGA survey}
\label{sec:fquench_compSAGA}

We expand on the comparison in the top panel of Fig.~\ref{fig:fquench} (pink annotations) by exploring, in the bottom right, the impact of various host and satellite selections on quenched fractions when constructing a SAGA-like host and satellite sample out of the TNG50 galaxy population. All curves denote the median quenched fractions of satellites stacked across their respective host selection. 

We start by applying the spatial selection criteria of SAGA-II satellites, i.e.~galaxies within projected $300~{\rm kpc}$ and a line-of-sight velocity within $\pm 275~{\rm km~s}^{-1}$, to our fiducial sample of TNG50 MW/M31-like hosts (dark purple curve), then further restrict this sample to the TNG50 SAGA-like hosts (purple curve, see \S\ref{sec:hostSelection_SAGAlike}). Then, we limit the satellite sample as in the top panel of Fig.~\ref{fig:fquench}, such that for each observed SAGA-II satellite, we select one random TNG50 galaxy around a TNG50 SAGA-like host with an absolute $r$-band magnitude within $\pm 0.5~{\rm mag}$ from the observed value (pink curve, as described in \S\ref{sec:satSelection}). This corresponds to the nominal selection explicitly given for the SAGA-II results \citep{Mao2021} and returns the quenched fraction curve already described and reported in the top panel of the same figure. Finally, we embrace the possibility proposed by \citet{Font2022} that SAGA-II galaxies include only those with surface brightness $\mu_{\rm eff, r} \lesssim 25~{\rm mag~arcsec}^{-2}$: we impose this cut after matching the samples (dashed light pink curve) but we further explore other implementations of this implicit selection in Appendix~\ref{sec:AppSurfBright}.

The more we restrict our fiducial sample to meet the selection criteria of the SAGA-II observations, the smaller the satellite quenched fractions predicted by the TNG50 model become. While the spatial SAGA-II satellite selection already results in lower quenched fractions compared to our fiducial, 3D radially-selected satellites due to the inclusion of fore- and background field galaxies, the quenched fractions are further reduced by $0-20$ percentage points when we exclusively consider the satellite populations of TNG50 SAGA-like hosts (dark purple vs. purple curves). The host mass range covered by SAGA-II focuses specifically on MW-like galaxies with absolute $K$-band magnitudes of $-23 > M_{\rm K} > -24.6$, which is centred on the fainter half of our sample of TNG50 MW/M31-like hosts \citep[see figure~1 of][]{Engler2021b}. As our fiducial MW/M31-like host sample includes more massive environments that can exert stronger effects on their satellites, it is reasonable for the satellite quenched fractions around the SAGA-like hosts to be smaller. We will further investigate trends with host mass in \S\ref{sec:quenchFracs_hostProps}. Matching the TNG50 satellites to the observed galaxies based on their $r$-band luminosity further decreases their quenched fractions, particularly below $10^8~\MSUN$. 

The dashed light pink curve represents the TNG50 results when implementing the additional surface brightness limit advocated by \cite{Font2022}. As anticipated in the Introduction, based on the results of the ARTEMIS simulations, the latter were able to attribute the low quenched fractions of the SAGA-II satellites to the survey's inherent surface brightness limit of $\mu_{\rm r} \sim 25~{\rm mag~arcsec}^{-2}$. At fixed $r$-band luminosity, fainter ARTEMIS satellites ($-12.3 < M_{\rm r} < -15$) with higher surface brightness are either mostly actively star forming or a mix of quenched and star forming galaxies. Faint and lower-surface brightness satellites ($\mu_{\rm r} > 25~{\rm mag~arcsec}^{-2}$), on the other hand, are almost exclusively quenched \citep[see figure~2 of][]{Font2022}. Thus, according to the ARTEMIS simulations, SAGA-II misses most quenched satellites at faint $M_{\rm r}$ and is simply biased to more actively star-forming galaxies. On the observational side, \cite{Carlsten2022} of the ELVES survey argued for incompleteness in the SAGA survey as well. Compared to ELVES hosts in the Local Volume (i.e.~within $12~{\rm Mpc}$), SAGA-II includes a significantly smaller number of red, early-type satellites -- even after only considering ELVES hosts of the same mass. This is particularly apparent in the inner regions of their hosts where ELVES satellites have a on-average fainter surface brightness. Now, whereas higher-surface brightness galaxies in the TNG50 simulation tend to be more actively star forming as well, there is still a significant number of quenched satellites with $\mu_{\rm r} < 25~{\rm mag~arcsec}^{-2}$ (see Appendix~\ref{sec:AppSurfBright} and Fig.~\ref{fig:surfBright_vs_Mr}). At intermediate to high masses of $M_* > 10^{8}~\MSUN$, the dashed light pink curve coincides with the results without the imposed surface brightness cut. However, such a limit clearly affects the selection of low-mass SAGA-like satellites, whose TNG50-predicted quenched fractions are now compatible with the low observational results of SAGA-II. It is important to point out that the number of SAGA-II satellites is overall not very large, 127 in total, i.e.~approximately $3-4$ per host, of which 59 are in the stellar mass range of $5 \times 10^6 - 10^8~\MSUN$. This is indeed well conveyed by the very large error bars in the reported quenched fractions of SAGA-II, but it also makes the results even more susceptible to incompleteness issues.

In conclusion, according to the IllustrisTNG galaxy formation model, the surface brightness limitation does not completely reconcile the discrepancies between the star formation activity of SAGA-II and TNG50 satellites. In this respect, our TNG50-based findings, whereby the quenched fractions of SAGA-like satellites at $M_* \sim 10^{8-9}~\MSUN$ do not change much when a surface brightness cut is applied {\it in addition} to the SAGA–II absolute magnitude cut, seem more in agreement with those by \cite{Karunakaran2022}. Still, accounting for the given observational error bars, the TNG50 and SAGA-II samples are consistent at the lowest-mass end considered here ($M_* \sim 5 \times 10^6 - 10^8~\MSUN$) but remain non-reconcilable around the mass scale of the SMC, i.e.~at $M_* \sim 10^{8-9}~\MSUN$. In Appendix~\ref{sec:AppSurfBright} we show that, however we construct SAGA-like satellite samples out of TNG50 based on luminosity, stellar mass, surface brightness or a combination thereof, the quenched-fraction predictions from TNG50 remain incompatible or only marginally compatible -- given the error bars -- with the SAGA-II results. 

Interestingly, the comparison of the solid pink curve vs. the dotted pink curve further shows that the precise definition of what quenched means may not be the culprit, with the former based on distance from the SFMS and the latter based on gas fraction. However, a full forward-modelling of simulated galaxies that would permit a quenched definition based on H$\alpha$ equivalent width as in SAGA-II will be a desirable future step. This would allow to conclusively quantify by how much the quenched fractions depend on definition and SF tracers, and to pin down the root cause of the SAGA-II discrepancy. \cite{Dickey2021} have undertaken a similar exercise by forward modelling TNG100 galaxies into H$\alpha$-based measures and quenched fractions: such a modelling would need to be repeated for TNG50 galaxies and in particular for MW/M31-like satellites.

\begin{figure*}
	\centering
	\includegraphics[width=.8\columnwidth]{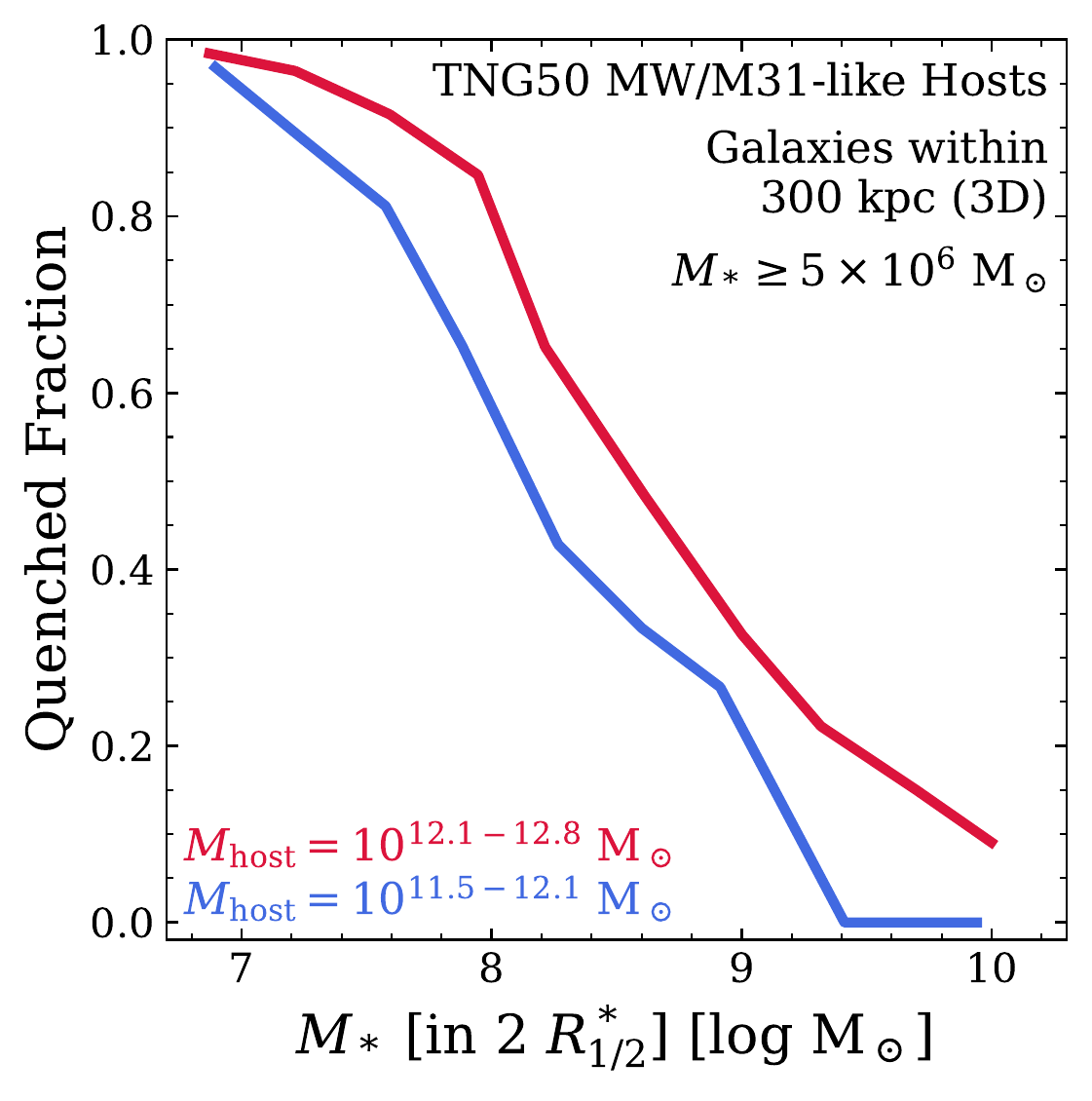}
	\includegraphics[width=.8\columnwidth]{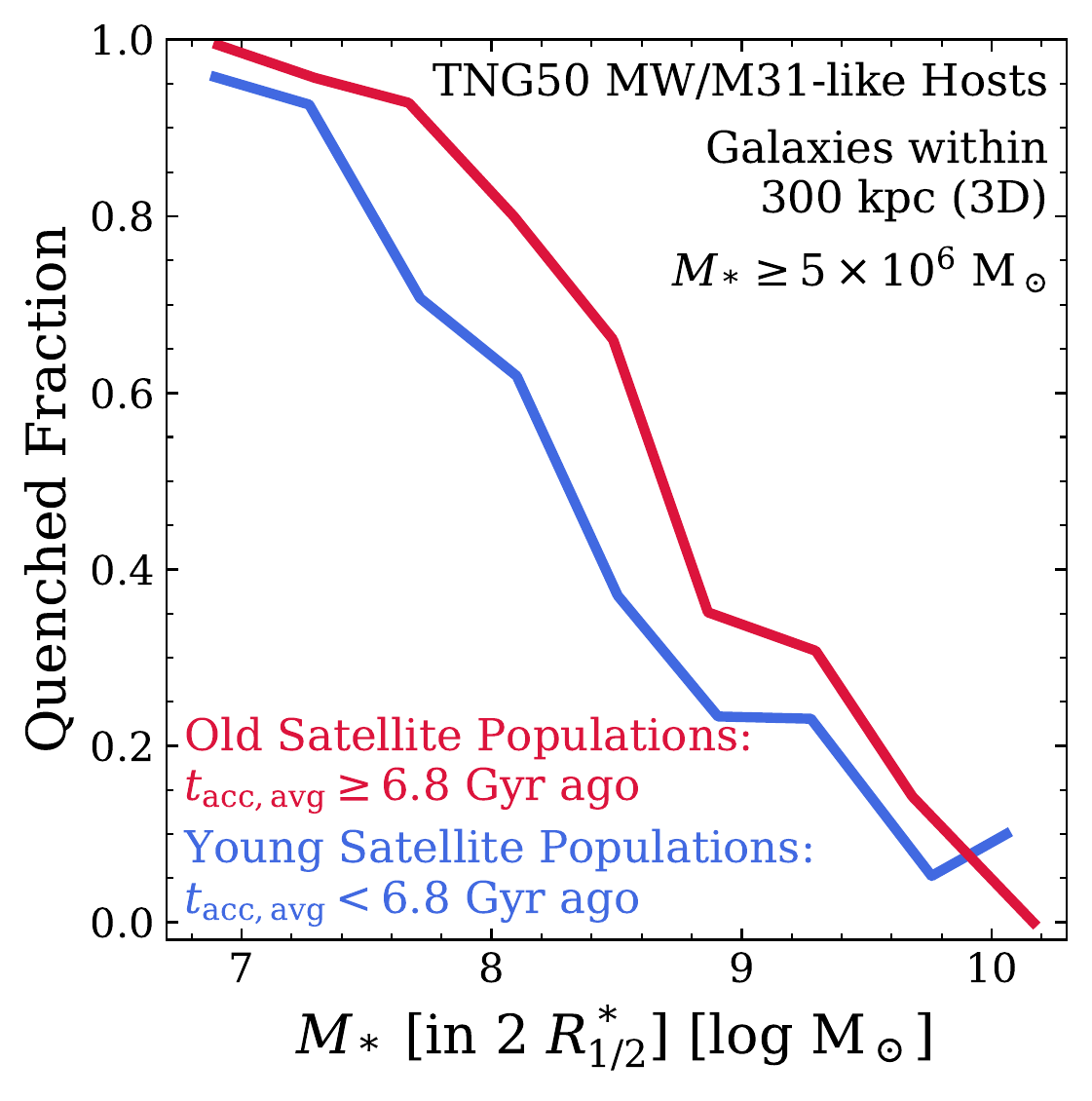}
	\includegraphics[width=.8\columnwidth]{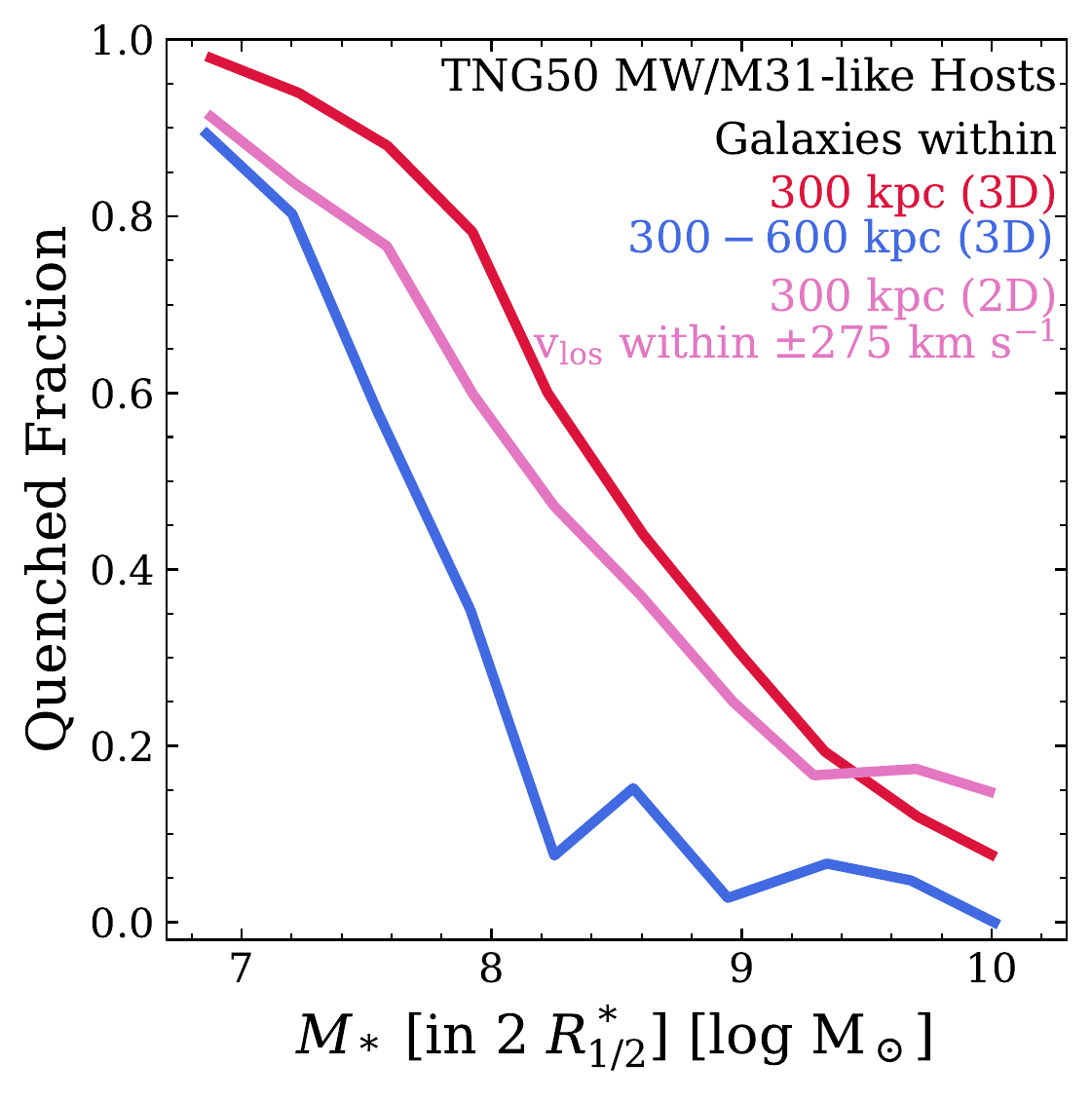}
	\includegraphics[width=.8\columnwidth]{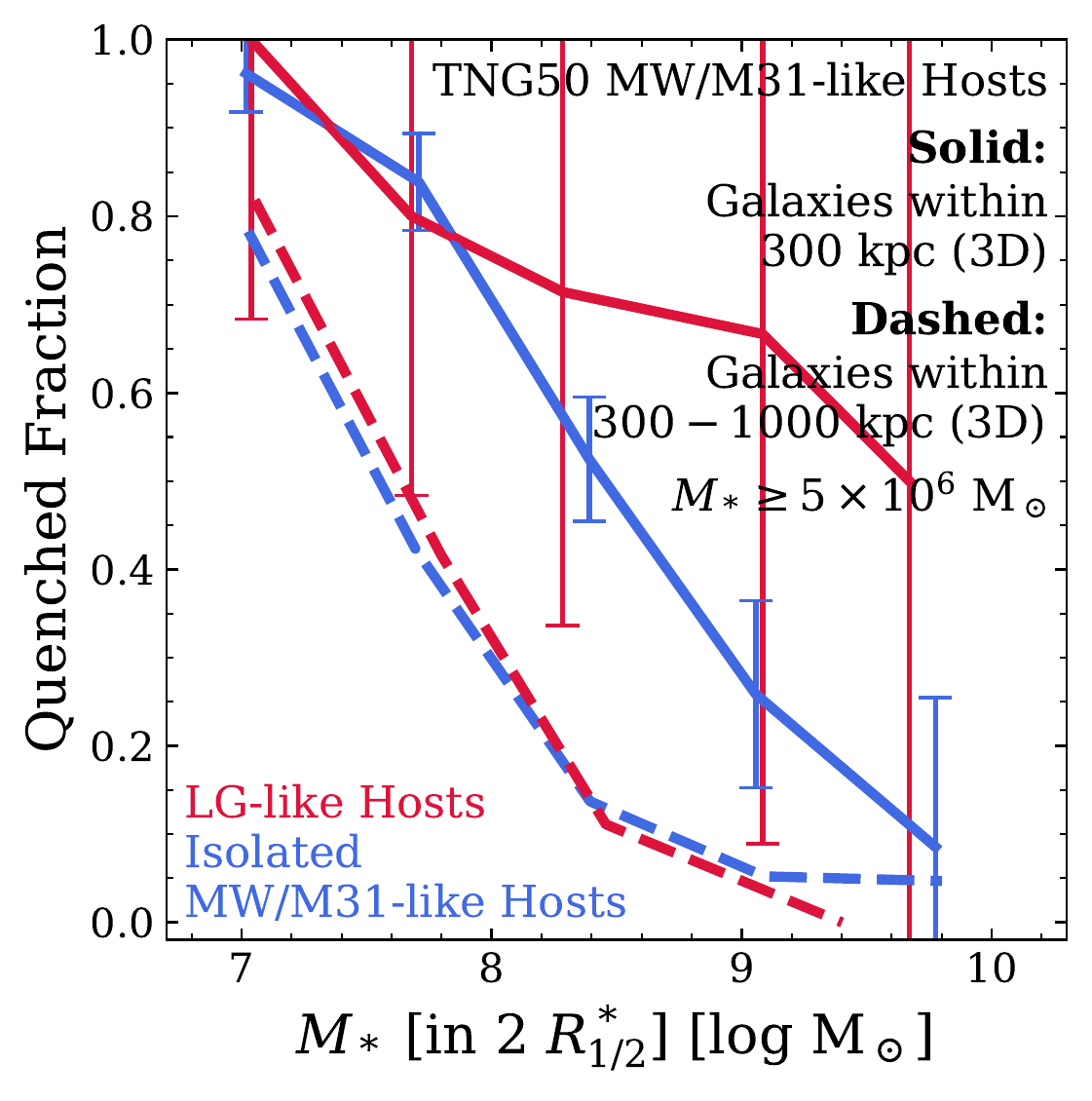}
    \caption{{\bf Dependence of satellite quenched fractions on host and satellite properties.} All panels display the median quenched fractions for TNG50 satellites with $M_* \geq 5 \times 10^6~{\rm M}_\odot$ stacked across all MW/M31-like hosts. We define galaxies to be quenched based on their distance to the SFMS. {\it Top left panel:} dependence on host halo mass $M_{\rm 200c}$ for more massive and less massive hosts (red and blue curve, respectively). {\it Top right panel:} dependence on the average accretion time of satellite systems $t_{\rm acc, avg}$ for satellites that were accreted early on average (red curve) and those with a later average infall (blue curve). The median average accretion time across our host sample is $6.8~{\rm Gyr~ago}$. {\it Bottom left panel:} satellites at different distances to their host. We compare the quenched fractions of TNG50 satellites within $300~\rm{kpc}$ of their host (red curve) to those at distances of $300-600~\rm{kpc}$ (blue curve), and an alternative selection of satellites that adopts the selection criteria of SAGA-II, i.e.~within $300~\rm{kpc}$ (2D) and with line-of-sight velocities of $\pm 275~\rm{km~s}^{-1}$ (pink curve). {\it Bottom right panel:} isolated MW/M31-like vs. LG-like hosts (see \S\ref{sec:hostSelection_LGlike} for their selection criteria). The error bars denote the Poissonian errors In addition to the quenched fractions of galaxies within $300~{\rm kpc}$ (solid curves), we illustrate the quenched fractions of galaxies at larger distances of $300-1000~{\rm kpc}$ to their host galaxy (dashed curves) as the satellites around LG-like hosts may become subject to further environmental effects in the intragroup medium.}
    \label{fig:fquench_hostProps}
\end{figure*}

\subsection{Dependence of satellite quenched fractions on host and satellite properties}
\label{sec:quenchFracs_hostProps}

With a statistical host sample of 198~MW/M31-like galaxies with over 1200 satellites, we are able to investigate the influence of host and satellite properties on their star formation activity and quenched fractions. This should also allow us to quantify whether the host-to-host variation of Fig.~\ref{fig:fquench}, bottom left panel, is completely stochastic or due to some physical effects.

In Fig.~\ref{fig:fquench_hostProps}, we examine the median quenched fractions of various stacked subpopulations of satellites split according to their total host mass $M_{\rm 200c}$ (top left panel), the average accretion time of each host's satellite population $t_{\rm acc,~avg}$ (top right panel), the distance to their host galaxy (bottom left panel), and whether their host is isolated or part of an LG-like system (bottom right, see \S\ref{sec:hostSelection_LGlike} for details on their selection). In all panels, we employ our fiducial selection of satellites (within 300 kpc from their host, \S\ref{sec:satSelection}) and quenched definition (distance from the SFMS).

We find clear trends with total host halo mass and the average accretion time of satellite populations (top panels of Fig.~\ref{fig:fquench_hostProps}). Such trends are significant despite the large host-to-host variations that remain within the two subsets of hosts. Since the satellite-to-host mass ratio in more massive TNG50 MW/M31-like hosts is on average higher, their satellite populations are more easily affected by environmental effects \citep[e.g.][]{DeLucia2012, Wetzel2012, Bahe2017, Davies2019}. Satellite populations with an earlier average accretion time, on the other hand, have been subject to environmental effects for longer, thus increasing their quenched fractions. This extends the results of \cite{Donnari2021a} who find the same trend across TNG300 group and cluster hosts with $M_{\rm 200c} = 10^{13} - 10^{15.2}~\MSUN$ for satellite galaxies with $M_* > 10^9~\MSUN$ (see their figures~3 and 4).

Furthermore, the median quenched fractions continuously decrease for satellites with increasing distance from their host (Fig.~\ref{fig:fquench_hostProps}, bottom left panel). Switching from our fiducial selection, i.e.~galaxies within $300~{\rm kpc}$ (3D, red curve) to the observational selection of SAGA-II using projected distances in combination with a line-of-sight velocity criterion (pink curve) already results in smaller quenched fractions due to the inclusion of star-forming foreground and background galaxies in the sample. Extending the satellite selection to outside the virial radius at $300-600~{\rm kpc}$ diminishes the quenched fractions even more, ranging from 90~to 0~per~cent. The environmental effects that cause satellite galaxies to quench clearly become more effective towards the inner regions of host haloes and satellites at smaller distances have spent longer periods of time exposed to environmental effects -- see a discussion on this degeneracy in \cite{Donnari2021a} and \cite{Joshi2021}.

It has been speculated that the difference between the average SAGA-II and the MW and M31 quenched fractions might occur due to the fact that the Galaxy and Andromeda are in a group-like system. In TNG50 we actually have three systems whose configuration is similar to the Local Group (see \S\ref{sec:hostSelection_LGlike}). While this is only a small amount, we nevertheless attempt to show how their quenched fractions compare to more isolated MW/M31-like hosts (Fig.~\ref{fig:fquench_hostProps}, bottom right panel, red vs. blue curves). We show quenched fractions for both satellites within $300~{\rm kpc}$ of their closest hosts (solid curves) and those at larger distances of $300-1000~{\rm kpc}$ (dashed curves). It would appear as if the environment within the virial radius of the main galaxy of an LG-like system may impact its massive satellites more than in isolated MW/M31-like hosts. However, the Poisson error bars show that this difference is not significant, as we only have a few massive satellites driving these conclusions. Moreover, we would have expected that both the host's companion galaxy and the intragroup medium between them in LG-like systems are able to affect satellites at larger distances and may already quench their star formation activity; this is not the case in Fig.~\ref{fig:fquench_hostProps} but we have checked (although do not show) that the qualitative conclusions change depending on the exact definition of LG analogues. This seems to indicate that we simply do not have sufficient statistics to make a robust statement. The TNG50 LG-like hosts will be further examined in \cite{Pillepich2023}, which introduces our sample of 198~MW/M31-like hosts in detail, whereas the circumgalactic medium of TNG50 MW-like hosts has been extensively characterised by \cite{Ramesh2023}.

We further inspected the satellite quenched fractions for correlations with the host formation time (i.e.~the time at which the host halo had built up 50~per~cent of its present-day total mass), but we do not show it here. There is a slight, apparent trend when considering all TNG50 MW/M31-like hosts across $0.7~{\rm dex}$ of stellar mass -- satellites in hosts that assembled later appear on average more quenched than those around hosts that formed earlier. However, this simply stems from the correlation of host assembly and total host mass \citep{Engler2021b}. If we instead consider smaller, fixed bins of $0.2~{\rm dex}$ in host mass, there is no clear difference between the quenched fractions of satellites around hosts that assembled earlier or later in time.

\begin{figure*}
	\centering
	\includegraphics[width=.8\textwidth]{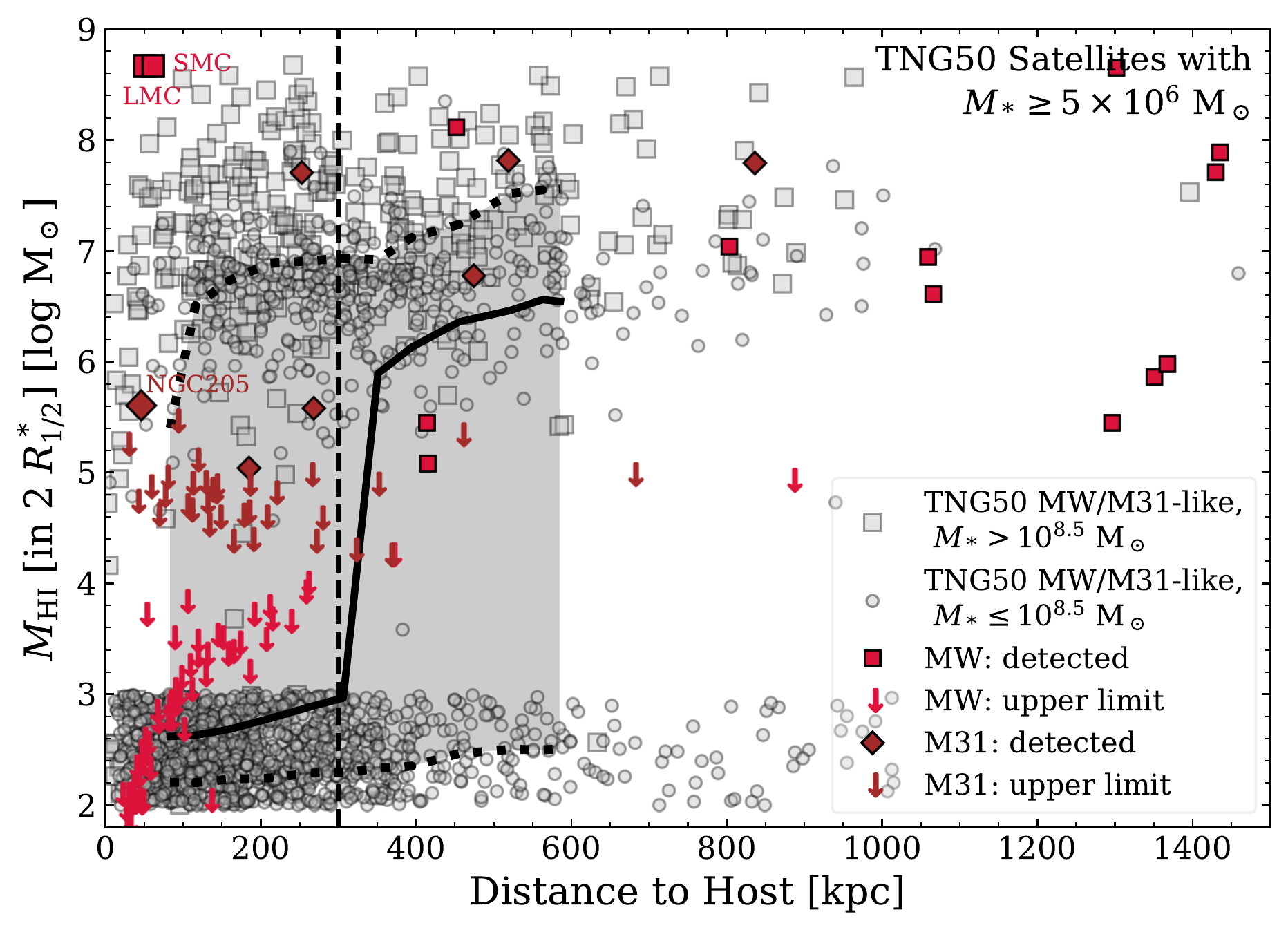}
    \caption{{\bf HI gas mass as a function of 3D distance to their host galaxy} for TNG50 satellites with a stellar mass of at least $M_* \geq 5 \times 10^6~{\rm M}_\odot$ around MW/M31-like hosts (grey dots). Massive satellite galaxies with $M_* > 10^{8.5}~\MSUN$ are shown with larger square symbols. We assign a random HI mass in the $10^{2-3}~{\rm M}_\odot$ range for satellites containing no gas whatsoever, i.e. below the resolution limit. The solid, black curve corresponds to the TNG50 median in bins of host-centric distance, while the dotted, black curves and grey, shaded area denote their scatter as $16^{\rm th}$ and $84^{\rm th}$ percentiles. We compare the TNG50 satellites to the observed satellites of the MW and M31 \protect\citep[3D distance, red squares and brown diamonds, respectively,][]{Putman2021}. While the filled squares and diamonds show detected HI mass measurements, the downward-facing arrows correspond to non-detections and simply denote upper limits. Note that there is no minimum stellar mass for the MW and M31 satellites reported here, i.e.~their samples include even ultra-faint satellite galaxies.}
    \label{fig:mHI_vs_distHost}
\end{figure*}

\section{Gas content of satellites around TNG50 MW/M31-like hosts}
\label{sec:gasContent}

As seen in the previous sections, satellite galaxies around MW/M31-like hosts -- particularly within their virial radius -- have mostly ceased to form stars. In the following, we expand upon the IllustrisTNG results for more massive hosts of \cite{Stevens2019, Stevens2021} and examine how the quenching of satellites after infall correlates with their gas content as fuel for star formation, with a special emphasis on their atomic and molecular hydrogen content (HI and H$_2$, respectively).

\subsection{HI mass vs. distance to host}

In Fig.~\ref{fig:mHI_vs_distHost}, we study the HI gas mass within two stellar half-mass radii $R_{1/2}^*$ of all satellites as a function of the 3D distance to their TNG50 MW/M31-like host. Here, we only consider satellites within the FoF region of their hosts. The HI masses are based on the work of \cite{Popping2019} who calculated the atomic and molecular hydrogen content of TNG50 gas cells in post-processing based on various theoretical recipes: here, we employ HI masses obtained using the metallicity-based approach of \cite{Gnedin2011}. As these HI masses cannot be computed for TNG50 satellites that contain no gas whatsoever (i.e.~contain less than a few $10^4\MSUN$ in gas, which is the target baryonic mass of the gas cells), we simply assign them a random mass of $M_{\rm HI} = 10^{2-3}~{\rm M}_\odot$, detached from the main relation. The grey circles in Fig.~\ref{fig:mHI_vs_distHost} denote individual TNG50 satellites with larger grey squares corresponding to massive simulated satellites with $M_* > 10^{8.5}~\MSUN$. The solid, black curve gives their median whereas the grey, shaded area and dotted, black curves correspond to their scatter as $16^{\rm th}$ and $84^{\rm th}$ percentiles. We compare the HI masses of TNG50 satellites to those of observed dwarfs around the MW and M31 \citep[red squares and brown diamonds, respectively,][]{Putman2021}. Filled symbols correspond to detected HI measurements while the downward-facing arrows have no detected HI content and merely represent upper limits according to the GALFA-HI \citep{Peek2011, Peek2018} and HI4PI \citep{HI4PICollab2016} surveys employed by \cite{Putman2021}. It should be noted that while the TNG50 satellites in Fig.~\ref{fig:mHI_vs_distHost} are restricted to $M_* \geq 5 \times 10^6~{\rm M}_\odot$, there is no stellar mass limitation on the observed satellite galaxies of the MW and M31, which include even ultra-faint dwarfs. We have checked (but do not show) that the scientific conclusions we give below also hold when the comparison is made to the observed satellites with $M_* \geq 5 \times 10^6~{\rm M}_\odot$ only.

While there are TNG50 satellites with and without significant HI content both within $300~{\rm kpc}$ of their host (dashed, vertical line; with the typical virial radii of these MW/M31-like galaxies being $205-270 {\rm kpc}$) and at larger distances out to $\gtrsim 1000~{\rm kpc}$, there is a very steep transition between gas-rich and gas-poor populations. Outside of $300-350~{\rm kpc}$, most satellite galaxies contain a significant amount of HI with $M_{\rm HI} \gtrsim 10^{6-7}~{\rm M}_\odot$. However, these gas masses drop very rapidly for satellites in the inner regions of TNG50 MW/M31-like hosts. Here, the majority of satellite galaxies contain no significant HI (i.e.~below the TNG50 mass resolution limit and below one hundredth of their stellar mass). Were we to distinguish between TNG50 MW-like and M31-like hosts (see \S\ref{sec:MWM31hosts} for details), this transition from gas-rich to gas-poor would be slightly shifted with respect to each other: satellites around more massive and larger M31-like hosts get deprived of their gas, as a population, at slightly larger distances than those around the less massive MW-like hosts.

\begin{figure*}
	\centering
	\includegraphics[width=.75\textwidth]{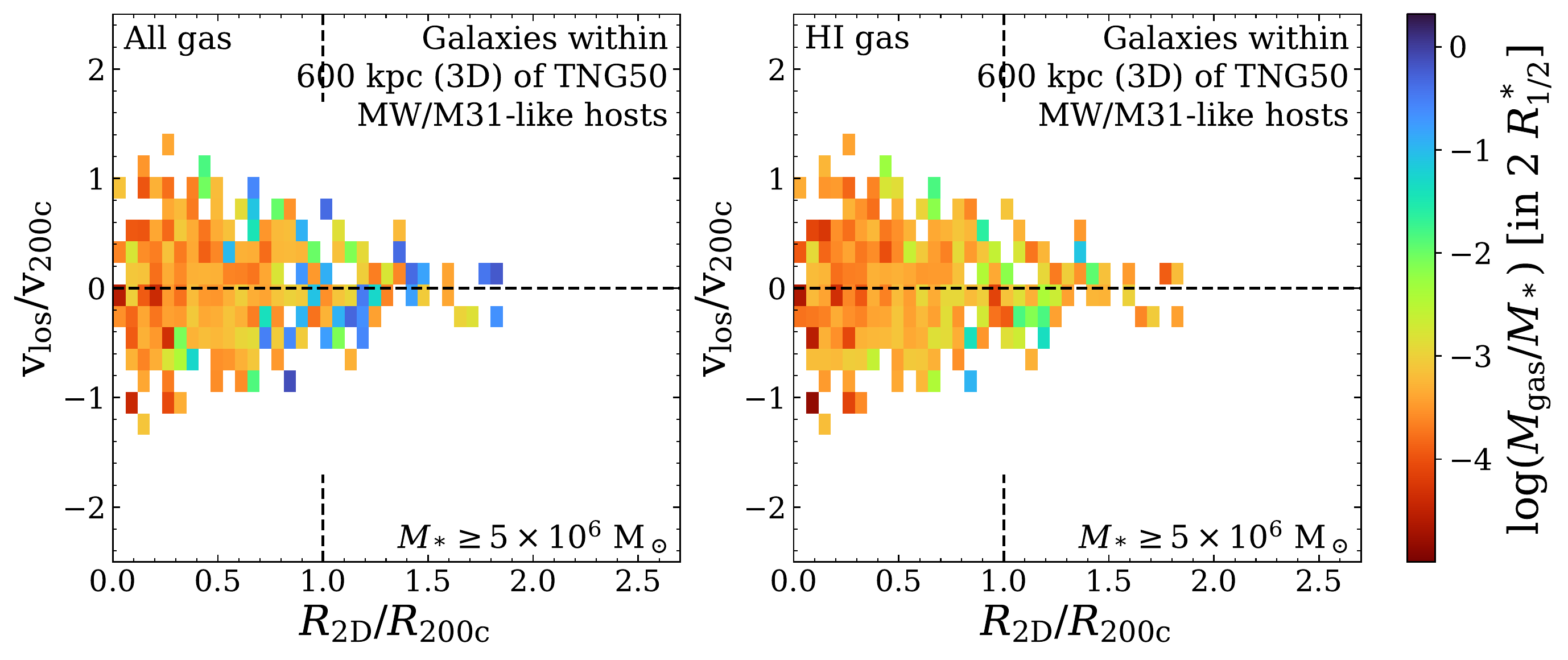}
	\includegraphics[width=.75\textwidth]{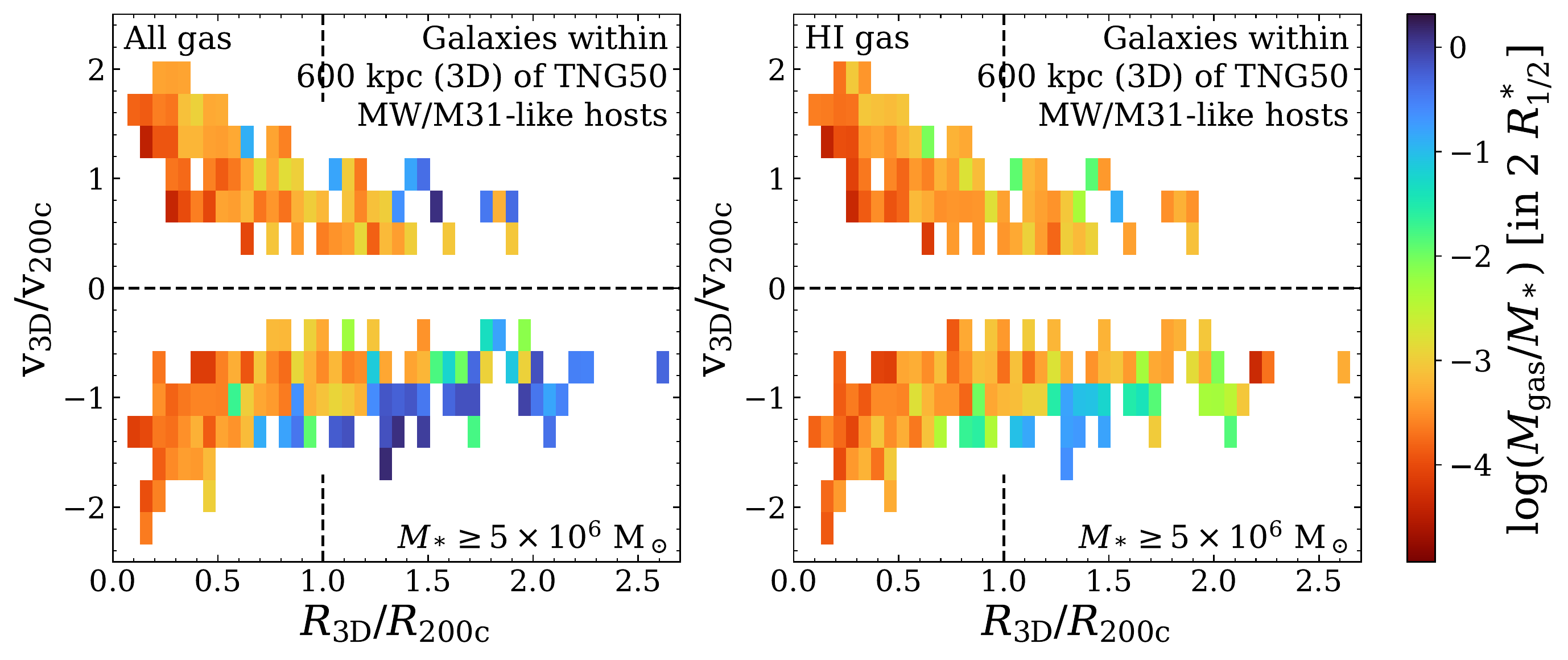}
    \caption{{\bf Phase-space distributions of satellites around TNG50 MW/M31-like hosts colour-coded by gas fractions.} We extend the sample beyond the virial radius (dashed, vertical lines) and include all galaxies with $M_* \geq 5 \times 10^6~{\rm M}_\odot$ within $600~{\rm kpc}$ of their host in the sample. Each bin contains a minimum of 5 satellites. The top panels show the phase-space distribution in a random projection using a 2D distance to their host and the corresponding line-of-sight velocity, while the bottom panels depict the 3D phase-space distribution. The left panels take all gas into account; the right panels are focused exclusively on HI~gas. TNG50 satellites that do not contain any gas are assigned a random gas mass of $10^{2-3}~{\rm M}_\odot$ (for either all gas or HI). It should be noted that while there is a visible gradient, most satellites (particularly within $300~{\rm kpc}$) barely contain gas with gas to stellar mass fractions $< 10$~per~cent, a typical proxy of quenched.}
    \label{fig:phaseSpace_gasFrac}
\end{figure*}

This behaviour is consistent with that of the observed dwarfs around the MW and M31 from \cite{Putman2021}. Most satellites with detected HI content are located outside of the virial radius of their host. Within $300 - 350~{\rm kpc}$, the observed satellite galaxies become dominated by non-detections. The closer they are to their host galaxy, the smaller their upper HI limits become. One notable exception from this are the two red squares in the upper left corner of Fig.~\ref{fig:mHI_vs_distHost} with $M_{\rm HI} \sim 10^{8.7}~{\rm M}_\odot$ at a distance of $\sim 50~{\rm kpc}$. These points correspond to the LMC and SMC and exhibit, at face value, larger HI masses than any of the TNG50 satellites that are located this close to their host galaxy. As the LMC and SMC are the most massive satellite galaxies of the MW and are most likely still on their first infall \citep{Besla2007, Besla2010, Boylan-Kolchin2011, Kallivayalil2013}, they are more resistant to environmental effects and are able to retain their gas more efficiently. The same holds for M31's NGC~205 residing at a distance of $\sim 50~{\rm kpc}$ and many massive TNG50 satellites. Almost all TNG50 satellites within $100~{\rm kpc}$ of their host that still contain significant amounts of HI are relatively massive ($M_* > 10^{8.5}~\MSUN$). Thus, they are more resistant to environmental effects and are able to retain their gas more easily. These results and comparisons, especially within $300~{\rm kpc}$ distance from the hosts, are consistent with the compatibility between the quenched fractions of TNG50 satellites and those observed for the MW and M31 shown in Fig.~\ref{fig:fquench}: the latter are, in fact, based on a gas fraction threshold (see \S\ref{sec:quenchFracs}) and both TNG50 MW/M31 analogues as well as the real MW and M31 exhibit very high quenched fractions (i.e.~high fractions of gas-poor satellites) towards lower-mass satellites.

\subsection{Satellite gas fractions in phase-space}

We extend the study of the gas content of satellite galaxies around TNG50 MW/M31-like hosts to the phase-space in Fig.~\ref{fig:phaseSpace_gasFrac}. Thus, not only their host-centric distance but also their velocity relative to their host are taken into account. We normalise the satellites' distances by their host's virial radius $R_{\rm 200c}$ and their velocities by their host's virial velocity $v_{\rm 200c}$ as a proxy for its velocity dispersion: $v_{\rm 200c} = \sqrt{G M_{\rm 200c} / R_{\rm 200c}}$.

We have verified that the satellites of TNG50 MW-like and M31-like hosts form overall consistent phase-space distributions compared to observed satellite systems of the MW, M31, and the SAGA survey \citep{Pillepich2023}. In order to show their evolution from gas-rich to gas-poor, all galaxies within $600~{\rm kpc}$ of their MW/M31-like hosts are included in the sample and are subsequently divided into 2D bins. Bins containing at least five satellite galaxies are colour-coded by their median gas fraction (blue to red from gas-rich to gas-poor). We consider both all gas (left panels) and specifically HI (right panels) within two stellar half-mass radii (i.e.~within the main body of the galaxy). For TNG50 satellites that contain no gas, we again assign a very low random gas mass of $10^{2-3}~{\rm M}_\odot$ for either all gas or HI. The top panels depict the phase-space distribution of satellites in a random projection using their 2D distance to their host galaxy and their corresponding relative line-of-sight velocity -- similar to how they might arise from observations. In the bottom panels, we employ 3D distance and velocities instead.

Overall, the same trend is clearly visible in all panels of Fig.~\ref{fig:phaseSpace_gasFrac}: gas-rich satellites are predominantly found in the outer parts of phase-space whereas the closer satellites are located towards its central regions, the smaller their average gas fractions are. However, it is important to note that while there is a visible gradient from gas-rich to gas-poor, most satellites -- particularly within $300~{\rm kpc}$ (vertical dashed line) -- barely contain gas, with gas fractions of $< 10$~per~cent. As a reminder, this is a typical gas mass fraction threshold to separate star-forming from quiescent galaxies (see \S\ref{sec:quenchFracs}). Even the galaxies in the outer regions of the phase-space at larger distances to their host with larger all-gas fractions are on average already gas-poor in HI (left vs. right panels). Considering the projected phase-space, no satellite bin contains HI gas fractions of more than 10~per~cent (top right panel); in 3D phase-space, on the other hand, such satellites are visible among infalling populations (i.e.~at negative velocities) outside the virial radius (bottom right panel and its bottom right region). While projection effects can wash out some details of this trend, satellites still display the same development. 

The evolution of the satellites' gas content is best visualised in the bottom left panel of Fig.~\ref{fig:phaseSpace_gasFrac}, which depicts satellite all-gas fractions in 3D phase-space. Gas-rich satellites are predominantly located at larger distances beyond the virial radius at negative, infalling velocities and -- to a lesser degree -- at positive, outgoing relative velocities. These backsplash galaxies may still contain some gas due to their own mass or a favourable orbital configuration. Following their trajectory through phase-space, satellites continue to lose gas to their host: the closer they travel to its central regions towards smaller distances and relative velocities, the gas-poorer they become. At fixed distance within the virial radius, slower moving satellites exhibit smaller gas fractions ($\sim$$10^{-4}$) than faster satellites ($\sim$$10^{-3}$) for both infalling and outgoing galaxies. This picture is consistent with the phase-space distribution of quenched and star-forming TNG300 satellites in \cite{Donnari2021a} who study galaxies with $M_* \geq 10^{9}~\MSUN$ in group and cluster hosts of $M_{\rm 200c} \geq 10^{13}~\MSUN$. Thus, we are able to extend their findings to both smaller galaxy and host masses.

\begin{figure*}
	\centering
	\includegraphics[width=.46\textwidth]{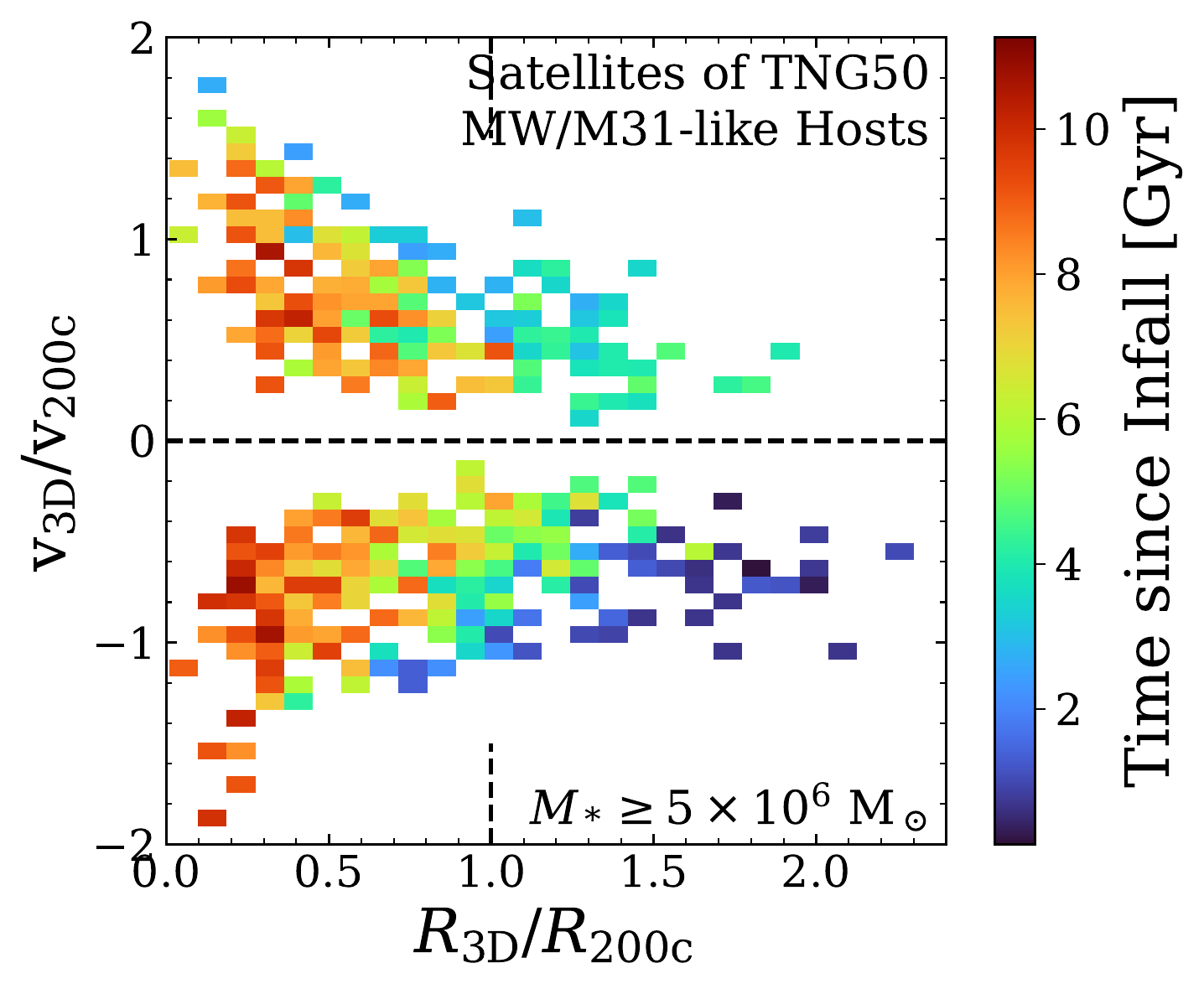}
	\includegraphics[width=.42\textwidth]{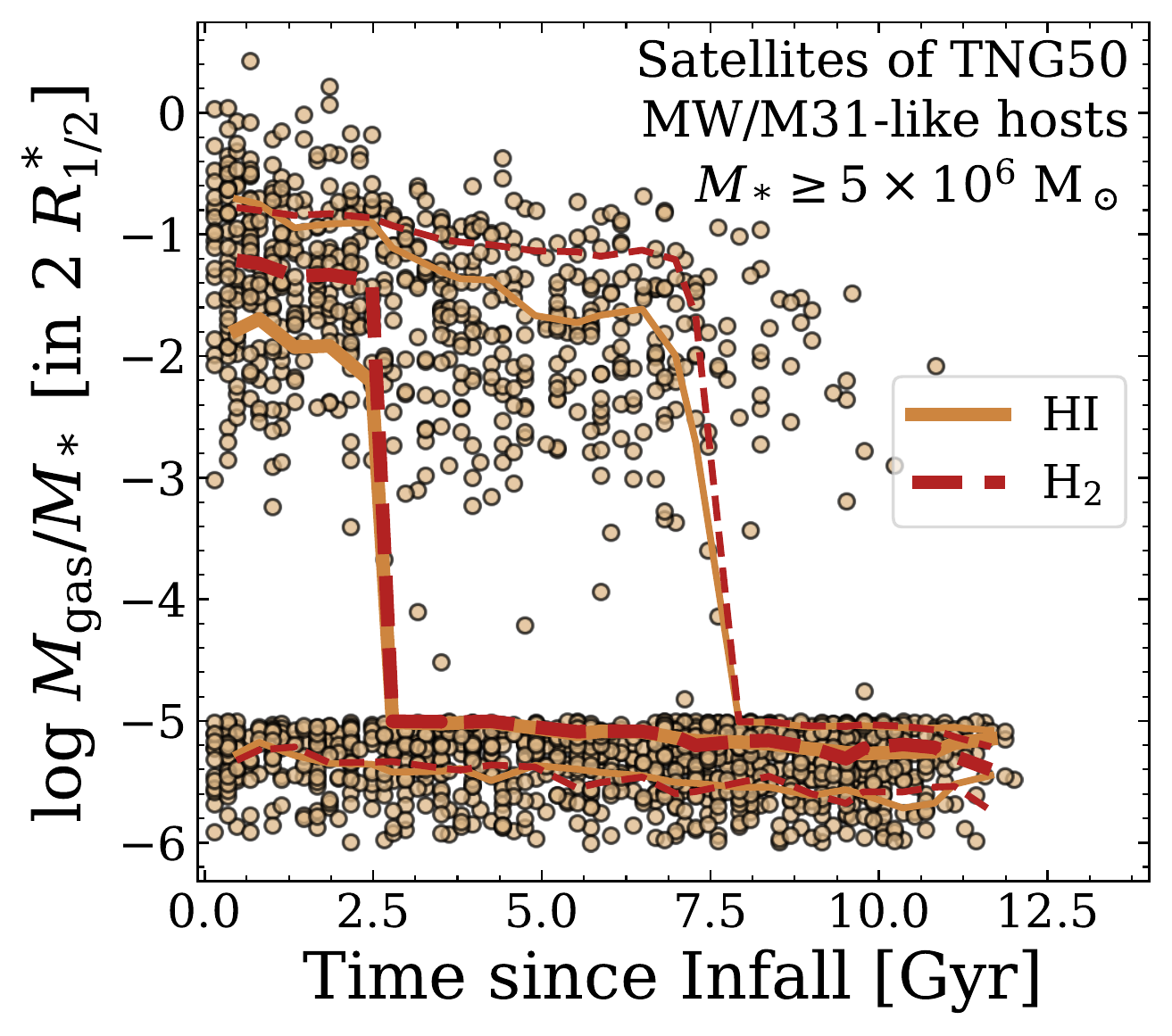}
    \caption{{\bf Correlations of infall time with satellite distribution in phase-space and their gas content.} {\it Left panel:} phase-space distribution of satellite galaxies with $M_* \geq 5 \times 10^6~{\rm M}_\odot$ around TNG50 MW/M31-like hosts, colour-coded by time since infall (red to blue for ancient to recent infall times). Each bin contains a minimum of 3 satellites. {\it Right panel:} satellite gas fractions within two stellar half-mass radii (i.e.~within the galaxy's main body) as a function of time since infall. Brown circles denote HI fractions; the thick, brown, solid and thick, red, dashed curves illustrate the running medians of HI and H$_2$ fractions, respectively, whereas the corresponding thin curves illustrate the scatter of HI and H$_2$ gas fractions as $16^{\rm th}$ and $84^{\rm th}$ percentiles.}
    \label{fig:infall_phaseSpace_gasFracs}
\end{figure*}

\subsection{Correlation of gas content and infall time}

We quantify how the 3D phase-space location of satellites with $M_* \geq 5 \times 10^6~{\rm M}_\odot$ within the FoF haloes of TNG50 MW/M31-like hosts correlates with their time since infall \citep[][with early to late infallers from red to blue]{Chua2017} in the left panel of Fig.~\ref{fig:infall_phaseSpace_gasFracs}. We define infall (or interchangeably, accretion) as the first time these galaxies became satellites of any host. For the majority of satellites, this corresponds to the infall into their present-day MW/M31-like host. As in Fig.~\ref{fig:phaseSpace_gasFrac}, their distribution is divided into 2D bins. Bins containing at least three satellites are colour-coded accordingly. 

The resulting phase-space distribution displays a clear correlation with median infall time: satellites that experienced their first infall only recently are found in the outskirts of phase-space at large distances and predominantly negative, infalling velocities. The average time since infall continuously shifts to earlier times towards the central regions of phase-space with the most ancient infallers closest to their host and with the smallest relative velocities. At fixed distance within the virial radius, satellites with slower velocities clearly fell into their host at an earlier time than faster moving satellites, mirroring the trend with gas fractions depicted in the bottom left panel of Fig.~\ref{fig:phaseSpace_gasFrac}. This correlation of phase-space position and infall time is in agreement with previous studies on the satellite populations of clusters in simulations and observations \citep{Rhee2017, Pasquali2019, Smith2019}. To our knowledge, this is the first study to show that this trend holds for less massive hosts with a statistically significant sample of both satellites and MW/M31-like hosts.

As satellite gas fractions and their time since infall exhibit very similar trends in phase-space, we quantify their correlation with each other directly in the right panel of Fig.~\ref{fig:infall_phaseSpace_gasFracs}, with respect to both HI and H$_2$ content. Again, satellites with no gas are assigned a random gas fraction of $10^{-5} - 10^{-6}$, detached from the main relation. The brown circles illustrate the HI content of individual satellites while the brown, solid and thinner curves denote their median and encompass their scatter ($16^{\rm th}$ to $84^{\rm th}$ percentiles), respectively, as a function of their time since infall. The red, dashed thick and thinner curves give the median and scatter of their H$_2$ gas fractions.

Satellites that fell into their host as recent as $2-3$ billion years ago exhibit a larger fraction of H$_2$ than HI within two stellar half-mass radii (note that, albeit we do not show it here, this trend is reversed when considering all gravitationally bound gas due to the larger abundance of atomic hydrogen in their gaseous haloes). However, both atomic and molecular gas fractions measured within the same aperture exhibit the same correlation for galaxies with intermediate and later times of infall: the longer a galaxy has spent as a satellite, the smaller are its gas fractions in terms of both HI and H$_2$. For satellites that have been accreted more than $2.5~{\rm Gyr}$ ago, the population-wide median gas fraction drops drastically and the satellite populations of TNG50 MW/M31-like hosts become dominated by satellites containing amounts of gas that are too small to resolve numerically -- the equivalent of observational non-detections.

\section{Stellar assembly of satellite galaxies}
\label{sec:stellarAssembly}

How did the $z=0$ star formation state of satellite galaxies around TNG50 MW/M31-like hosts come about? Throughout this section, we address this question by focusing on our fiducial satellite selection: simulated galaxies with $M_* \geq 5 \times 10^6~\MSUN$ within $300~{\rm kpc}$ (3D) of their MW/M31-like hosts.

\subsection{Cumulative star formation histories}
\label{sec:SFHs}

In this section, we investigate the build-up of stellar mass in satellites and their assembly throughout cosmic time using normalised, cumulative star formation histories~(SFHs).
We follow the approach previously adopted by \cite{Joshi2021} and for each galaxy, we construct a histogram of the formation times of all gravitationally bound stellar particles, weighted by their present-day mass, in bins of $50~{\rm Myr}$. The resulting SFHs are characterised by specific stellar assembly times $\tau_{10}$, $\tau_{50}$, and $\tau_{90}$, which we obtain by interpolation. They correspond to the lookback times by which a given satellite had formed 10, 50, and 90~per~cent of its present-day stellar mass. 

\begin{figure*}
    \centering
    \includegraphics[width=0.89\textwidth]{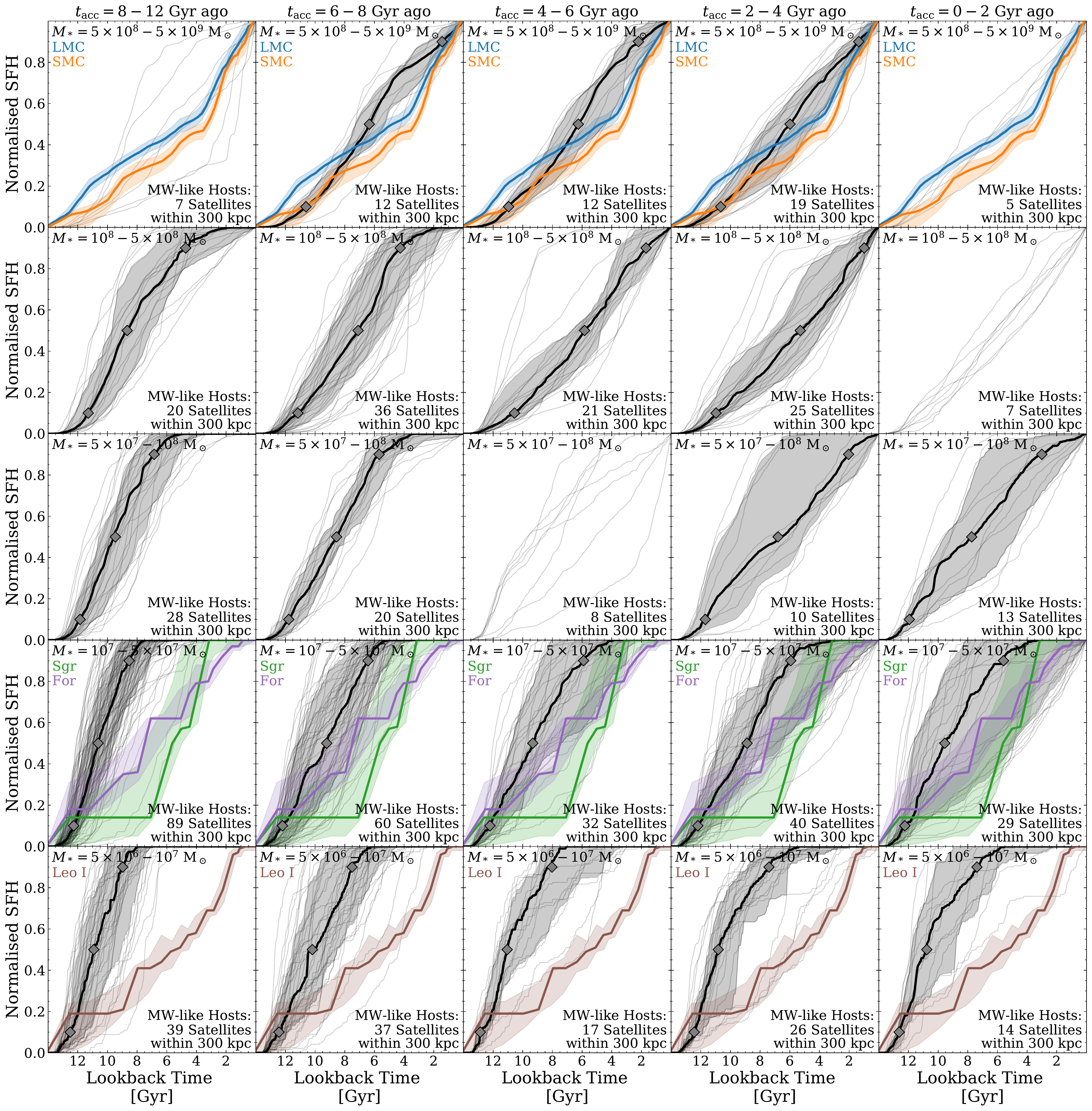}
    \caption{{\bf Cumulative star formation histories (SFHs) of satellites within $\mathbf{300~{\rm \bf kpc}}$ of TNG50 MW-like hosts (i.e.~with $\mathbf{M_{\boldsymbol{*}} \boldsymbol{=} 10^{10.5} \boldsymbol{-} 10^{10.9}~{\rm \bf M}_{\boldsymbol{\odot}}}$)}. Each row depicts a different bin in satellite stellar mass (decreasing from top to bottom) whereas each column corresponds to a different bin of time since infall (accretion time $t_{\rm acc}$, with early to late infallers from left to right). The thin, grey curves show the cumulative SFHs of individual TNG50 satellites. For bins containing at least ten satellite galaxies, we compute the median SFH and its scatter as $16^{\rm th}$ and $84^{\rm th}$ percentiles (black curves and grey shaded areas, respectively). The grey diamonds denote the median stellar assembly times $\tau_{10}$, $\tau_{50}$, and $\tau_{90}$, i.e.~ the times at which 10, 50, and 90~per~cent of the present-day stellar mass has been assembled. Furthermore, we compare the normalised, cumulative SFHs of TNG50 satellites to those of observed satellites around the MW (coloured curves): the LMC and SMC \protect\citep{Weisz2013}, as well as Sagittarius, Fornax, and Leo~I \protect\citep{Weisz2014a} using the best fit SFHs and their 68~per~cent confidence intervals due to random and systematic uncertainties.}
    \label{fig:sfh_MWlike}
\end{figure*}

\begin{figure*}
    \centering
    \includegraphics[width=0.89\textwidth]{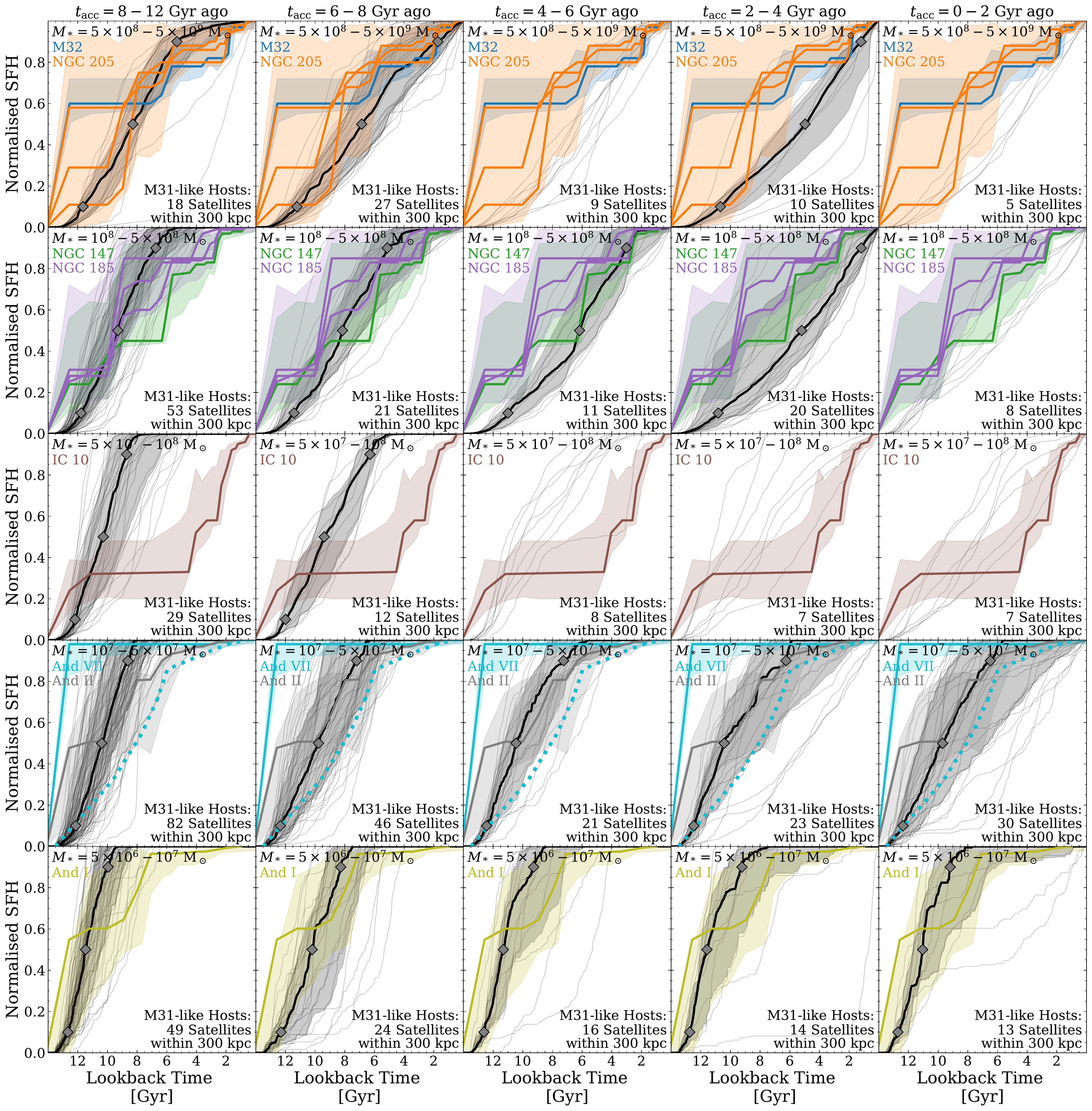}
    \caption{{\bf Cumulative SFHs of satellites within $\mathbf{300~{\rm \bf kpc}}$ of TNG50 M31-like hosts (i.e.~with $\mathbf{M_{\boldsymbol{*}} \boldsymbol{=} 10^{10.9} \boldsymbol{-} 10^{11.2}~{\rm \bf M}_{\boldsymbol{\odot}}}$)}. Annotations are as in Fig.~\ref{fig:sfh_MWlike}. We compare the TNG50 SFHs to those of observed satellites around M31 \protect\citep[coloured curves,][]{Weisz2014a, Skillman2017} using the best fit SFHs and their 68~per~cent confidence intervals due to random and systematic uncertainties. For And~VII, we additionally show an alternative SFH \protect\cite[dotted cyan curve,][]{Navabi2021}. As the observations of NGC~205 and NGC~185 both cover multiple fields, the SFH for each field is shown separately for these dwarfs.}
    \label{fig:sfh_M31like}
\end{figure*}

Figs.~\ref{fig:sfh_MWlike} and~\ref{fig:sfh_M31like} depict the normalised, cumulative SFHs -- i.e.~the fraction of their present-day stellar mass as a function of lookback time -- of TNG50 satellites within $300~{\rm kpc}$ of 138 MW-like and 60 M31-like hosts, respectively (see \S\ref{sec:MWM31hosts}), as a function of both their own stellar mass (decreasing from top to bottom panels) and their time since infall (or accretion; early to late infallers from left to right columns). In each panel, the thin, grey curves depict the SFHs of individual satellites while the thick, black curves and grey shaded areas (in bins containing at least ten satellites) illustrate their median and scatter as $16^{\rm th}$ and $84^{\rm th}$ percentiles. The grey diamonds correspond to their median stellar assembly times $\tau_{10}$, $\tau_{50}$, and $\tau_{90}$. There are 24 (11) galaxies around TNG50 MW-like (M31-like) hosts whose SFHs are not shown in Figs.~\ref{fig:sfh_MWlike} and~\ref{fig:sfh_M31like} as their stellar masses exceed $5 \times 10^9~\MSUN$ and thus the most massive bin. There are no present-day surviving galaxies around TNG50 MW/M31-like hosts that were accreted earlier than $12~{\rm Gyr}$ ago.

Overall, the median SFHs in both Figs.~\ref{fig:sfh_MWlike} and~\ref{fig:sfh_M31like} display the same trends with satellite stellar mass and infall time: more massive satellites and those that experienced a later infall exhibit more extended SFHs. As less massive satellites and early infallers are affected by their environment to a stronger degree, they quench earlier than their more massive or late-infall counterparts. Thus, their stellar assembly ends earlier, resulting in a shorter SFH. While these trends hold for the SFHs of individual satellites as well, individual galaxies exhibit a significant degree of diversity. In fact, many satellites display a stellar assembly that is substantially different to the median of a given stellar mass and infall time bin \citep[see also][]{Joshi2021}. Furthermore, we find more extended SFHs for satellites at greater distances of $300-600~{\rm kpc}$ to their host galaxy (see Appendix~\ref{sec:AppSFHs_600kpc} and Fig.~\ref{fig:sfh_in600kpc}). As their environmental effects become weaker outside of their virial radius, galaxies can still continue to form stars (see also the bottom left panel of Fig.~\ref{fig:fquench_hostProps}). This is in agreement with previous findings from the FIRE-2 simulations on the SFHs of satellites around isolated MW-like and LG-like hosts \citep{GarrisonKimmel2019}. The scatter of individual SFHs, as well as their correlations with satellite stellar mass, their time since infall, and the distance to their host are consistent with the results of \cite{Joshi2021} for the general TNG50 dwarf population with $M_* = 10^{7} - 10^{10}~\MSUN$ in hosts of $M_{\rm 200c} = 10^{12} - 10^{14.3}~\MSUN$. \cite{Joshi2021} also find more extended SFHs for satellites that inhabit outer phase-space regions.

In Fig.~\ref{fig:sfh_MWlike}, we compare the TNG50 SFHs to those of observed dwarfs around the MW that satisfy the same selection criteria (i.e.~with $M_* \geq 5 \times 10^6~\MSUN$ and within $300~{\rm kpc}$ of the MW): the Large and Small Magellanic Clouds \citep[LMC and SMC, respectively,][]{Weisz2013}, as well as Sagittarius, Fornax, and Leo~I \citep{Weisz2014a}. Whereas other MW dwarf spheroidals (dSph), such as Ursa Minor or Draco, would represent more ``typical'' dSphs, they are less massive than our satellite mass range and are thus not included in this comparison. The SFHs of MW dwarfs, based on analyses of their colour-magnitude diagrams, are shown across all accretion time bins. Due to the different methods of obtaining cumulative SFHs between observed and simulated satellites, some differences are to be expected -- especially at early cosmic times since the construction of observational SFHs in this range can be particularly challenging. Similarly, in Fig.~\ref{fig:sfh_M31like}, we compare the SFHs of the satellite populations of our TNG50 M31-like hosts to those of the observed dwarfs around M31, mostly obtained by colour-magnitude diagram analyses by \cite{Weisz2014a}. For And~II and And~I, we employ the more recently published SFHs from the Initial Star formation and Lifetimes of Andromeda Satellites (ISLAndS) project \citep{Skillman2017}; for And~VII, we include a second SFH based on long-period variable stars \citep[dotted cyan curve,][]{Navabi2021}. In both cases, we replicate the SFHs from observations in all columns, i.e. without assuming any prior on the accretion time of observed satellites. 

Similar to the case of individual TNG50 galaxies, the SFHs of the  observed MW and M31 satellites appear, in most cases, distinct from the median TNG50 SFHs at the corresponding mass. While there is some overlap between the stellar assembly of the LMC and SMC with the median and $1\sigma$~scatter of the TNG50 SFHs across all infall time bins, the three less massive MW satellites exhibit a shift to a more recent assembly compared to the TNG50 median in a given bin of infall time. Nevertheless, they are in agreement with the SFHs of individual TNG50 satellites. Analogously, within the large sample of TNG50 simulated galaxies, it is possible to identify SFHs of individual TNG50 galaxies that are similar to those of the observed M31 satellites.

\subsection{Hints on the infall times of observed satellites}
\label{sec:infalltimes}

Thanks to the compatibility between TNG50 and observed SFHs and despite the large galaxy-to-galaxy variation, Figs.~\ref{fig:sfh_MWlike} and~\ref{fig:sfh_M31like} suggest that we can roughly estimate the infall times of observed satellites based on their stellar assembly. Namely, we can utilise the offsets between the typical SFHs of simulated galaxies of a given infall time (different columns) and those of observed satellites to infer an accretion period for the observed MW and M31 satellite galaxies. The particularly extended SFHs of the LMC and SMC are best in agreement with a late infall $0-4~{\rm Gyr}$ ago. This is consistent with previous studies suggesting the Magellanic Clouds to be on a first infall scenario and inferring an accretion time of $\sim 2~{\rm Gyr}$ ago \citep{Besla2007, Besla2010, Laporte2018a}. For both Sagittarius and Fornax, estimating their infall times is less clear: comparing their SFHs to the TNG50 median SFHs and their $1\sigma$~scatter suggests a relatively recent infall, however, they are consistent with individual SFHs of TNG50 across all accretion time bins up to $8-12~{\rm Gyr}$ ago, albeit slightly less extended for the earliest infallers. In fact, previous studies favour an intermediate to early infall for Sagittarius around $z \sim 1$ since it is a tidal remnant in the process of being disrupted \citep{Dierickx2017a, Dierickx2017b, Laporte2018b} whereas the infall time of Fornax can range between $2-9~{\rm Gyr}$ ago \citep{Rocha2012, Wang2016} with recent observations suggesting a first infall scenario as it contains a young stellar population formed $100~{\rm Myr}$ ago with an associated HI cloud \citep{Bouchard2006, DeBoer2013, Yang2022}. The extended stellar assembly of the Leo~I dwarf galaxy hints at a more recent accretion of $0-4~{\rm Gyr}$ ago, consistent with previous estimates of $2~{\rm Gyr}$ ago as it is only weakly gravitationally bound to the MW \citep{Rocha2012, Bajkova2017, Fillingham2019}.

In the case of M31's satellites, compared to the stellar assembly of TNG50 satellites, the recent SFH of M32 hints at an infall into the halo of M31 around $4-8~{\rm Gyr}$ ago. Previous studies estimate that M32's progenitor started to interact with M31 $\sim 5~{\rm Gyr}$ ago \citep[e.g.][]{DSouza2018} -- consistent with our infall estimate. The three orange curves of NGC~205 illustrate the SFHs obtained from observations of different fields. While there is a significant degree of scatter between these SFHs -- particularly at early times where the extraction of SFHs is difficult -- they suggest an infall of $6-8~{\rm Gyr}$ ago. As NGC~205 is not actively star-forming despite its large stellar mass, previous estimates have attributed an early infall of several Gyr or even more than $9.5~{\rm Gyr}$ ago \citep{Wetzel2015, Angus2016}. Both NGC~147 and the three observed fields of NGC~185 imply an accretion at early to intermediate times. NGC~147 exhibits a slightly later assembly and its SFH is well in agreement with the TNG50 median of the satellites that were accreted $4-6~{\rm Gyr}$ ago. NGC~185, on the other hand, probably experienced an earlier infall: while its early assembly agrees well with the TNG50 median in the $t_{\rm acc} = 8-12~{\rm Gyr}$ ago bin, its later assembly is similar to the median SFH of satellites with $t_{\rm acc} = 6-8~{\rm Gyr}$ ago. Previous works estimated a similar infall of around $5~{\rm Gyr}$ ago for NGC~147 and $8~{\rm Gyr}$ ago for NGC~185 \citep[e.g.][]{Geha2015}. Proposed to be on its first infall \citep{Wetzel2015a}, the dwarf irregular IC~10 exhibits a relatively recent assembly of stellar mass, most consistent with an infall $0-2~{\rm Gyr}$ ago according to TNG50. The two observed SFHs for And~VII from \cite{Weisz2014a} (solid curve) and \cite{Navabi2021} (dotted curve) depict two very distinct assemblies. While according to the \cite{Weisz2014a} SFH And~VII built up practically all of its stellar mass earlier than $12~{\rm Gyr}$ ago -- earlier than for any TNG50 satellite -- they argue that their extraction of And~VII's SFH based on its colour-magnitude diagram was not appropriate as their observations are too shallow thus excluding essential features such as the red clump or the horizontal branch. And~VII's SFH based on long-period variable stars from \cite{Navabi2021}, on the other hand, is consistent with several SFHs of individual TNG50 satellites. Due to the significant degree of scatter between the TNG50 SFHs, it is hard to constrain the infall time of And~VII, And~II, and And~I. While there is a level of agreement with the TNG50 medians and their $1\sigma$ scatter, as well as with the SFHs of individual TNG50 satellites from the most recent infall up to $8~{\rm Gyr}$ ago, previous works expect them to be accreted at least $5~{\rm Gyr}$ ago due to their quenched state \citep{Wetzel2015a, Fouquet2017}.

\subsection{Correlation of stellar assembly and satellite luminosity}

\begin{figure*}
    \centering
    \includegraphics[width=.85\textwidth]{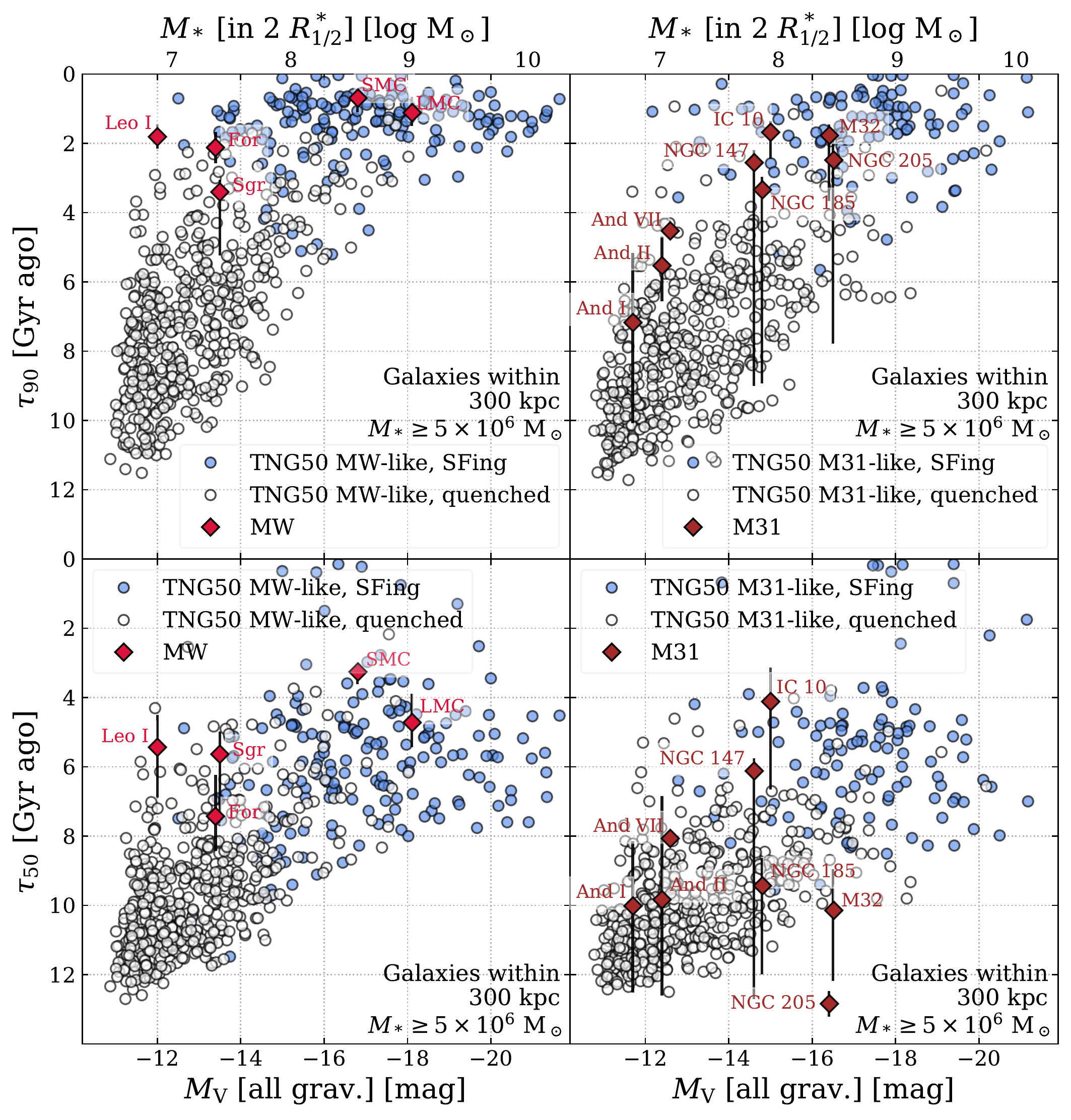}
    \caption{{\bf Satellite stellar assembly times as a function of absolute $V$-band magnitude and stellar mass.} We define satellites (circles) as all galaxies with $M_* \geq 5 \times 10^6~\MSUN$ within $300~{\rm kpc}$ of either TNG50 MW-like hosts (left panels) or M31-like hosts (right panels) and define their stellar assembly in terms of $\tau_{90}$ (top panels) and $\tau_{50}$ (bottom panels), i.e.~the lookback times at which the satellites had assembled 90 and 50~per~cent of their present-day stellar mass. Filled blue circles denote galaxies that are star-forming or in the green valley at $z=0$, according to their distance to the SFMS; empty circles represent quenched satellites. Furthermore, we compare our TNG50 satellites to the observed dwarfs of the MW and M31 \protect\citep[red and brown diamonds, respectively,][]{Weisz2013, Weisz2014a, Navabi2021}.}
    \label{fig:stellarAssembly_vs_Vmag}
\end{figure*}

We investigate the SFH shapes of TNG50 satellites more quantitatively by inspecting their stellar assembly times $\tau_{90}$ and $\tau_{50}$. Fig.~\ref{fig:stellarAssembly_vs_Vmag} illustrates $\tau_{90}$ (top panels) and $\tau_{50}$ (bottom panels) as a function of absolute $V$-band magnitude $M_{\rm V}$ and stellar mass for star-forming and quenched satellites (blue and empty circles, respectively), separately for MW-like (left panels) and M31-like hosts (right panels, see \S\ref{sec:MWM31hosts} for details). We compare the TNG50 satellites to the corresponding observed dwarfs of the MW and M31 (red and brown diamonds, respectively) with $\tau_{90}$ and $\tau_{50}$ extracted from the cumulative SFHs from \citealt{Weisz2013, Weisz2014a} and \citealt{Navabi2021}, and $M_{\rm V}$ taken from \citealt{McConnachie2012}.

Overall, all panels in Fig.~\ref{fig:stellarAssembly_vs_Vmag} depict the same average trend for TNG50 satellites: brighter and more massive satellites exhibit a more extended stellar assembly than fainter satellites, both in terms of $\tau_{90}$ and $\tau_{50}$ -- consistent with the previous findings of \cite{Joshi2021} regarding the general dwarf population in TNG50. However, there is a significant degree of scatter. Even the faintest satellites in our sample with $M_{\rm V} \sim -12$ may reach similarly late assembly times as their brighter counterparts of $\tau_{50} \sim 4-8~{\rm Gyr}$ ago or $\tau_{90} \sim 0-4~{\rm Gyr}$ ago. For brighter satellites with $M_{\rm V} \sim -18$ to~$-22$, the relation of stellar assembly and luminosity for satellites around TNG50 MW-like hosts flattens as these galaxies are still actively star-forming. 
Comparing MW-like and M31-like hosts (left vs. right panels of Fig.~\ref{fig:stellarAssembly_vs_Vmag}), we find slightly different shapes. At fixed luminosity, satellites around M31-like hosts exhibit earlier assembly times both in terms of $\tau_{\rm 90}$ and $\tau_{\rm 50}$ -- i.e.~they cease to form stars earlier -- than satellites in MW-like hosts. This could also be recognised by comparing the median SFHs of Figs.~\ref{fig:sfh_MWlike} and \ref{fig:sfh_M31like} in stellar mass and infall time bins. Thus, the relation between stellar assembly time and satellites' mass or magnitude in MW-like systems is steeper than in M31-like systems, particularly at fainter luminosities. All this is reasonable considering the larger mass of M31 analogues and consistent with our findings in \S\ref{sec:gasContent} and Fig.~\ref{fig:mHI_vs_distHost}. As satellites in more massive and more extended hosts are deprived of their gas reservoirs sooner after infall, the build-up of their stellar mass is shorter than in less massive hosts. As for the observed MW and M31, the evolution of satellite populations may vary significantly even between similar-mass hosts \citep{Weisz2014c}. While \cite{Joshi2021} have demonstrated that satellites in more massive hosts exhibit less extended SFHs across a larger range of host masses of $10^{12}-10^{14.3}~\MSUN$, this trend even holds within a narrower mass range for MW/M31-like hosts.

Comparing the simulation and observations, the stellar assembly times of the observed MW and M31 satellites are realised by the TNG50 model and fall within the loci occupied by the TNG50 satellites. The MW satellites included here all exhibit relatively extended SFHs -- consistent with the results of \cite{Skillman2017} regarding MW dSphs of the ISLAndS project -- with the latest stellar assembly times (both in terms of $\tau_{90}$ and $\tau_{50}$) for the LMC and SMC as the brightest satellites. While the M31 satellites show the same trend for $\tau_{90}$ as the MW's, their intermediate assembly $\tau_{50}$ as a function of $V$-band luminosity exhibits a larger degree of scatter.

It should be kept in mind that the results of Fig.~\ref{fig:stellarAssembly_vs_Vmag} stem from many simulated satellites across many MW/M31-like hosts. We have considered, although do not show, the distribution of satellite systems of {\it individual} TNG50 MW/M31-like hosts across these relations: systems like the observed MW and M31 -- i.e.~an exclusively late assembly for both $\tau_{90}$ and $\tau_{50}$ for its more massive satellites with $M_* \geq 5 \times 10^6~\MSUN$ in the case of the MW and an on average late assembly for $\tau_{90}$ and a diverse assembly for $\tau_{50}$ in the case of M31 -- are fairly rare. While some of the TNG50 hosts are surrounded by satellites with a similar stellar assembly to those observed for MW and M31 dwarfs, their satellite abundance is either smaller with only one or two satellites in total, or they typically contain at least one other satellite galaxy with an intermediate or early stellar assembly. If we were to extend our satellite mass range towards smaller stellar masses, the same would be visible for the observed dwarfs of the MW since its less massive dSphs almost exclusively exhibit an early assembly. The opposite case, in which all satellites form their stellar mass relatively early in time, is actually more common throughout our sample of TNG50 MW/M31-like hosts -- similar to the more typical, less massive MW dSphs not included in this comparison. Finally, we have found no discernible trends according to TNG50 between the stellar assembly times of satellites and the distance to their MW-like or M31-like host.

\subsection{Quenching times}
\label{sec:quenchingtimes}

Fig.~\ref{fig:stellarAssembly_vs_Vmag} allows us to extract numbers for the time since quenching predicted by TNG50 for satellites of MW/M31-like galaxies. There we distinguish quenched (empty) vs. star-forming (filled blue circles) satellites according to their sSFR and distance from the SFMS at $z=0$ \citep{Pillepich2019}. Assuming that $\mathbf{\tau_{90}}$ is a good proxy for the time since quenching for currently quenched galaxies, according to TNG50, the average MW/M31-like satellites with $M_* \geq 5 \times 10^{6}~\MSUN$ quenched $7.5\substack{+2.0\\-2.5}$~Gyr ago ($16^{\rm th}-84^{\rm th}$ percentiles). This compares to $6.9\substack{+2.5\\-3.3}$ Gyr when using the quenching times of \cite{Joshi2021}. Satellites in M31-like hosts quench on average $1.1~{\rm Gyr}$ earlier than those in MW-like hosts. On the other hand, looking at the lower panels of Fig.~\ref{fig:stellarAssembly_vs_Vmag}, $\mathbf{\tau_{50}}$ is not a good predictor for the satellites' present-day star formation state.

\subsection{Correlation of stellar assembly and infall}

\begin{figure}
    \centering
    \includegraphics[width=\columnwidth]{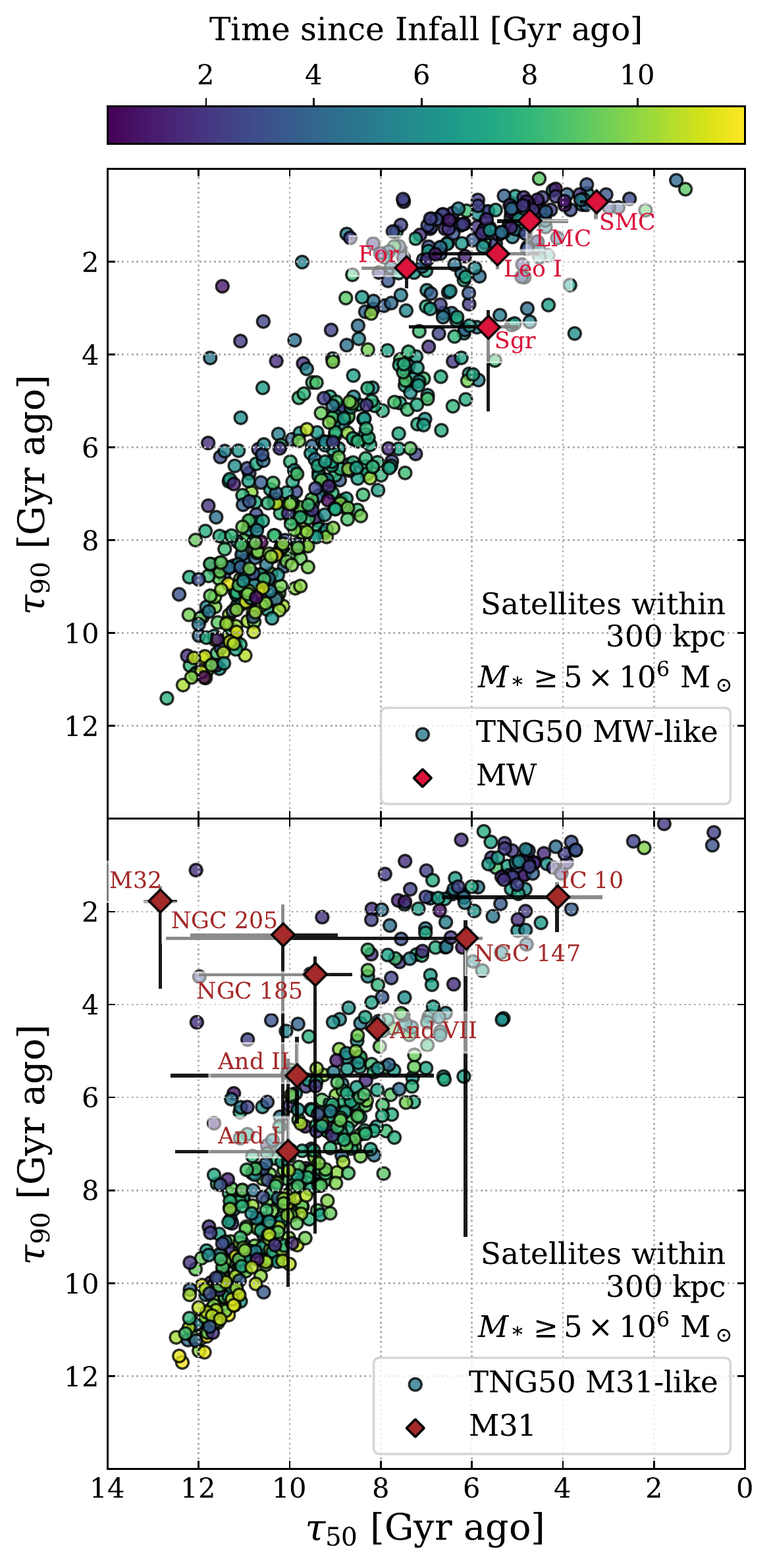}
    \caption{{\bf Relation between stellar assembly times $\mathbf{\tau_{90}}$ and $\mathbf{\tau_{50}}$ and time since infall.} We consider satellites with $M_{*} \geq 5 \times 10^6~{\rm M}_{\odot}$ within $300~{\rm kpc}$ of TNG50 MW-like hosts (top panel) and M31-like hosts (bottom panel) separately, colour-coded by their time since infall (purple to yellow circles). The stellar assembly times of observed MW and M31 satellites (red and brown diamonds) are as in Fig.~\ref{fig:stellarAssembly_vs_Vmag}.}
    \label{fig:t90_vs_t50}
\end{figure}

The relationship between late assembly time $\tau_{90}$, intermediate assembly time $\tau_{50}$, and time since infall is examined in Fig.~\ref{fig:t90_vs_t50}\footnote{There is a less populated region around $\tau_{90} \sim 2-4~{\rm Gyr}$ ago for satellites of both TNG50 MW-like and M31-like hosts, and visible also in Fig.~\ref{fig:stellarAssembly_vs_Vmag}, top panels. As we only consider satellites that currently reside within $300~{\rm kpc}$ of their host, this excludes backsplash galaxies, i.e.~galaxies that already experienced their first infall into their host environment but temporarily reside outside of our selection aperture. This results in a bimodal distribution of infall times \citep{Yun2019, Engler2021a} and an apparent gap in the relation of stellar assembly times.}.

Overall, satellites of TNG50 MW-like and M31-like hosts form the same relation of $\tau_{90}$ and $\tau_{50}$ as well as depict a clear correlation of stellar assembly with their time since infall: satellites that formed their stellar mass particularly early -- with $\tau_{50} \sim 12~{\rm Gyr}$ ago and $\tau_{90} \sim 10 - 12~{\rm Gyr}$ ago -- predominantly fell into their hosts at very early times while galaxies that reached 90~per cent of their present-day stellar mass within the last $3~{\rm Gyr}$ only became satellites in recent times. However, late infallers can be found across the whole range of $\tau_{50}$ and $\tau_{90}$; those with earlier stellar assembly times quenched before they became satellites.

Furthermore, the relation of satellites around MW-like hosts exhibits a larger scatter than for the satellites in TNG50 M31-like systems. Whereas the deeper potentials of the more massive M31-like hosts have more similar impacts on their satellite populations, SFHs of satellites around MW-like galaxies exhibit a larger diversity. Due to their shallower host potentials, other aspects such as the satellites' orbital parameters or their own mass may play a more significant role in shaping their SFHs. Yet, we have checked but do not show that, according to TNG50, there is no residual correlation of satellite stellar assembly times with host halo mass, and merely a slight but not significant trend with early and intermediate host assembly.

As already noted for the case of Fig.~\ref{fig:stellarAssembly_vs_Vmag}, the stellar assembly times of observed MW and M31 dwarfs are well represented by analogue TNG50 satellites, i.e. the observed assembly times fall within the relation predicted by TNG50. While the five most massive MW satellites all exhibit a relatively recent build-up of their stellar mass with $\tau_{50} < 8~{\rm Gyr}$ ago and $\tau_{90} < 4~{\rm Gyr}$ ago --  thereby implying a recent infall on average (see \S\ref{sec:SFHs}) -- the satellites of M31 display a larger degree of scatter with a much earlier stellar assembly. This includes several galaxies that assembled the majority of their stars at $\tau_{50} > 9~{\rm Gyr}$ ago but only reached 90~per~cent of their present-day stellar mass in the recent $4~{\rm Gyr}$. It should be emphasised, however, that the recent assembly noted above only holds for the most massive MW satellites; most other, less massive MW dwarfs actually exhibit an early stellar assembly and are dominated almost exclusively by old stellar populations. The ISLAndS project on M31 dSphs \citep{Skillman2017} previously had found no dwarfs with early $\tau_{50}$ and late $\tau_{90}$. While we limit our selection to the brighter half of the classical satellite regime, \cite{Weisz2019} were able to find even more ancient systems consistent with exponentially declining SFHs by studying fainter satellites of M31 and the MW ($M_{\rm V} < -6.1~{\rm mag}$).

\section{Discussion and further remarks}
\label{sec:discussion}

The compatibility that we have quantified in this paper between the TNG50 outcome and many observations of dwarf galaxies in the Local Volume confirms the underlying IllustrisTNG galaxy formation model and the large statistical power of the simulation. Moreover, it lends credibility to the numerous trends and secondary effects that we have uncovered and quantified in this paper. However, we have not yet pinned down in detail the physical causes for which satellites of MW/M31-like hosts across our mass range quench. In the following, we note a couple of considerations that could serve as useful starting points and inspiration for future analyses. Moreover, we close by discussing the implications of our results on the underlying galaxy formation model in comparison to previous ones in the literature.

\begin{figure*}
    \centering
    \includegraphics[width=.9\textwidth]{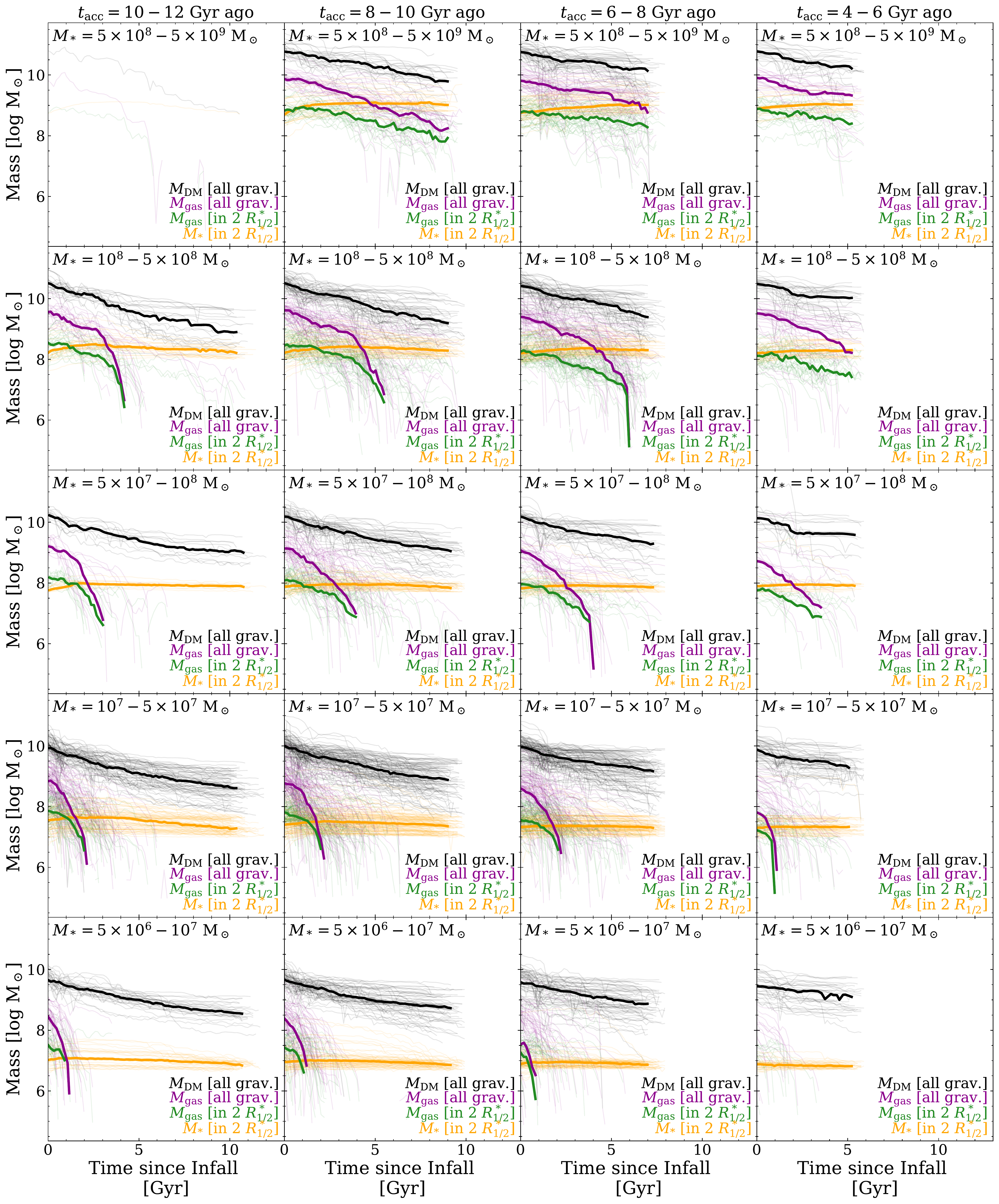}
    \caption{{\bf Evolution of mass components after infall for present-day satellite galaxies around TNG50 MW/M31-like hosts.} We divide satellite populations by their stellar mass and their time of first infall into any host. For the majority of satellites, however, this corresponds to the infall into their present-day MW/M31-like host. Each column corresponds to a different infall time period (with early to late infall from left to right), each row denotes a different satellite stellar mass range (decreasing from top to bottom). The most massive bin in the top row spans a larger range in stellar mass than the others to account for the smaller abundance of massive satellites around MW/M31-like hosts. Each panel illustrates the evolution of all gravitationally bound dark matter (i.e.~across the halo; black curves), all gravitationally bound gas (purple curves), gas within $2 R_{1/2}^*$ (i.e.~within the main body of the galaxy; green curves), and stellar mass within $2 R_{1/2}^*$ (orange curves). Thin curves in the background correspond to the mass evolution of individual satellites while the thick curves depict their median. Since the top left panel only contains a single galaxy, we only show its individual evolutionary tracks. While the median curves for the evolution of dark matter and gas are relatively smooth, many evolutionary tracks of individual satellites exhibit multiple sharp drops and increases, the latter of which may depict instances of reaccretion of gas or dark matter. In the case for all gravitationally bound particles, they may alternatively be caused due to different identifications of the subhalo surrounding the satellite between snapshots by the {\sc subfind} algorithm (see also \protect\citealt{Engler2021a} for a discussion on halo finder limitations).}
    \label{fig:massEvo}
\end{figure*}

\begin{figure*}
    \centering
    \includegraphics[width=.245\textwidth]{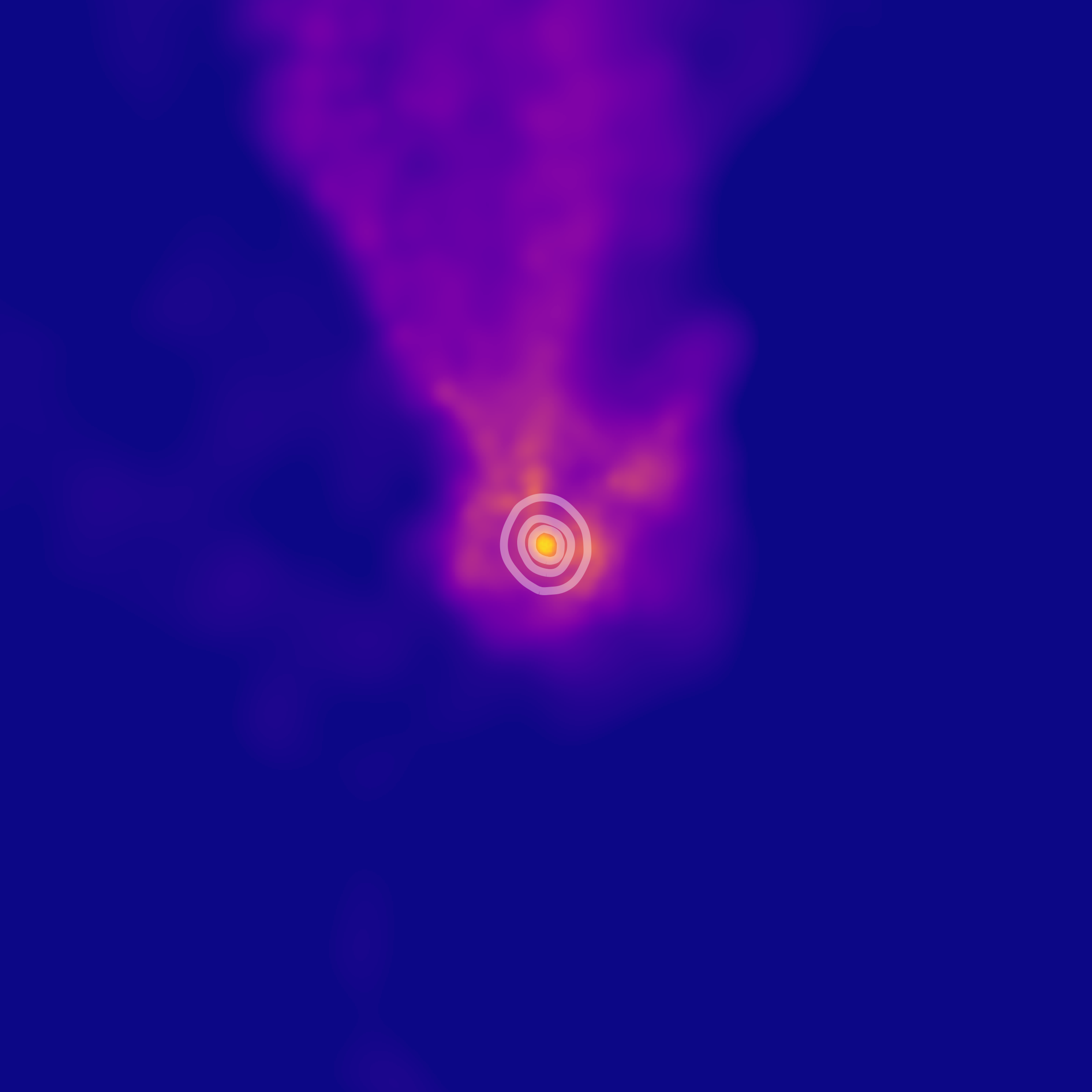}
    \includegraphics[width=.245\textwidth]{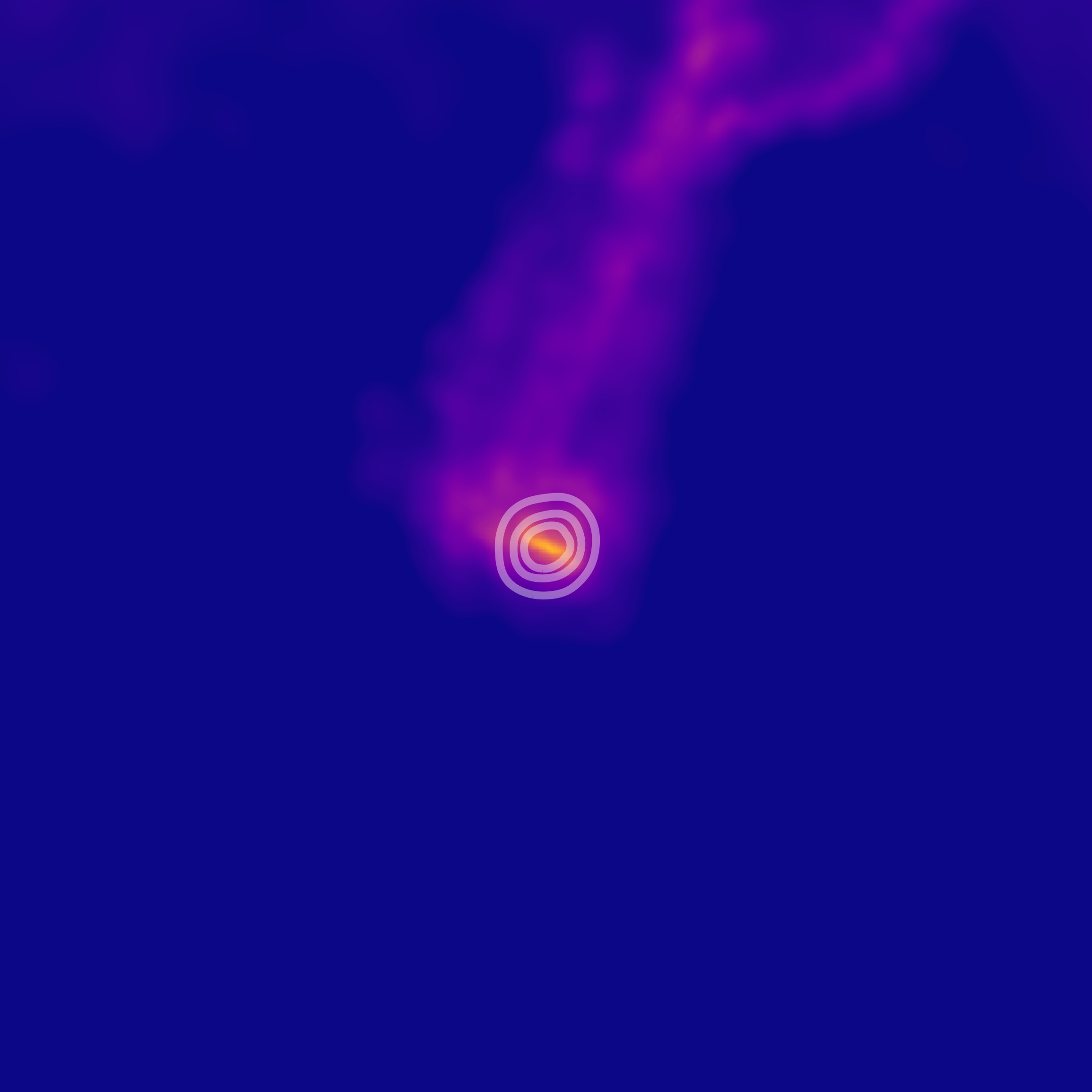}
    \includegraphics[width=.245\textwidth]{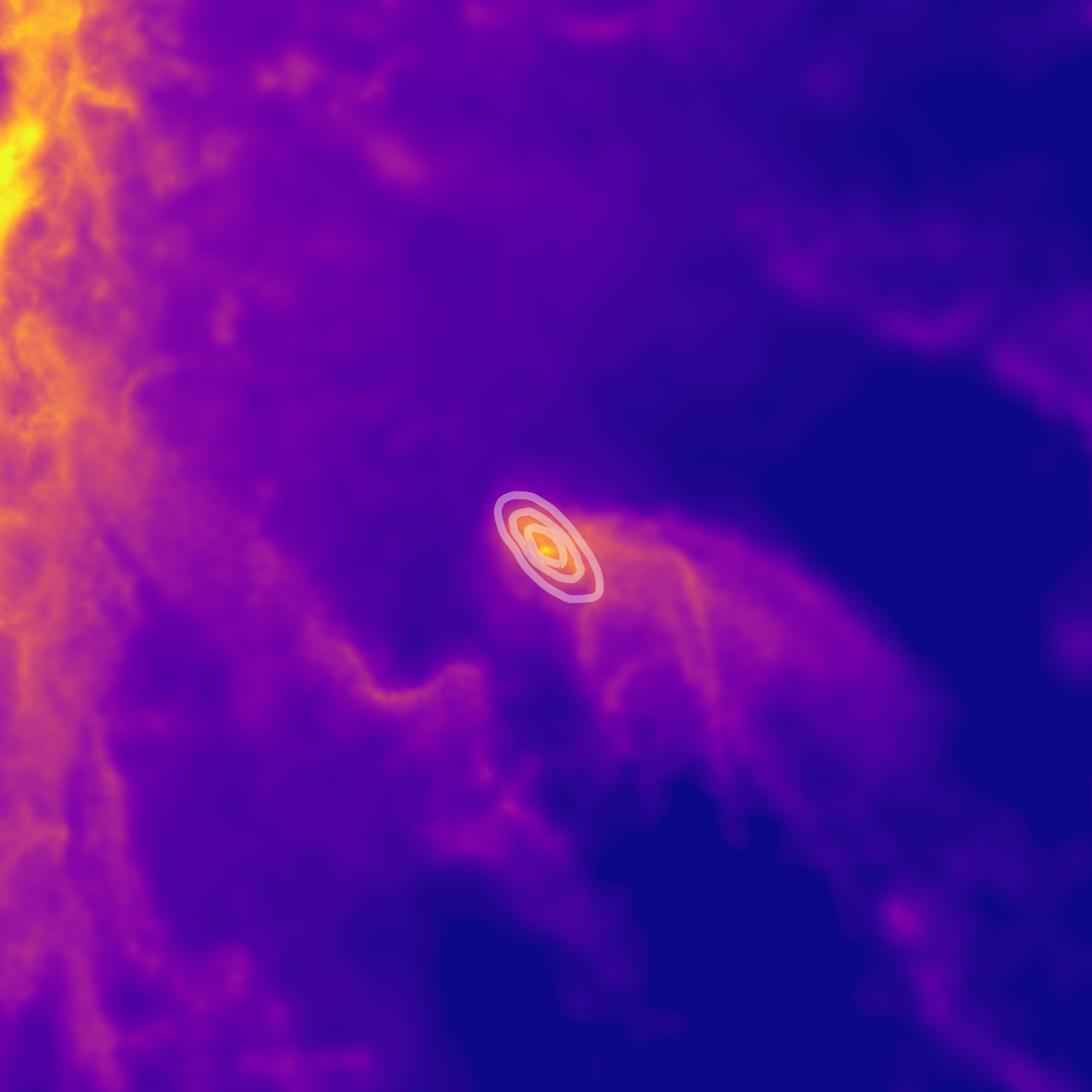}
    \includegraphics[width=.245\textwidth]{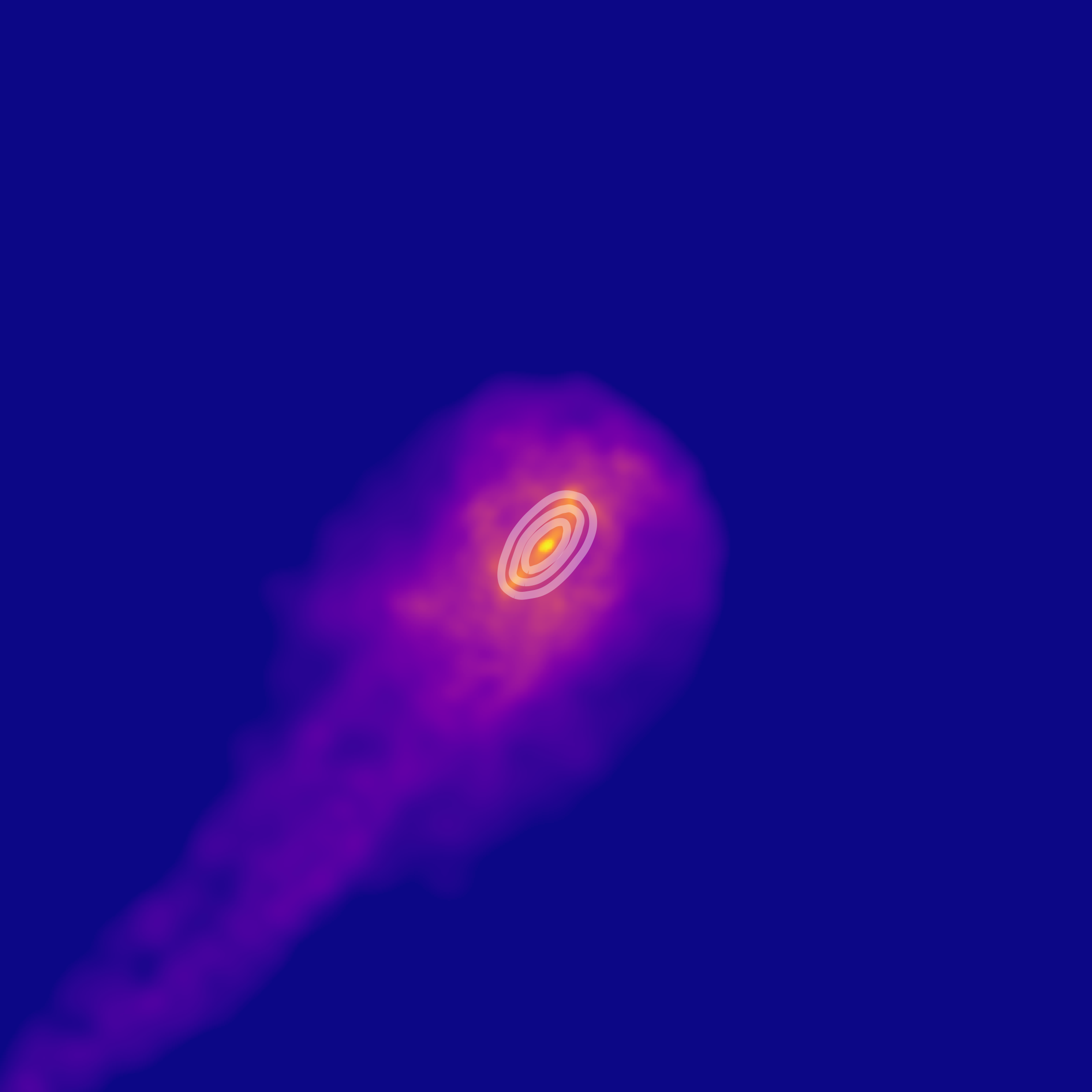}
    \caption{Images of gas column density for four massive satellites around TNG50 MW/M31-like hosts that exhibit, at the present-day, a clear jellyfish-like morphology, i.e.~are undergoing ram pressure stripping. These have $M_* > 10^{8.3}~\MSUN$, similar to the Magellanic Clouds around the MW or M32, NGC~205, and Triangulum around M31, and have been chosen randomly among about $70-90$ TNG50 MW/M31-like galaxies that we have identified to have clear signatures of ram-pressure stripping. The contours in the centre of the images trace the stellar column density of the main body of the satellite galaxy. Their jellyfish status was further confirmed by the Zooniverse citizen science project Cosmological Jellyfish \citep{Zinger2023}.}
    \label{fig:jellyfish}
\end{figure*}

\subsection{Evolution of mass components after infall}
\label{sec:chapSatGas_massEvo_postInf}

Fig.~\ref{fig:massEvo} illustrates the evolution of the satellite mass components as a function of their time since infall. This may shed some light on the physical processes determining the ultimate quenched fractions and star formation histories of the previous sections. Barring stellar mass, the net loss of all matter components that galaxies undergo after they become satellites of a MW/M31-like host is substantial.

Dark matter mass (black curves) decreases with time due to tidal stripping inside the gravitational potential of their new host and remains the dominant mass component for all satellites until the present day -- interestingly, even after $10 - 12~{\rm Gyr}$ of active tidal stripping. 
%
The stellar mass component of satellites (orange curves) barely evolves after accretion, at the net of star formation and tidal stripping, consistent with previous results on the galaxy-halo connection of IllustrisTNG satellites across hosts of $M_{\rm 200c} = 10^{12} - 10^{15.2}~\MSUN$ \citep{Engler2021a}. While there might still be a slight increase in stellar mass during the first $1-2~{\rm Gyr}$ after infall, particularly for more massive satellites, we know from the previous sections that active star formation ceases eventually in most cases. Still, most satellites in our sample do not show indications of having been significantly affected by tidal stripping of their stars, yet, despite of having lost large fractions of their dark matter halo. If we were to look at all gravitationally bound stellar mass instead, there would be a slight decrease in stellar mass towards present-day times\footnote{It should be noted that the picture on the evolution of stellar mass in satellites in Fig.~\ref{fig:massEvo} is a somewhat biased view as we only consider the present-day, surviving satellite populations of TNG50 MW/M31-like hosts. The number of all satellites ever accreted is significantly larger by a factor of $4-5$ at a given stellar mass \citep{Engler2021b}. Most satellites that would exhibit a significant amount of stellar mass loss within the main body of their respective galaxy, would be quickly disrupted at this point \citep{Bahe2019} and would thus not be included in the satellite sample depicted in Fig.~\ref{fig:massEvo}.}.

More directly related to quenching and the results of the previous sections is the rate of gas mass loss, which is similar to the rate of dark matter loss for the most massive satellites (top panels). Lower-mass galaxies, on the other hand, lose their gas much faster (from top to bottom). In fact, both the gas mass throughout the satellite haloes (purple curves) and the gas within their inner galaxy regions (green curves) are reduced simultaneously and rapidly. Consistent with our findings in \S\ref{sec:gasContent} on the very-low if not vanishing gas fractions, only the most massive satellites are able to retain their gas until the present day as they are more resistant to environmental effects (ram pressure or tidal stripping) due to their deeper gravitational potential wells. The simultaneous removal of gas across satellites' haloes and inner regions is somewhat surprising and a novel insight from the IllustrisTNG simulations \citep[see also][]{Rohr2023}: instead, previous (semi-analytic) models have generally assumed that the cold gas in satellite disks can only be removed {\it after} the hot halo of the satellite has been stripped \citep[e.g.][]{Font2008, Kang2008, Weinmann2009, Kimm2011, Starkenburg2012, Xie2020, Jiang2021}.

\subsection{Jellyfish galaxies around MW/M31-like hosts}
\label{sec:jellyfish}
 
Whether the removal of gas in MW/M31-like satellites is driven by ram pressure or tidal stripping cannot be directly determined from the analysis described so far. Considering how quickly low-mass dwarfs lose their gas during or after infall into a MW/M31-like host and that the rate of gas mass loss is similar for differently-bound gas (Fig.~\ref{fig:massEvo}), it is fair to speculate that ram pressure stripping plays a dominant role in the evolution of MW/M31-like satellites in TNG50 and probably ultimately leads to their quenching \citep[see also][]{Fillingham2015, Fillingham2016, Simpson2018, Akins2021, Samuel2022}. 

These considerations are not at odds with the fact that, in truth, only a few percent ($70-90$ among $\sim$1200) of the TNG50 MW/M31-like satellites with $M_* \geq 5 \times 10^6~\MSUN$ appear as jellyfish galaxies at $z=0$ (see Fig.~\ref{fig:jellyfish}). We have determined this estimate by visually inspecting gas column density maps following the approach of \cite{Yun2019} and of \cite{Zinger2023} in the Zooniverse citizen science project Cosmological Jellyfish\footnote{\url{https://www.zooniverse.org/projects/apillepich/cosmological-jellyfish}}. As seen in the previous sections, the majority of satellites have little to no gas in the first place by $z=0$. Thus, they are not appropriate candidates for objects that show some signs of undergoing ram pressure stripping\footnote{In particular, two inspectors have visually evaluated 351 of our 1237 satellites at $z=0$ that still hold some amount of gas (namely, $M_{\rm gas}/M_* > 0.01$): the requirement for some gas to still be present already reduces the fraction of jellyfish candidates to less than a third of all satellites that could be seen around MW/M31-like galaxies above a certain stellar mass or magnitude limit. These candidates span the entire host mass range of the 198 MW/M31-like hosts.}.

However, given that such satellites did have larger amounts of gas at and prior to infall, it is to be expected that some or many of them may have been jellyfish in the past while they were in the process of losing their gas reservoirs. We confirm this claim by following the main progenitor branches of a subset of our 1237~fiducial satellites to investigate if any of them have been identified as jellyfish according to the Cosmological Jellyfish Zooniverse project, which covers all 33~snapshots of TNG50 between $z=0-0.5$ and the specific snapshots at $z=0.7, 1, 1.5, 2$. Note that this only covers satellites that have or had a stellar mass of $M_* > 10^{8.3}~\MSUN$ at the time of inspection. An additional few tens of our present-day massive satellites at $z=0$ are currently or have been jellyfish galaxies in the past according to the Cosmological Jellyfish Zooniverse project. These qualitative findings -- which will be followed up further in dedicated studies -- further indicate that ram pressure stripping is an important mechanism that can affect satellites of MW/M31-like galaxies, even in the past when the hosts were less massive and their circumgalactic medium less dense.

\subsection{Additional remarks and looking ahead}
\label{sec:implications}

The consistency between the outcome of TNG50 and the inferences on the observed systems studied in this paper is encouraging. However, to pin down the specific elements and choices of the underlying IllustrisTNG galaxy formation model that are responsible for this agreement is not straightforward and is beyond the scope of this paper. In the following, we nevertheless offer some considerations that might be useful to better contextualise our results.

The reason for which it is not apparent to isolate successful model elements is manifold. On the one hand, galaxy formation models implemented in cosmological hydrodynamical simulations such as IllustrisTNG, EAGLE, AURIGA and FIRE-2 (just to cite a few: see \S\ref{sec:intro} and \citealt{Pillepich2023}), comprise many physical processes, which act across spatial and time scales and which couple in highly non-linear manners. A possible and common way forward to do so is by inspecting the same observable -- e.g.~the quenched fractions (Fig.~\ref{fig:fquench}) or the SFHs (Figs.~\ref{fig:sfh_MWlike} and \ref{fig:sfh_M31like}) of MW/M31-like satellites -- and varying the underlying model choices, as typically done when models are motivated and presented in first place \citep[e.g.][for IllustrisTNG]{Pillepich2018a, Weinberger2017}. The difficulty of the physical interpretation with this approach, however, lies in the fact that, firstly, such model variations are usually performed on smaller simulations and thus smaller galaxy sample sizes and, secondly, often such model variations lead to galaxy populations that are simply not realistic at all -- and hence may lead to physical conclusions that do not apply in the real Universe or a realistic implementation thereof.

More importantly, in fact, we have shown that, at least according to TNG50 and the other IllustrisTNG simulations, the galaxy-to-galaxy diversity is staggering also at {\it fixed} model (and at fixed galaxy mass and redshift) and even more so for the stellar populations and star formation properties of dwarf galaxies \citep{Joshi2021}. We have de facto quantified the enormous dwarf-to-dwarf diversity both in this paper (Figs.~\ref{fig:fquench}, \ref{fig:mHI_vs_distHost}, \ref{fig:sfh_MWlike}, \ref{fig:sfh_M31like}, \ref{fig:stellarAssembly_vs_Vmag}) and in previous ones \citep[e.g. ][]{Engler2021b, Joshi2021}, whereby the expected scatter at the dwarf scale is further magnified for the case of satellites, e.g.~by the fact that their hosts are subject to similar galaxy-to-galaxy variations and in turn affect the evolution of their satellite populations differently. Whereas part of such scatter is systematic and we understand it in light of physical and large-scale structure formation processes (see e.g.~the dependencies of quenched and gas fractions uncovered in Figs.~\ref{fig:fquench_hostProps} and \ref{fig:t90_vs_t50} or of the shapes of the star formation histories in Figs.~\ref{fig:sfh_MWlike} and \ref{fig:sfh_M31like} as well as in \citealt{Joshi2021}), some non-negligible residual stochasticity seems to be inherent to galaxy formation and evolution \citep[see e.g.][for discussions on the butterfly effect]{Keller2019, Genel2019,Borrow2022}.

In light of the reflections above, the innovation and edge of this paper and of simulations like TNG50, compared to previous or less competitive ones, lie precisely on having addressed comparisons to observational statements with hundred-(thousand-)strong samples of (satellite) galaxies. All this also implies that comparisons of simulated samples to only two individual galaxies, such as the MW and M31 and their satellite systems, or of individual simulated systems to larger observed samples, would not alone suffice to demonstrate the realism of the underlying galaxy formation and physical models in general. In the case of the MW and M31 and their satellites, this in turn calls for further efforts on both the theory and observation sides. On the one hand, as a numerical community we need to keep progressing simultaneously on both the richness of the physical models and resolution but also on the sizes of the simulated samples, in the wake of the efforts currently championed by programs like TNG50 \citep{Pillepich2019, Nelson2019b}, AURIGA \citep{Grand2017}, ARTEMIS \citep{Font2020} and more recently FIREbox \citep{Feldmann2022}: see fig.~1 and table~1 of \cite{Pillepich2023}. On the other hand, additional efforts are required to extend the way paved by surveys like SAGA-II \citep{Geha2015,Geha2017, Mao2021} and ELVES \citep{Carlsten2021, Carlsten2022} towards even larger and, more importantly, less biased and less incomplete samples of MW-like galaxies and their satellite systems.

\section{Summary \& conclusions}
\label{sec:summary}

Throughout this study, we have analysed the star formation activity of present-day satellites around 198~MW/M31-like hosts predicted by the cosmological magnetohydrodynamical galaxy simulation TNG50. Its combination of cosmological volume and zoom-in-like resolution has allowed us to study statistical samples of both hosts and satellites, and to reliably characterise about 1200 satellites down to stellar masses of $M_* = 5 \times 10^6~{\rm M}_\odot$. In a previous paper \citep{Engler2021b}, we have shown that the satellite abundance and stellar mass and luminosity functions predicted by TNG50 are well compatible with those inferred from observations. Here we have expanded upon that analysis and studied the quenched fractions and stellar mass assembly of TNG50 satellites and compared them to observations in the Local Group, as well as within and beyond the Local Volume. Thanks to the large galaxy samples available with TNG50, we have been able to search for possible correlations predicted by the simulation model between satellite quenched fractions and host and satellite properties. We have explored the satellites' gas content with a specific focus on atomic and molecular hydrogen as the fuel for star formation, and have related their lack of gas to both their location within the host system and their time since infall. Finally, we have quantified the effects of MW/M31-like environments on the stellar assembly times and the cumulative SFHs of satellites as a function of both their own stellar mass and their time since infall according to TNG50.

The results of this paper are summarised as follows.

\begin{itemize}
    \item The quenched fractions of satellites around MW/M31-like hosts in TNG50 is a strong function of their own stellar mass: below $M_{*}\sim 10^7~{\rm M}_\odot$, most satellite galaxies have ceased to form stars (Fig.~\ref{fig:fquench}), with quenched fractions reaching up to $95-100$~per~cent for galaxies within $300~{\rm kpc}$ of their host and 90~per~cent for satellites outside the virial radius at distances of $300-600~{\rm kpc}$. For the latter population, i.e.~for more distant satellites, quenched fractions are consistently lower across the studied satellite mass range (Fig.~\ref{fig:fquench_hostProps}).\\
    
    \item The average, i.e.~stacked, quenched fractions of TNG50 MW/M31-like satellites are in agreement with those of the observed satellites around the MW and M31 of \cite{Wetzel2015}. This holds whether TNG50 galaxies are defined quenched based on their distance from the star-forming main sequence or on a gas fraction threshold. However, the variations in quenched fractions across the many individual TNG50 MW/M31-like hosts are very large (Fig.~\ref{fig:fquench}).\\
    
    \item While the quenched fractions of satellites around Local Volume hosts from the ELVES survey \citep{Carlsten2022} lie between those of the LG \citep{Wetzel2015} and SAGA-II \citep{Mao2021}, they are overall consistent with TNG50 after accounting for their host and satellite selection criteria. On the other hand, the satellites of the SAGA-II survey exhibit consistently lower quenched fractions than the LG, ELVES, and TNG50 satellites across all stellar masses. This inconsistency remains in place for satellites with $M_{*}\sim 10^{8-9}~{\rm M}_\odot$ even when we apply the selection criteria of SAGA-II observations for both hosts and satellites to TNG50, match observed and simulated satellites based on their $r$-band magnitudes, and account for the observationally-derived error bars. Differently than for the ARTEMIS simulations \citep{Font2022}, accounting for the possible inherent incompleteness due to the observational surface brightness detection limits does not fully resolve the tension between SAGA-II and TNG50 results, if not at the lowest mass end studied ($M_*\sim 5 \times 10^6 - 10^8~\MSUN$; Fig.\ref{fig:fquench}).\\
    
    \item According to TNG50, more massive hosts and those harbouring satellite populations with an earlier average accretion time exhibit systematically larger satellite quenched fractions. Moreover, satellites at larger distances to their host form stars more actively (Fig.~\ref{fig:fquench_hostProps}). Even though TNG50 reproduces three MW/M31-like pairs in LG-like configurations, these are not sufficient to quantitatively assess the impact of LG-like environments on the quenched fractions of satellites at various distances from their hosts, as the qualitative conclusions change depending on the definition of LG analogues.\\
    
    \item The gas content and HI masses of TNG50 satellites around MW/M31-like systems are a strong function of host-centric distance, decreasing the closer they get to their host galaxy (Fig.~\ref{fig:mHI_vs_distHost}). While most satellites outside the virial radius of their host still contain atomic hydrogen, their average HI content drops drastically around and below distances of $300-350~{\rm kpc}$. Within $300~{\rm kpc}$, most TNG50 satellites contain no gas whatsoever (i.e.~too little gas to resolve numerically and overall lower than one per cent of the satellite's stellar mass). This trend is qualitatively consistent with the behaviour of observed satellites around the actual MW and M31 galaxies \citep{Putman2021}.\\
    
    \item The satellites' gas content is further correlated with their position in phase-space. Satellites in the outer regions of phase-space, i.e.~at larger distances to their host and with larger infalling relative velocities (and to a lesser degree for outgoing velocities) tend to be, on average, gas-richer than those in the inner regions of phase-space. At small host-centric distances and small relative velocities, satellites exhibit the smallest gas fractions (Fig.~\ref{fig:phaseSpace_gasFrac}). On average, the gas-to-stellar mass fractions of those within the virial radius lie below $1-10$~per~cent. This holds for all gas in general and for HI specifically, in both projected and three-dimensional phase-space.\\
    
    \item The satellites' times of first infall are well correlated with their phase-space position and, therefore, with their gas fractions (Fig.~\ref{fig:infall_phaseSpace_gasFracs}). Whereas satellites with recent infall times ($0-4~{\rm Gyr}$ ago) are on average located in the outer regions of phase-space, ancient infallers ($8-12~{\rm Gyr}$ ago) inhabit its central regions. This clear trajectory throughout phase-space is consistent with the findings of previous studies of satellite populations in galaxy clusters \citep{Rhee2017, Pasquali2019, Smith2019}. Furthermore, satellite gas fractions in terms of atomic and molecular hydrogen are correlated with infall time and depict a bimodal distribution. While recent infallers still contain significant amounts of gas, gas fractions drop notably and rapidly for satellites whose first infall occurred more than $2.5~{\rm Gyr}$ ago. These TNG50 satellites mostly contain no gas at all (i.e.~below a few $10^4~\MSUN$).\\
    
    \item Satellite galaxies around TNG50 MW/M31-like hosts exhibit a significant degree of diversity regarding their stellar assembly times and their individual SFHs. On average, median SFHs become more extended for more massive and brighter satellites, for satellites that experienced a more recent accretion, and -- to a lesser degree -- for satellites around less massive hosts (Figs.~\ref{fig:sfh_MWlike}, ~\ref{fig:sfh_M31like},~\ref{fig:stellarAssembly_vs_Vmag}, and~\ref{fig:t90_vs_t50}).\\
    
    \item TNG50 MW-like and M31-like hosts affect the stellar assembly of their satellite populations differently. At fixed luminosity, satellites around TNG50 M31-like hosts exhibit a larger scatter in their stellar assembly times $\tau_{50}$ and $\tau_{90}$ than those around the less massive MW-like hosts -- either due to the stronger environmental effects in the more massive M31-like systems, or simply because there are more satellites present around more massive hosts (Fig.~\ref{fig:stellarAssembly_vs_Vmag}). At fixed $\tau_{90}$, there is a larger scatter in $\tau_{50}$ for satellites around MW-like hosts than for those around M31-like hosts as properties such as their orbital parameters might have more of an influence on the shape of their SFH in less massive hosts (Fig.~\ref{fig:t90_vs_t50}).\\
    
    \item Finally, TNG50 returns SFHs of individual satellites that are consistent with those of observed dwarfs around the MW and M31 (Figs.~\ref{fig:sfh_MWlike} and~\ref{fig:sfh_M31like}). Overall, the stellar assembly times of the observed MW and M31 satellites are reproduced by the TNG50 model and fall within the relations formed by the numerous TNG50 analogue satellites (Figs.~\ref{fig:stellarAssembly_vs_Vmag} and~\ref{fig:t90_vs_t50}). According to TNG50, MW/M31-like satellites with $M_* \geq 5 \times 10^{6}\MSUN$ quenched $6.9\substack{+2.5\\-3.3}$~Gyr ago ($16^{\rm th}-84^{\rm th}$ percentiles; \S\ref{sec:quenchingtimes}). The infall times of observed MW and M31 satellites that we can infer based on TNG50 SFHs (\S\ref{sec:infalltimes}) are also in agreement with previous works and provide a complementary constraint.\\
\end{itemize}

In conclusion, we have quantitatively investigated the quenched fractions, gas content, and stellar mass assembly of satellite galaxies around MW/M31-like hosts in TNG50. The TNG50 results are consistent with those inferred for the real MW and M31 and for most similar hosts in the local Universe, lending credibility to the underlying galaxy formation model. Moreover, we have shown that, within the virial radius of their host, most satellites rapidly lose their gas reservoirs over the course of a few~Gyr, and subsequently cease to form stars and become quenched. This process is particularly effective for low-mass satellites. Since more massive satellites are more resistant to their environment, they manage to retain their gas for longer and may continue to form stars even up to the present day. MW/M31-like hosts in TNG50 assert environmental effects on their satellites to a similar degree as the observed MW and M31 systems, with a sharp drop in satellite gas content for populations within the virial radius -- for all gas in general, as well as HI and H$_2$ specifically. Thus, the TNG50 model produces a realistic population of satellites in MW/M31-like systems with respect to their star formation activity and gas content. This opens up new opportunities with TNG50: for example, among others, we have shown that the shape of the SFHs of observed MW and M31 satellites can be used to further constrain their infall times.

\section*{Acknowledgements}
CE acknowledges support by the Deutsche Forschungsgemeinschaft (DFG, German Research Foundation) -- Project-ID 138713538 -- SFB~881 (``The Milky Way System'', subproject A01) and through the DFG Project 394551440, and thanks Gerg\"{o} Popping for sharing post-processed hydrogen catalogues, as well as Nicolas Martin for useful discussions and input.
DN acknowledges funding from the DFG through an Emmy Noether Research Group (grant number NE~2441/1-1).
The primary IllustrisTNG simulations were realised with compute time granted by the Gauss Centre for Super-computing (GCS): TNG50 under GCS Large-Scale Project GCS-DWAR (2016; PIs Nelson/Pillepich), and TNG100 and TNG300 under GCS-ILLU (2014; PI Springel) on the GCS share of the supercomputer Hazel Hen at the High Performance Computing Center Stuttgart (HLRS). Additional simulations for this paper were carried out on the Draco and Cobra supercomputers at the Max Planck Computing and Data Facility (MPCDF).

\section*{Data Availability}
Data of the TNG50 simulation series are publicly available from the IllustrisTNG repository: \url{https://www.tng-project.org} and are described in detail by \citet{Nelson2019a}. Data directly referring to content and figures of this publication is available upon request from the corresponding author.



\bibliographystyle{mnras}
\bibliography{MWM31Sats_gasContent.bib} 



\appendix

\section{Satellite surface brightness vs. luminosity}
\label{sec:AppSurfBright}

\begin{figure*}
    \centering
    \includegraphics[width=.95\columnwidth]{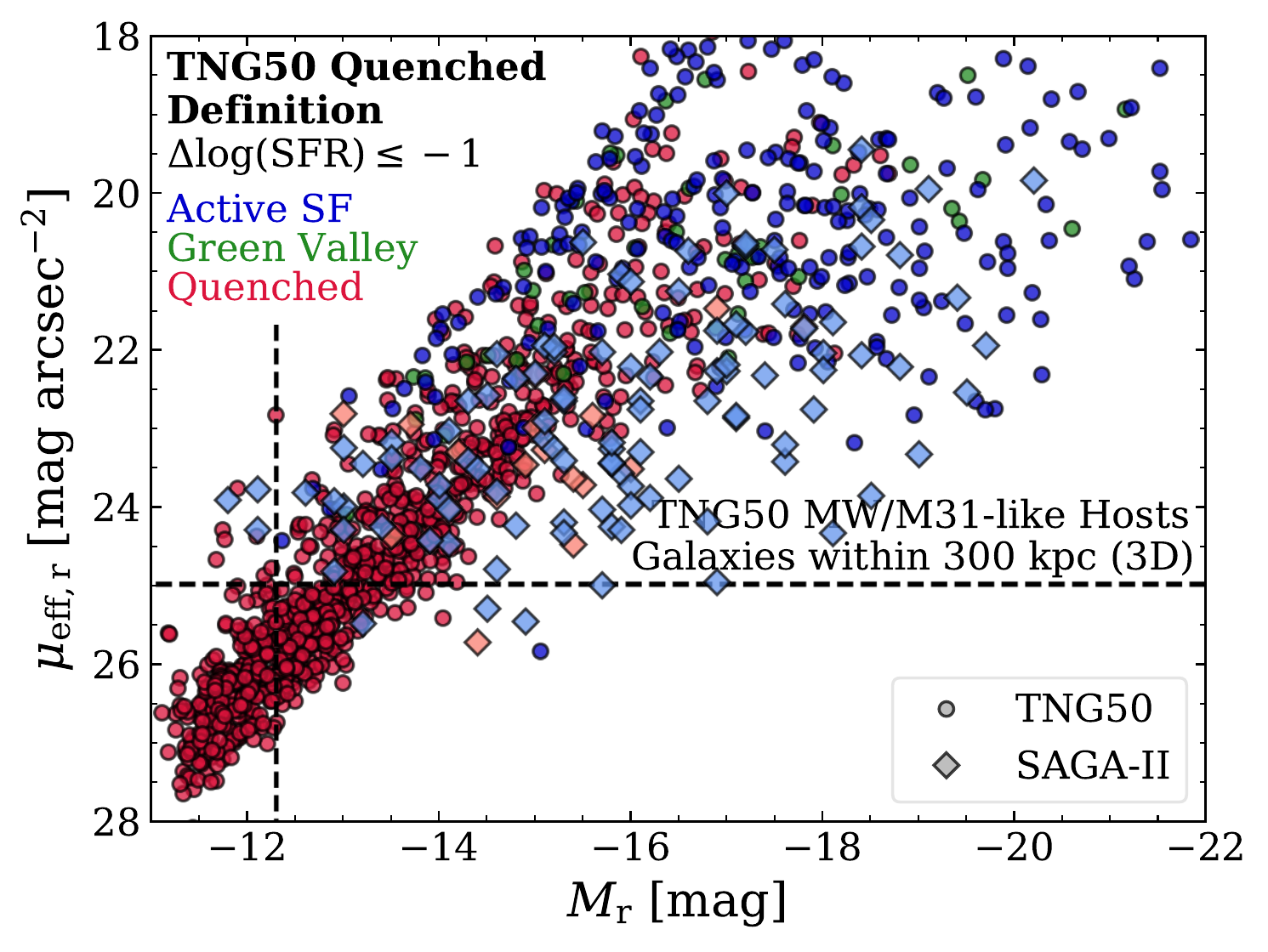}
    \includegraphics[width=.95\columnwidth]{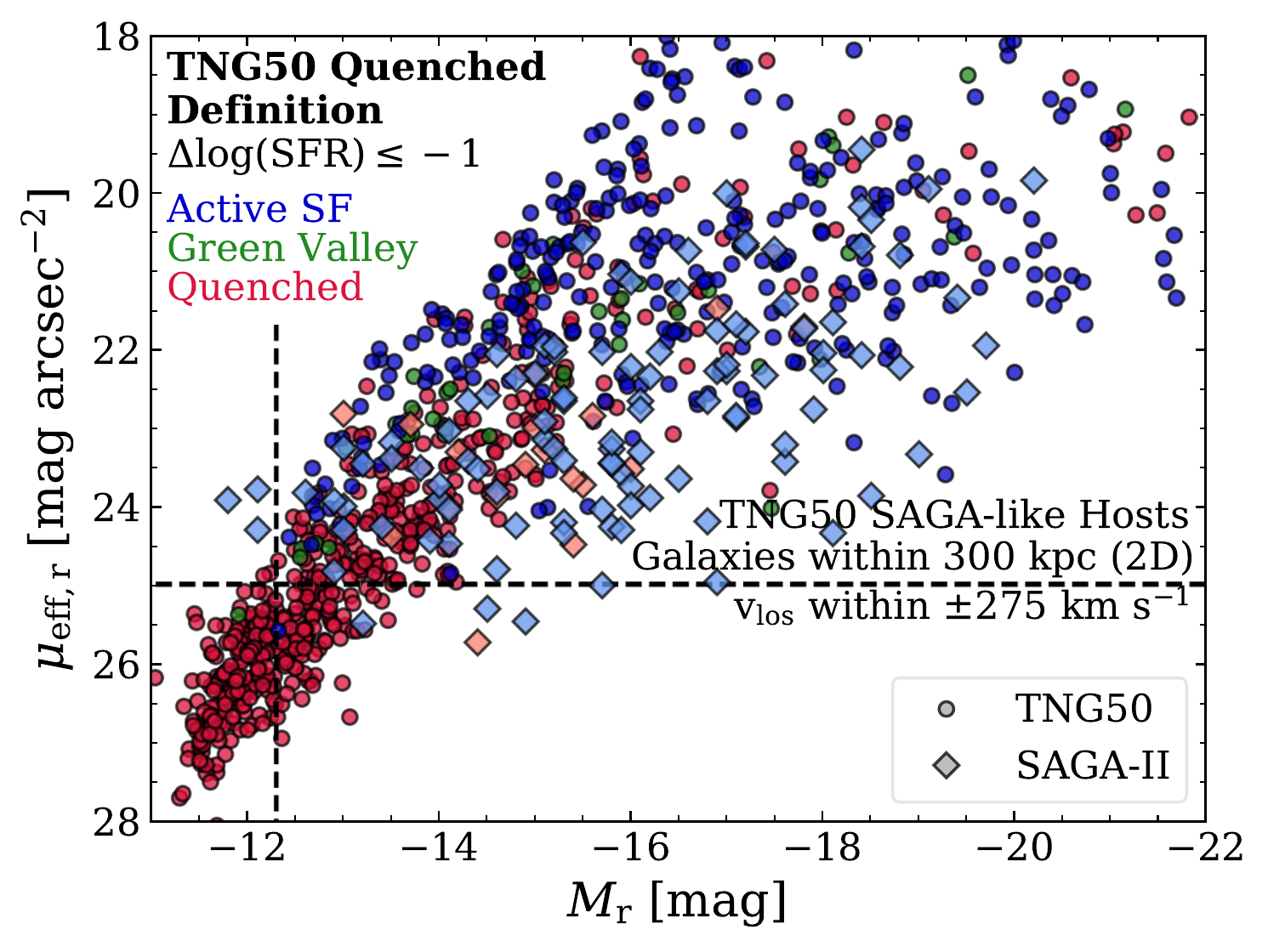}
    \caption{{\bf Effective $\mathbf{r}$-band surface brightness $\mathbf{\mu_{\rm \bf eff, r}}$ as a function absolute $\mathbf{r}$-band magnitude $\mathbf{M_{\rm \bf r}}$ for satellites of TNG50 and SAGA-II colour-coded by star formation activity.} The fiducial TNG50 quenched definition is based on their distance to the SFMS and characterises satellites as star-forming (blue dots), green valley (green dots), and quenched (red dots). The 127~satellites of SAGA-II are defined to be star-forming or quenched according to their H$\alpha$ equivalent width ${\rm EW(H}\alpha) < 2~\text{\r{A}}$ \protect\citep[blue and red diamonds, respectively][]{Mao2021}. The dashed lines denote the observational limitations of SAGA-II: their explicit $r$-band magnitude limit of $-12.3~{\rm mag}$ and their implicit surface brightness limit of $25~{\rm mag~arcsec}^{-2}$ according to \protect\cite{Font2022}. {\it Left panel:} for our sample of 198~TNG50 MW/M31-like hosts employing our fiducial satellite selection, i.e.~all galaxies within $300~{\rm kpc}$~(3D) of their host. {\it Right panel:} for our sample of 108~TNG50 SAGA-like hosts (see \S\ref{sec:hostSelection_SAGAlike} for details) adopting the basic satellite selection criteria of SAGA-II, i.e.~all galaxies within {\it projected} $300~{\rm kpc}$ and with line-of-sight velocities of $\pm 275~{\rm km~s}^{-1}$ of their host galaxy. The sample of matched, SAGA-like satellites we employ in \S\ref{sec:quenchFracs} is drawn from this distribution.}
    \label{fig:surfBright_vs_Mr}
\end{figure*}

\begin{figure}
    \centering
    \includegraphics[width=.95\columnwidth]{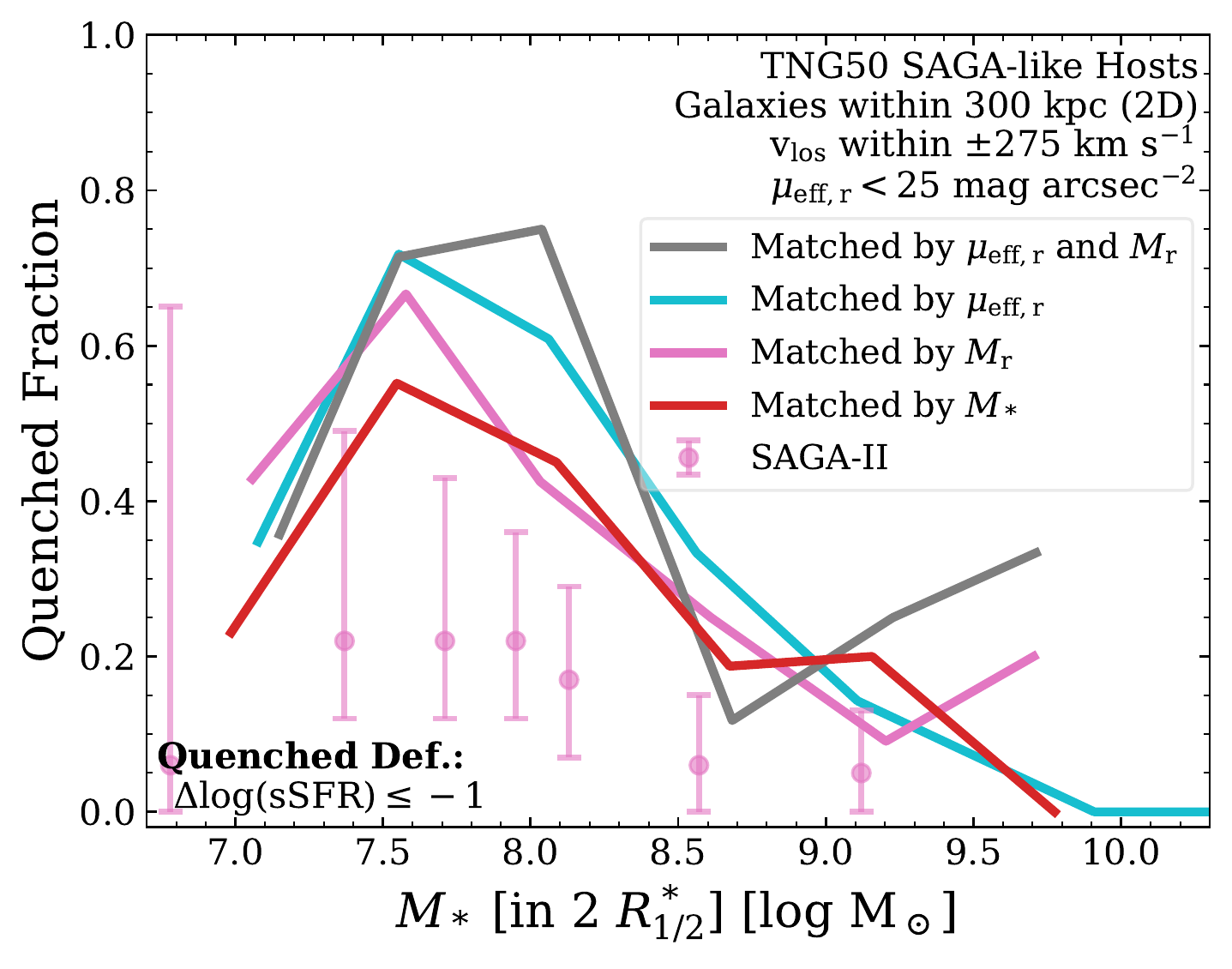}
    \caption{{\bf Satellite quenched fractions for various matched selections to SAGA-II.} All satellite samples are based on the TNG50 SAGA-like hosts and the selection criteria of SAGA-II. Unlike in Fig.~\ref{fig:fquench}, we apply the implicit surface brightness limitation of $\mu_{\rm eff, r} < 25~{\rm mag~arcsec}^{-2}$ {\it before} matching the TNG50 satellites to their observed counterparts based on their stellar mass, surface brightness, absolute $r$-band magnitude, or a combination thereof. The pink circles denote the SAGA-II quenched fractions.}
    \label{fig:fquench_compMatchedSatsSAGAII}
\end{figure}

In \S\ref{sec:quenchFracs}, we discuss the differences in quenched fractions between the satellite populations of our TNG50 MW/M31-like hosts and the observed MW-like hosts of SAGA-II. As SAGA-II is subject to an implicit limitation in surface brightness according to the ARTEMIS simulations \citep[][see their figure~2]{Font2022} that separates star-forming and quenched galaxies, we examine the relation of effective $r$-band surface brightness $\mu_{\rm r}$ and absolute $r$-band magnitude $M_{\rm r}$ of TNG50 satellites in Fig.~\ref{fig:surfBright_vs_Mr}. Its left panel depicts our fiducial sample -- galaxies within $300~{\rm kpc}$ (3D) of our 198~MW/M31-like hosts -- whereas the right panel shows satellites around our 108~TNG50 SAGA-like hosts according to the spatial SAGA-II selection: galaxies within $300~{\rm kpc}$ (2D) with a line-of-sight velocity of $\pm 275~{\rm km~s}^{-1}$ relative to their host galaxy. The latter sample is the basis of the SAGA-like satellites adopted in \S\ref{sec:quenchFracs} with which we find the best agreement with SAGA-II (albeit still with an offset of $\sim$20~percentage points). Fig.~\ref{fig:surfBright_vs_Mr} compares the star formation activity of simulated TNG50 and observed SAGA-II satellites {\it at face value}, i.e.~we adopt our fiducial quenched definition based on the satellites' distance to the SFMS \citep{Pillepich2019} for TNG50 galaxies (blue, green, and red dots for star-forming, green valley, and quenched satellites), while quenched definition of SAGA-II satellites is based on their H$\alpha$ equivalent width \citep[blue and red diamonds, respectively,][]{Mao2021}. The dashed lines indicate the explicit luminosity and implicit surface brightness limit of SAGA-II: $M_{\rm r} < -12.3$ and $\mu_{\rm eff, r} < 25~{\rm mag~arcsec}^{-2}$.

While the TNG50 satellites exhibit the same qualitative trend as SAGA-II -- brighter satellites, both in terms of $M_{\rm r}$ and $\mu_{\rm eff, r}$, are more actively forming stars -- satellites from SAGA-II form a slightly different relation with more star-forming galaxies brighter than $M_{\rm r} \sim -14$ with $\mu_{\rm eff, r} = 23 - 25~{\rm mag~arcsec}^{-2}$. For our fiducial selection in the left panel of Fig.~\ref{fig:surfBright_vs_Mr}, the TNG50 satellite population in this surface brightness range is already dominated by quenched galaxies. Whereas star-forming TNG50 satellites are more common at $\mu_{\rm eff, r} < 23~{\rm mag~arcsec}^{-2}$, there is still a significant amount of quenched satellites, especially at magnitudes fainter than $-17~{\rm mag}$. This discrepancy is somewhat alleviated when employing our sample of TNG50 SAGA-like hosts and adopting the satellite selection criteria of SAGA-II in the right panel of Fig.~\ref{fig:surfBright_vs_Mr}. For satellites brighter than $-15.5~{\rm mag}$, the relation's quenched sequence is less defined and more galaxies with $\mu_{\rm eff, r} = 23 - 25~{\rm mag~arcsec}^{-2}$ are still actively forming stars. However, unlike the SAGA-II satellites, quenched galaxies still dominate the satellite population in this surface brightness range for TNG50. While the smaller host mass range of the TNG50 SAGA-like hosts in combination with the observational selection criteria allow for the presence of more star-forming galaxies compared to our fiducial sample of MW/M31-like hosts -- either due to weaker environmental effects or the inclusion of interloper galaxies from the fore- and background -- they do not fully reproduce the behaviour of the observed SAGA-II satellites.

This is further illustrated in Fig.~\ref{fig:fquench_compMatchedSatsSAGAII} using the median stacked quenched fractions for multiple TNG50 SAGA-like satellite samples that were matched to the observed SAGA-II satellites based on their surface brightness, luminosity, or stellar mass. However, as opposed to the matched satellites in Fig.~\ref{fig:fquench}, we impose the implicit surface brightness limit of $\mu_{\rm eff, r} < 25~{\rm mag~arcsec}^{-2}$ {\it before} matching. Nevertheless, all matched TNG50 samples still exhibit consistently higher quenched fractions than SAGA-II.

\section{Star formation histories at larger distances}
\label{sec:AppSFHs_600kpc}

\begin{figure*}
    \centering
    \includegraphics[width=0.89\textwidth]{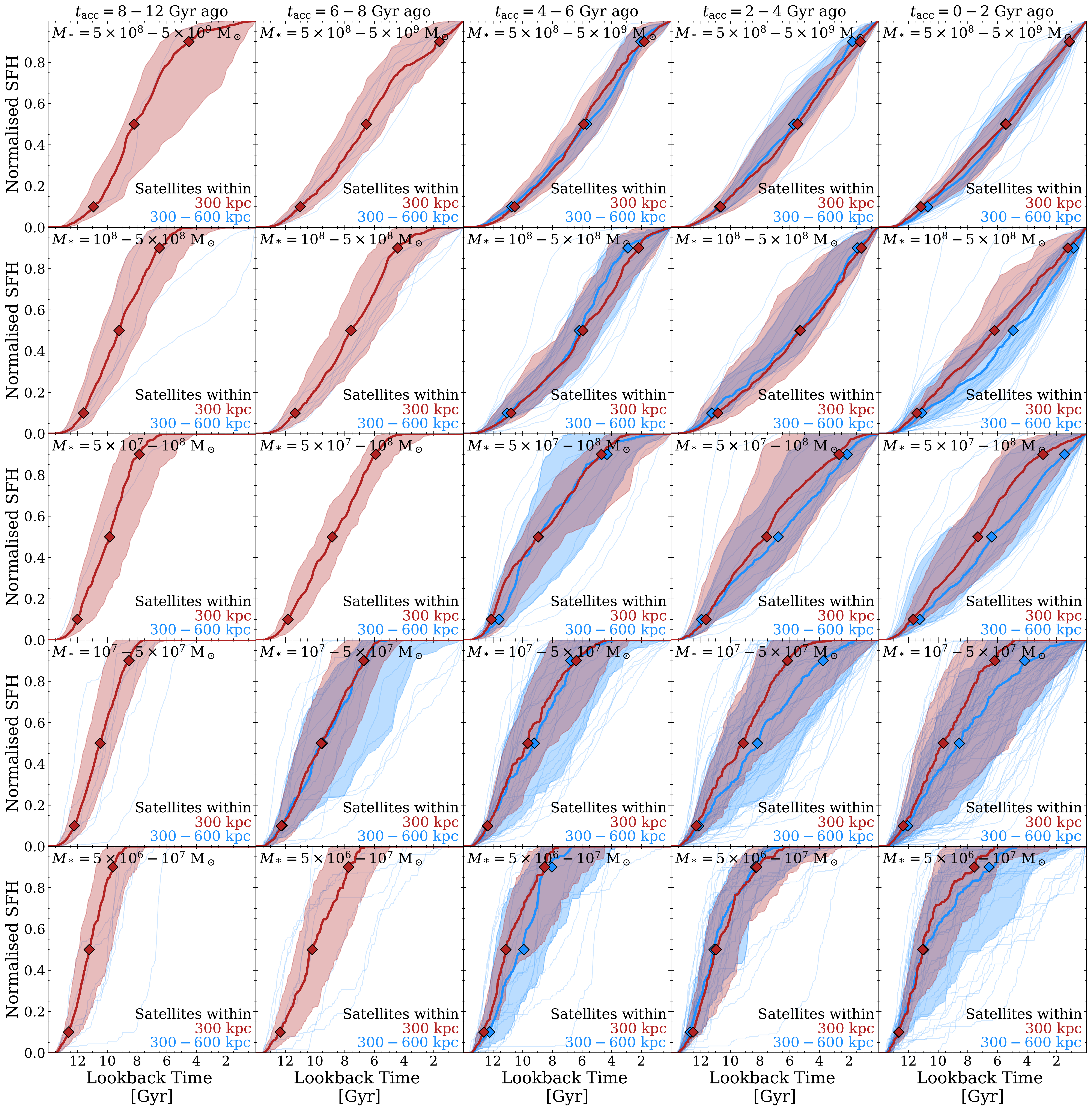}
    \caption{{\bf Cumulative SFHs of satellites around TNG50 MW/M31-like hosts out to $\mathbf{600~{\rm \bf kpc}}$}. As in Figs.~\ref{fig:sfh_MWlike} and~\ref{fig:sfh_M31like}, satellite stellar mass varies across rows (decreasing from top to bottom) whereas the columns depict different bins of accretion time $t_{\rm acc}$ (with early to late infallers from left to right). We illustrate the median SFHs of satellites within $300~{\rm kpc}$ of their host (red curves) as well as the median and individual SFHs for satellites at larger distances of $300-600~{\rm kpc}$ (thick and thin blue curves, respectively). We compute the median and scatter ($16^{\rm th}$ and $84^{\rm th}$ percentiles, shaded areas) for bins containing at least ten satellites. The red and blue diamonds denote the median stellar assembly times $\tau_{10}$, $\tau_{50}$, and $\tau_{90}$.}
    \label{fig:sfh_in600kpc}
\end{figure*}

Fig.~\ref{fig:sfh_in600kpc} extends the analysis of satellite SFHs (see \S\ref{sec:SFHs}) beyond the virial radius by comparing the stellar assembly of satellites within $300~{\rm kpc}$ of TNG50 MW/M31-like hosts (red curves) to those at $300-600~{\rm kpc}$ (blue curves). Thick curves illustrate the median cumulative SFHs, the shaded areas show their scatter ($16^{\rm th}$ to $84^{\rm th}$ percentiles), and the thin blue curves depict the SFHs of individual satellites at $300-600~{\rm kpc}$. Furthermore, Fig.~\ref{fig:sfh_in600kpc} gives the satellites' stellar assembly as a function of stellar mass across rows (decreasing from top to bottom) and of their time since accretion $t_{\rm acc}$ across columns (with early to late infallers from left to right).

Whereas there is no particular difference between the most massive satellites within $300~{\rm kpc}$ and at $300-600~{\rm kpc}$ (top row) due to their inherent resistance to environmental effects, less massive satellites at larger distances mostly exhibit more extended cumulative SFHs than their counterparts within the virial radius. The median SFHs and stellar assembly times reflect this in most bins; some others merely display an increased scatter or individual SFHs (e.g.~for early infallers) that are shifted towards a later assembly compared to the median SFHs within $300~{\rm kpc}$. This is consistent with the results of \cite{GarrisonKimmel2019} who studied the SFHs of satellites around isolated MW- and LG-like hosts in the FIRE-2 simulations.


\bsp	
\label{lastpage}
\end{document}